\documentclass[13pt,a4paper,nofootinbib,superscriptaddress]{revtex4-2}

\pdfoutput=1
\usepackage[utf8]{inputenc}
\usepackage{graphicx,bm,bbm}
\usepackage{slashed,verbatim}
\usepackage{amssymb,graphicx,epstopdf}
\usepackage{subfigure}
\usepackage{amsmath,mathtools,mathrsfs}
\usepackage{soul}
\usepackage[normalem]{ulem}
\usepackage{cancel}
\usepackage{xcolor}
 \usepackage[colorlinks=true,linktocpage=true,linkcolor=blue,citecolor=blue]{hyperref}
\allowdisplaybreaks

\allowdisplaybreaks

\def\be{\begin{equation}}
\def\ee{\end{equation}}
\def\bea {\begin{eqnarray}}
\def\eea {\end{eqnarray}}
\def\bc {\begin{center}}
\def\ec {\end{center}}
\def\nn {\nonumber}
\def\a{\alpha}
\def\b{\beta}

\def\mn {\mu\nu}
\def\({\left(}
\def\){\right)}
\def\[{\left[}
\def\]{\right]}

\DeclareGraphicsExtensions{.jpg,.pdf,.eps}

\usepackage{scalerel}

\newcommand*{\paral}{\stretchrel*{\parallel}{\perp}}

\def\Th{\hat{T}}
\def\psib{\overline{\psi}}
\def\kpar{k_{\paral}}
\def\slkpar{k_{\paral}\!\!\!\!\!\!/}
\def\qpar{q_{\paral}}
\def\ppar{p_{\paral}}
\def\vpar{v_{\paral}}
\def\vp{\varpi}

\begin{document}
\title{Collective modes of  gluons in an anisotropic thermo-magnetic medium  }

\author{Bithika Karmakar}
\email{karmakarbithika.93@gmail.com}
\affiliation{Institute of Physics Belgrade, University of Belgrade, Belgrade, Serbia}
\author{Ritesh Ghosh}
\email{ritesh.ghosh@saha.ac.in}
\affiliation{Saha Institute of Nuclear Physics, 1/AF Bidhannagar, Kolkata - 700064, India}
\affiliation{Homi Bhabha National Institute, 2nd floor, BARC Training School Complex, Anushaktinagar, Mumbai, Maharashtra 400094, India}
\author{Arghya Mukherjee}
\email{arbp.phy@gmail.com}
\affiliation{School of Physical Sciences, National Institute of Science Education and Research, HBNI, Jatni, Khurda 752050, India}

\begin{abstract}
We  study the collective modes of  gluons in an anisotropic thermal  medium in the presence of a constant background magnetic field using the hard-thermal loop (HTL) perturbation theory. The  momentum space anisotropy of the medium has been incorporated through the generalized $`$Romatschke-Strickland' form of the distribution function,  whereas, the magnetic modification arising from the quark loop contribution has been taken into account in the lowest Landau level approximation. We consider two special cases: (i) a spheroidal anisotropy with the anisotropy vector orthogonal to the external magnetic field and (ii) an ellipsoidal anisotropy with two mutually orthogonal vectors describing aniostropies along and orthogonal to the field direction. The general structure of the polarization tensor in both cases are equivalent and consists of six independent basis tensors. We find that the introduction of momentum anisotropy ingrains azimuthal angular dependence in  the thermo-magnetic collective modes. Our study suggests that the presence of a strong  background magnetic field can significantly reduce the growth rate of the unstable modes which may have important implications in the equilibration of magnetized quark-gluon plasma. 

\end{abstract}

\maketitle 

\section{Introduction}
The incredible success of hydrodynamical approaches  in describing the enormous data of heavy ion collision experiments \cite{Busza:2018rrf,Pasechnik:2016wkt}, including the nuclear supression factor, radial flow and elliptic flow measurements provides  compelling evidence  to believe that the produced hot and dense matter   thermalizes within about 0.6 fm/c after the initial impact~\cite{Heinz:2004pj,Huovinen:2001cy,Hirano:2002ds}. However, in the seminal work by Baier, Mueller, Schiff and Son in Ref.~\cite{Baier:2000sb}, the thermalization time via scattering processes in the weak-coupling limit has been estimated theoretically to be 2.5 fm/c or above. Recent studies \cite{Berges:2020fwq,Epelbaum:2013ekf,Berges:2013fga} in weak coupling limit have improved our understanding of quark-gluon-plasma(QGP) equilibration to a great extent. On the other hand, several attempts also has been  made to study the thermalization at strong coupling limit within AdS/CFT formulation~\cite{Strickland:2013uga,Chesler:2008hg,Chesler:2009cy,Heller:2011ju,Casalderrey-Solana:2011dxg}. The modern formulations of relativistic fluid dynamics  suggests that neither local near-equilibrium nor near-isotropy is required in order to have a successful hydrodynamical  description of the
experimental results \cite{Romatschke:2016hle}. Several efforts have been  made over the years in the development of relativistic viscous hydrodynamics \cite{Romatschke:2017ejr} which systematically  incorporates the dissipative effects~\cite{Florkowski:2017olj,Jaiswal:2016hex,Jeon:2015dfa,Kovtun:2012rj}. The viscous hydrodynamics concludes that at time $\sim$ 2 fm/c, the QGP created in ultra relativistic heavy-ion collisions (URHIC) has different longitudinal and transverse pressures~\cite{Strickland:2013uga}. This occurs due to the rapid expansion of the QCD matter along the longitudinal direction (beam direction) which gives rise to a large local rest frame momentum space anisotropy ~\cite{Strickland:2013uga,Mandal:2013jkd,Romatschke:2003ms} in the $p_T-p_L$ plane. This anisotropic momentum distribution can cause plasma instabilities in the system which  contribute in the thermalization and isotropization process of the QCD plasma ~\cite{Arnold:2003rq,Mrowczynski:1988dz,Mrowczynski:1993qm,Mrowczynski:2000ed}. It is found that the exponential growth of the unstable modes  plays an important role in the dynamics of the system in weak coupling limit~\cite{Randrup:2003cw}. In the hydrodynamic side, the anisotropic hydrodynamics (aHydro) framework  is formulated to efficiently take into account the large momentum space anisotropy of the system \cite{Strickland:2014pga,Alqahtani:2017mhy}. On the other hand, the hard-thermal-loop perturbation theory \cite{Ghiglieri:2020dpq,Su:2012iy} has been employed~\cite{Romatschke:2003ms,Nopoush:2017zbu} to systematically study the properties of anisotropic QCD plasma. Typically,  one uses a specific distribution function of light quarks and gluons which is widely known as the  $`$Romatschke-Strickland' (RS) form~\cite{Romatschke:2003ms,Romatschke:2004jh}. There has been a concerted effort to study the effect of the  momentum-space anisotropies on the heavy-quark potential~\cite{Nopoush:2017zbu,Dumitru:2007hy,Burnier:2009yu}, bottomonia suppression~\cite{Strickland:2011mw,Strickland:2011aa,Krouppa:2015yoa}, photon and dilepton production rates~\cite{Schenke:2006yp,Bhattacharya:2015ada}, wake potential~\cite{Mandal:2013jla} and so on. The generalized RS form of the distribution function \cite{Tinti:2013vba} that takes into account  the azimuthal momentum-space anisotropy has  been recently investigated in Refs.~\cite{Kasmaei:2018yrr,Ghosh:2020sng,Carrington:2021bnk}.  

On the other hand, the production of  strong  magnetic fields~\cite{Kharzeev:2007jp,Skokov:2009qp} at early stages of the non-central heavy-ion collisions has triggered enormous research interest in the theoretical, phenomenological and experimental understanding of the strongly interacting matter under extreme conditions \cite{Huang:2015oca,Miransky:2015ava}.  The time dependence of the produced  magnetic field   has remained a subject of debate for a long period of time in the heavy-ion collision community~\cite{McLerran:2013hla,Roy:2017yvg,Huang:2015oca}. At the early stages of the collision, the system is dominated  mainly by the gluons. Subsequently, a large number of quarks and anti-quarks are produced and the system evolves towards the equilibrium. Therefore, the system is believed to be much less conducting in the early times. Considering the  Pb+Pb collision at $\sqrt{s}=2.76$ TeV, it is found in Ref.~\cite{Roy:2017yvg} that for an insulating medium, a magnetic field of strength $\sim$ 100 $m_\pi^2$  rapidly decays to a very low value  within around 0.1 fm/c after the initial impact \cite{Huang:2015oca}. The rapid decrease in the field strength follows the $1/t^3$ behavior. However, the electrical conductivity of the medium significantly influences the  time evolution of the electromagnetic fields in the late stage when the system reaches near the equilibrium state. It is shown in the Refs.~\cite{Tuchin:2013apa,Tuchin:2013ie} that the electrical conductivity of the medium can resist the decay of the magnetic field, at least to some extent. Intense research works have been performed to study the properties of the QCD matter in presence of such strong magnetic background which resulted in several interesting findings like chiral magnetic effect~\cite{Fukushima:2008xe,Kharzeev:2007jp}, magnetic catalysis~\cite{Lee:1997zj}, inverse magnetic catalysis~\cite{Bali:2011qj,Ayala:2014iba}, non-trivial magnetic modifications of chiral symmetry broken/restored phases~\cite{Andersen:2012dz,Avancini:2016fgq}, photon and dilepton production rate~\cite{Wang:2020dsr,Tuchin:2013bda,Bandyopadhyay:2016fyd,Das:2021fma,Ghosh:2018xhh,Hattori:2020htm}, thermodynamic properties~\cite{Bali:2011qj,Rath:2017fdv,Karmakar:2019tdp,Bandyopadhyay:2017cle}, heavy quark potential~\cite{Singh:2017nfa}, transport coefficients~\cite{Kurian:2018qwb,Kurian:2018dbn} and so on. 

The production of  strong magnetic field at early stages of collision naturally motivates one to investigate the magnetic field effects on anisotropic QGP. In presence of external magnetic field ( with intensity $B$ ), one can define a  hierarchy of energy scales  as $\sqrt{|eB|}\gg T\gg g_sT$ which essentially determines the regime of validity of  the strong magnetic field approximation. Here $e$ denotes the electric charge of proton  and $g_s$ is the  strong coupling constant.  In this regime, the quarks occupy only the lowest Landau level and the dynamics becomes 1+1 dimensional. 
 In this article, we restrict ourselves to the lowest Landau level approximation and investigate the gluon collective modes in presence of anisotropic momentum distribution.  For this purpose, the one loop gluon self energy is obtained in the  HTL approximation using the real time formalism of thermal field theory. We note here that the general structure of the polarization tensor plays an important role in the determination of the effective propagator and the collective modes.  The thermo-magnetic collective modes has  been studied recently in Refs.~\cite{Hattori:2017xoo,Karmakar:2018aig}. The direction of the  external magnetic field brings in an  anisotropy in the system and naturally breaks the spherical symmetry. It also appears among the available four vectors that has to be taken into account for the construction of the general structure.  The situation is  similar to the spheroidal momentum space anisotropy. Thus,  it is interesting to compare the two scenarios: one is the anisotropy due to the background field  and the  other is the anisotropy that arises due to the modeling of the non-equilibrium distribution function from the equilibrium distribution by suitable stretching or squeezing.  In the present study we systematically address this issue.
Throughout the article, we use the following convention:  $\eta_{\paral}^{\mu\nu}={\rm diag}(1,0,0,-1)$ and $\eta_{\perp}^{\mu\nu}={\rm diag}(0,-1,-1,0)$ with $\eta^{\mu\nu}=\eta_{\paral}^{\mu\nu}+\eta_{\perp}^{\mu\nu}$  where the Lorentz indices $\{\mu, \nu\}\in\{0,1,2,3\}$. For a generic four vector $a^\mu$, we define $a_{\paral}^\mu=(a^0,0,0,a^3)=(a_0,0,0,a_z)$ and $a_\perp^\mu=(0,a^1,a^2,0)=(0,a_x,a_y,0)$. The corresponding scalar products  are defined as   $(a_{\paral}\cdot b_{\paral})=a^0b^0-a^3b^3$ and  $(a_\perp\cdot b_\perp)=-a^1b^1-a^2b^2$.

\section{Formalism}
In this section we obtain the one loop  gluon self energy in presence of anisotropic thermo-magnetic medium within HTL approximation. For this purpose we follow the  real-time Schwinger-Keldysh formalism \cite{Dumitru:2009fy,Carrington:1997sq,Carrington:1998jj,Mrowczynski:2000ed,Mrowczynski:2016etf} based on contour Green's functions which is applicable for non-equilibrium field theories.  The basic formalism to obtain the retarded, advanced and the Feynman  self-energies is  reviewed in \cite{Mrowczynski:2016etf,Nopoush:2017zbu} in a self-contained manner. Here we briefly recall the essential steps to obtain the  retarded  part of the gluon  self-energy in an anisotropic  background. 
 In the real time Keldysh formalism,  the Green's functions for the quark field of a given flavour $\psi^i_\a$ and gluon field $A^a_\mu$ can be expressed as
\begin{align}
i\[S(x,y)\]^{ij}_{\a\b}&=\left\langle\Th\[\psi^i_\a(x)\psib^j_\b(y)\]\right\rangle~,\nn\\
i\[D(x,y)\]^{ab}_{\mu\nu}&=\left\langle\Th\[A^a_\mu(x)A^b_\nu(y)\]\right\rangle~.
\end{align}
where the  spinor indices are represented by the set $\{\a,\b\}\in\{1,2,3,4\}$ and the color indices  in fundamental and adjoint representations of $SU(N_c)$ group with $N_c=3$ are represented by the sets $\{i,j\}\in\{1,2,3\}$ and $\{a,b\}\in\{1,2\cdots 8\}$   respectively. Here the angular bracket notation $\left\langle\cdots\right\rangle$ denotes the expectation value, and the time ordering  $\Th$ of two generic fields $\Phi_1$ and   $\Phi_2$ has the usual meaning 
\begin{align}
\Th\[\Phi_1(x)\Phi_2(y)\]&=\Theta(x^0-y^0)\Phi_1(x)\Phi_2(y)\pm\Theta(y^0-x^0)\Phi_2(y)\Phi_1(x)~,
\end{align}
where $\Theta$ denotes the Heaviside step function and the upper (lower) sign corresponds to the bosonic (fermionic) nature of the $\Phi$ fields.  At one loop level, the gluon polarization function has three different contributions arising namely from  the gluon tadpole and loop diagrams,  the ghost loop diagram and  the quark loop diagram. 
In presence of external magnetic field, the contributions from gluon and ghost remain unmodified whereas  corrections appear in the  quark loop contribution.  Moreover, in the HTL approximation, the expressions of the photon and gluon self-energy  differ only in the definition of the Debye mass. Thus, to find the net contribution of the  gluon  and ghost loops in presence of anisotropic momentum distribution, it is convenient to obtain the photon polarization function first without the external magnetic field, and then  replace the  QED Debye mass  by the corresponding QCD  expression for pure glue \cite{Nopoush:2017zbu}. The retarded self energy so obtained,   is given by \cite{Kasmaei:2018yrr,Ghosh:2020sng}:
\begin{align}
\tilde{\Pi}_{ab}^{\mu\nu}(\omega,{\bm p},\xi)&=\delta_{ab}~\tilde{m}_D^2\int\frac{d\Omega_{\bm v}}{4\pi}v^\mu\frac{v^l+\xi_1({\bm v}\cdot {\bm a_1})a_1^l+\xi_2({\bm v}\cdot {\bm a_2})a_2^l}{(1+\xi_1({\bm v}\cdot {\bm a_1})^2+\xi_2({\bm v}\cdot {\bm a_2})^2)^2}\left.\Big[\eta^{\nu l}-\frac{v^\nu p^l}{\omega-{\bm p}\cdot {\bm v}+i0^+}\Big]\right\vert_{{\scriptscriptstyle l \in\{1,2,3\}}},\label{glughost_part}
\end{align}
where the strong coupling constant  $g_s$ appears explicitly  in the expression of 
$\tilde{m}_D^2=\frac{g_s^2\Lambda_T^2}{3}N_c$  which corresponds to the QCD Debye mass  with $N_f=0$, and the scale $\Lambda_T$  in the equilibrium limit corresponds to the temperature.  It should be mentioned here that the dependence on the parameters $\Lambda_T$, $\xi_1$ and $\xi_2$, originates from the modeling of  the non-equilibrium distribution function  following the $`$Romatschke-Strickland' procedure. The anisotropic  distribution function for the gluons and ghosts  is constructed from the bosonic equilibrium distribution function  as \cite{Kasmaei:2016apv}
\begin{align}
f^{\rm B}_{\mbox{aniso}}({\bm k})\equiv f^{\rm B}_{\mbox{iso}}\Bigg(\frac{1}{\Lambda_T'}\sqrt{{\bf k}^2+\xi_x ({\bf k}\cdot\hat{\bf x})^2+\xi_y({\bf k}\cdot\hat{\bf y})^2+\xi_z({\bf k}\cdot\hat{\bf z})^2}\Bigg)\,.
\end{align}
In the conformal limit, with a simple rearrangement of the parameters given by 
\bea
 \frac{1+\xi_x}{1+\xi_y}\rightarrow 1+\xi_1\,,~~~
 \frac{1+\xi_z}{1+\xi_y}&\rightarrow&1+\xi_2\,,~~~
 \Lambda_T\rightarrow\frac{\Lambda_T'}{\sqrt{1+\xi_y}}\label{redefine}\,,
 \eea
one can characterize the ellipsoidal anisotropic distribution function in terms of  the anisotropy tuple $\xi=(\xi_1,\xi_2)$ and  the scale $\Lambda_T$. Thus, the nonequilibrium bosonic distribution relevant for the present scenario is given by \cite{Nopoush:2017zbu,Ghosh:2020sng}
\begin{align}
f^{\rm B}_{\mbox{aniso}}({\bm k})\equiv f^{\rm B}_{\mbox{iso}}\Bigg(\frac{\sqrt{{\bm k}^2+\xi_1({\bm k}\cdot{\bm a_1})^2+\xi_2({\bm k}\cdot{\bm a_2})^2}}{\Lambda_T}\Bigg).
\end{align}

In this work, the spatial anisotropy vectors ${\bm a_1}$, ${\bm a_2}$ are chosen along $\hat{x}=(1,0,0)$ and $\hat{z}=(0,0,1)$  directions respectively whereas the spatial components $v^l$ of the parton four velocity $v^\mu=(1,{\bm v})$  as well as the external momentum vector ${\bm p}$ are chosen in the spherical polar coordinates with angles $(\theta_k, \phi_k)$ and $ (\theta_p, \phi_p)$ respectively. To obtain the quark loop contribution of the retarded self energy in real time, here we  recall the required definitions of the four Green's functions  based on the propagation along the contour:
\begin{align}
i\[S^>(x,y)\]^{ij}_{\a\b}&=\left\langle\psi^i_\a(x)\psib^j_\b(y)\right\rangle~,\nn\\
i\[S^<(x,y)\]^{ij}_{\a\b}&=-\left\langle\psib^j_\b(y)\psi^i_\a(x)\right\rangle~,\nn\\
i\[S^{\bar{c}}(x,y)\]^{ij}_{\a\b}&=\left\langle\Th^{\bar{c}}\[\psi^i_\a(x)\psib^j_\b(y)\]\right\rangle~,\nn\\
i\[S^{\bar{a}}(x,y)\]^{ij}_{\a\b}&=\left\langle\Th^{\bar{a}}\[\psi^i_\a(x)\psib^j_\b(y)\]\right\rangle~.
\label{sdef}
\end{align}
Here $\Th^{\bar{c}}$ is same as the usual time-ordering operator  $\Th$ defined earlier whereas the anti-time-ordering operator $\Th^{\bar{a}}$ is defined as 
\begin{align}
\Th^{\bar{a}}\[\Phi_1(x)\Phi_2(y)\]&=\Theta(y^0-x^0)\Phi_1(x)\Phi_2(y)\pm\Theta(x^0-y^0)\Phi_2(y)\Phi_1(x)~,
\end{align}
where the upper (lower) sign corresponds to the bosonic (fermionic) nature of the generic $\Phi$ fields. The  Green's function $S^{\bar{c}/\bar{a}}(x,y)$ is the same as the  time ordered propagator $S(x,y)$ with both $x^0$ and $y^0$ chosen on the 
upper/lower branch of the contour where the contour runs  along the  forward/backward time direction. On the other hand $S^{<}(x,y)$ and $S^{>}(x,y)$ is same as $S(x,y)$ with $x^0$ on the upper  and $y^0$ on the lower  branch and vice versa. To avoid clutter in the notations, let us first consider the   photon self-energy in presence of magnetic background which is given by
\begin{align}
i\Pi^{\mu\nu}(x,y)&=-e^2 {\rm Tr}\[\gamma^\mu  S(x,y)\gamma^\nu S(y,x)\] 
\end{align}
where $e$ is the magnitude of the electron charge and $S(x,y)$ represents the electron propagator in presence of magnetic field. From the similar definitions as given in  \eqref{sdef}, one can easily express the polarization tensor as a sum of $\Pi_{\mu\nu}^>$ and $\Pi_{\mu\nu}^<$ where
\begin{align}
i\Pi_{\mu\nu}^>(x,y)&=-e^2 {\rm Tr}\[\gamma_\mu  S^>(x,y)\gamma_\nu S^<(y,x)\] ~,\nn\\
i\Pi_{\mu\nu}^<(x,y)&=-e^2 {\rm Tr}\[\gamma_\mu  S^<(x,y)\gamma_\nu S^>(y,x)\] ~.
\label{pi_grt_less}
\end{align}
Now,  the retarded self energy is defined  as 
\begin{align}
\Pi_{\mu\nu}^R(x,y)&=\theta(x^0-y^0)\big[\Pi_{\mu\nu}^>(x,y)-\Pi_{\mu\nu}^<(x,y)\big]~.
\label{retarded}
\end{align}
It should be noted here that the fermion propagator in presence of a background magnetic field possess a multiplicative  phase factor which  spoils translational invariance \cite{ Schwinger:1951nm}.   However, in the one loop photon polarization, the phase factor arising from the two fermion propagator cancels each other and only the translationally invariant parts of the propagators contribute. The same argument also applies for the quark loop in the  gluon polarization tensor that we are interested in. Thus, from here on out, it is useful to decompose the fermion propagator as \cite{Shovkovy:2012zn} $S(x,y)=e^{i\Phi(x_{\perp},y_{\perp})}\overline{S}(x-y)$ and  consider only the invariant $\overline{S}(x-y)$ part in the self energy. In that case,  we are free to choose $y=0$ because of translational invariance and obtain the retarded self-energy as
\begin{align}
i\Pi^{\mu\nu}_R(x)&=-\frac{e^2}{2} {\rm Tr}\[\gamma^\mu \overline{S}_F(x) \gamma^\nu \overline{S}_A(-x)+\gamma^\mu \overline{S}_R(x) \gamma^\nu \overline{S}_F(-x)\]~.
\end{align}
Note that, in the above expression, the $S^>$ and  $S^<$ propagators that arise from 
Eq.\eqref{pi_grt_less} and Eq.\eqref{retarded},  have been expressed in terms of  the Feynman, advanced and retarded propagators which are  defined respectively as 
\begin{align}
S_F(x,y)&=S^>(x,y)+S^<(x,y)~,\nn\\
S_A(x,y)&=-\theta(y^0-x^0)\[S^>(x,y)-S^<(x,y)\]~,\nn\\
S_{R}(x,y)&=\theta(x^0-y^0)\[S^>(x,y)-S^<(x,y)\]~.
\end{align}
 In the momentum space one obtains
\begin{align}
i\Pi^{\mu\nu}_R(p)&=-\frac{e^2}{2}\int\frac{d^4k}{(2\pi)^4} {\rm Tr}\[\gamma^\mu \overline{S}_F(k) \gamma^\nu \overline{S}_A(q)+\gamma^\mu \overline{S}_R(k) \gamma^\nu \overline{S}_F(q)\]~,\label{momentum_space_pi}
\end{align}
where $q=k-p$. In the mass-less limit, the invariant part of the  propagators  with lowest Landau level approximation are given by \cite{Shovkovy:2012zn}
\begin{align}
\overline{S}_R(k)&=\slkpar~\left(1+s_\perp i\gamma^1\gamma^2\right) \Delta_R(k)=\slkpar~\left(1+s_\perp i\gamma^1\gamma^2\right)\frac{\exp\big(-\frac{k^2_\perp}{|e_f B|}\big)}{\kpar^2+i\epsilon~ {\rm sgn}(k^0)}~, \nn\\
\overline{S}_A(k)&=\slkpar~\left(1+s_\perp i\gamma^1\gamma^2\right) \Delta_A(k)=\slkpar~\left(1+s_\perp i\gamma^1\gamma^2\right)\frac{\exp\big(-\frac{k^2_\perp}{|e_f B|}\big)}{\kpar^2-i\epsilon~{\rm sgn}(k^0)} ~, \nn\\
\overline{S}_F(k)&=\slkpar~\left(1+s_\perp i\gamma^1\gamma^2\right) \Delta_F(k)=\slkpar~\left(1+s_\perp i\gamma^1\gamma^2\right)\Big[(2\pi i)\big[-1+2 f_{\rm F}(k_z)\big]\Big] \delta(\kpar^2) \exp\Big(-\frac{k^2_\perp}{|e_f B|}\Big)~,
\end{align}
where $s_\perp={\rm sgn}(e_f B)$ with `${\rm sgn}$' representing sign function and  the electric charge of the fermion is denoted as $e_f=q_f e$ which is equal to $-e$ for the electron.   Also we note that the expressions of the fermion propagator used here is derived for a constant magnetic field intensity $B$ along the $\hat{z}$ direction which is same as the direction of the anisotropy vector ${\bm a_2}$. It should be noted that  in presence of a background magnetic field, the energy eigenvalue  for the free fermion only depends on  the  longitudinal momentum (say $k_z$) and the Landau level index (say $n$)   as these are the  conserved quantum numbers independent of the gauge choice. On the other hand, the transverse momentum, which appears in the  expression of the propagators, should be considered only as a  conjugate variable arising from the Fourier transform of the translationally invariant part and  it does not appear in the energy eigenvalue.  Hence, in the lowest landau level approximation ($n=0$), we construct the  nonequilibrium fermion distribution function  from  the equilibrium Fermi-Dirac distribution function $f_{\rm F}(k_z)$ as 
\begin{align}
f^{\rm F}_{\rm aniso}(k_z)&\equiv f^{\rm F}_{\rm iso}\Big(\sqrt{k^2_z+\xi_z({\bf k}\cdot\hat{\bf z})^2 }/\Lambda_T'\Big)= f^{\rm F}_{\rm iso}\Big(|k_z|/\lambda_T\Big)~,\label{fermi_aniso}
\end{align}
where the newly defined momentum scale $\lambda_T$ is related to $\Lambda_T$ given in Eq.\eqref{redefine}  as  $\lambda_T=\Lambda_T/\sqrt{1+\xi_2}$.
 Now, using the definition of the  propagators in Eq.\eqref{momentum_space_pi}, the spinor trace can be performed and one obtains
\begin{align}
i\Pi^{\mu\nu}_R(p)&=-4e^2\int\frac{d^4k}{(2\pi)^4}\Big[\kpar^\mu\qpar^\nu+\kpar^\nu\qpar^\mu-\eta^{\mu\nu}_{\paral}(\kpar\cdot\qpar)\Big]\Big[\Delta_R(k)\Delta_F(q)+\Delta_F(k)\Delta_A(q)\Big]~,\nn\\
&=-8e^2\int\frac{d^4k}{(2\pi)^4}\Big[\kpar^\mu\qpar^\nu+\kpar^\nu\qpar^\mu-\eta^{\mu\nu}_{\paral}(\kpar\cdot\qpar)\Big]\Delta_F(k)\Delta_A(q)~,
\end{align}
where in the last step we have used $\Delta_F(-k)=\Delta_F(k)$ and $\Delta_R(-k)=\Delta_A(k)$. As we are interested in the medium effects, here we only consider  the medium modified part of $\Delta_F(k)=4\pi i f_{\rm F}(k_z)e^{-\frac{k^2_\perp}{|e_f B|}}\delta(k_{\paral}^2)$. Moreover, with this structure of the propagator, the longitudinal and the transverse part of the integrals gets separated and one can easily  perform the integral over the transverse momentum as
\begin{align}
\int\frac{d^2k_\perp}{(2\pi)^2}\exp\Big(-\frac{k^2_\perp}{|e B|}\Big)\exp\Big(-\frac{q^2_\perp}{|e B|}\Big)&=\frac{|eB|}{8\pi}\exp\Big(-\frac{p^2_\perp}{2|eB|}\Big)~.
\end{align}
The polarization function now becomes
\begin{align}
\Pi^{\mu\nu}_R(p)&=-4 e^2|eB|\exp\Big(-\frac{p^2_\perp}{2|eB|}\Big)\int\frac{d^2k_{\paral}}{(2\pi)^2}f_{\rm F}(k_z)\Bigg[\frac{2\kpar^\mu\kpar^\nu-(\kpar^\mu\ppar^\nu+\kpar^\nu\ppar^\mu)+\eta^{\mu\nu}_{\paral}(\kpar\cdot\ppar)}{\ppar^2-2(\kpar\cdot\ppar)-i\epsilon~ {\rm sgn}(k^0-p^0)}\Bigg]\delta(\kpar^2)~.
\end{align}
In the HTL approximation we consider the external momentum to be `soft' that is $p\sim e\Lambda_T$ and the internal momentum is `hard' that is $k\sim\Lambda_T$. With this hierarchy, a Taylor series expansion of   the terms inside the square brackets    can be performed which up to second order is given as
\begin{align}
 &\frac{2\kpar^\mu\kpar^\nu-(\kpar^\mu\ppar^\nu+\kpar^\nu\ppar^\mu)+\eta^{\mu\nu}_{\paral}(\kpar\cdot\ppar)}{\ppar^2-2(\kpar\cdot\ppar)-i\epsilon~ {\rm sgn}(k^0-p^0)}=\frac{2\kpar^\mu\kpar^\nu-(\kpar^\mu\ppar^\nu+\kpar^\nu\ppar^\mu)+\eta^{\mu\nu}_{\paral}(\kpar\cdot\ppar)}{-2(\kpar\cdot\ppar)-i\epsilon~ {\rm sgn}(k^0)}\Bigg[1-\frac{p_{\paral}^2}{2(\kpar\cdot\ppar)+i\epsilon~ {\rm sgn}(k^0)}\Bigg]^{-1}~,\nn\\
 &\approx\frac{2\kpar^\mu\kpar^\nu}{-2(\kpar\cdot\ppar)-i\epsilon~ {\rm sgn}(k^0)}-\frac{\eta_{\paral}^{\mu\nu}}{2}+\frac{\kpar^\mu\ppar^\nu+\kpar^\nu\ppar^\mu}{2(\kpar\cdot\ppar)+i\epsilon~ {\rm sgn}(k^0)}-\ppar^2\frac{2 \kpar^\mu \kpar^\nu }{\big[2(\kpar\cdot\ppar)+i\epsilon~ {\rm sgn}(k^0)\big]^2}~.
\end{align}
As in the thermal case, the first term in the expansion does not contribute. Integrating over the $k^0$ variable using the delta function property 
\begin{align}
\delta(k^2_{\paral})&=\frac{\delta(k^0-|k_z|)+\delta(k^0+|k_z|)}{2|k_z|}~,
\end{align}
one  obtains  
\begin{align}
\Pi^{\mu\nu}_R(p)&= e^2\frac{|eB|}{\pi}\exp\Big(-\frac{p^2_\perp}{2|eB|}\Big)\int\frac{dk_z}{2\pi}\frac{f_{\rm F}(k_z)}{|k_z|}\left.\Bigg[\eta_{\paral}^{\mu\nu}-\frac{\kpar^\mu\ppar^\nu+\kpar^\nu\ppar^\mu}{(\kpar\cdot\ppar)+i\epsilon}+\frac{\ppar^2\kpar^\mu \kpar^\nu }{\big[(\kpar\cdot\ppar)+i\epsilon\big]^2}\Bigg]\right\vert_{k^0=|k_z|}~.
\end{align}
The term in the square braces can be related to a total derivative term as
\begin{align}
\left.\Bigg[\eta_{\paral}^{\mu\nu}-\frac{\kpar^\mu\ppar^\nu+\kpar^\nu\ppar^\mu}{(\kpar\cdot\ppar)+i\epsilon}+\frac{\ppar^2\kpar^\mu \kpar^\nu }{\big[(\kpar\cdot\ppar)+i\epsilon\big]^2}\Bigg]\right\vert_{k^0=|k_z|}&=-|k_z|\frac{\partial}{\partial k_z}\left.\Bigg[p_z\frac{\kpar^\mu\kpar^\nu}{|k_z|(\kpar\cdot\ppar+i \epsilon)}-\frac{\kpar^\mu\eta_{\paral}^{\nu 3}}{|k_z|}\Bigg]\right\vert_{k^0=|k_z|}~.
\end{align}
After performing an integration by parts with the assumption $\lim_{k_z\rightarrow\pm\infty}f(k_z)=0$ one  obtains
\begin{align}
 \Pi^{\mu\nu}_R(p)&= e^2\frac{|eB|}{\pi}\exp\Big(-\frac{p^2_\perp}{2|eB|}\Big)\int\frac{dk_z}{2\pi}\frac{\partial f_{\rm F}(k_z)}{\partial k_z}\left.\Bigg[p_z\frac{\kpar^\mu\kpar^\nu}{|k_z|(\kpar\cdot\ppar+i \epsilon)}-\frac{\kpar^\mu\eta_{\paral}^{\nu 3}}{|k_z|}\Bigg]\right\vert_{k^0=|k_z|}~.
\end{align}
As in the thermal case, the above expression can further be simplified by expressing the magnitude and the angular integrals separately. Considering the anisotropic distribution function as given in Eq.~\eqref{fermi_aniso}     one obtains 
 \begin{align}
\Pi^{\mu\nu}_R(p)&=\frac{m^2_{D,e}}{2}\exp\Big(-\frac{p^2_\perp}{2|eB|}\Big)\sum_{{\rm sgn}(k_z)=\pm1}\left.\frac{v_{\paral}^\mu v_{\paral}^l}{1+\xi_2}\Bigg[\eta_{\paral}^{\nu l}-\frac{\vpar^\nu p^l}{(\vpar\cdot\ppar+i \epsilon)}\Bigg]\right\vert_{l=3}~,\label{electron_loop}
\end{align}
where the Debye mass is defined  as 
\begin{align}
m^2_{D,e}&=-\frac{e^2}{\pi^2}|e B|\int d|k_z| \frac{\partial f^{\rm iso}_{\rm F}(|k_z|)}{\partial |k_z|}=e^2\frac{|eB|}{2\pi^2}~.
\end{align}
One can observe that, as a consequence of dimensional reduction in the strong field approximation, the solid angle integral with the $4\pi$ angular average in Eq.~\eqref{glughost_part}, now reduces to a  summation along with an average over two possible directions. It should be noticed that unlike the thermal case, the self-energy is independent of the momentum scale $\Lambda_T$  and the anisotropy parameter  appears only in a multiplicative factor without any directional dependence. However, the implicit dependence on the momentum scale is present due to the running of the coupling constant.  Now, incorporating the flavor sum and the color factor, the quark loop contribution  in the retarded gluon polarization tensor  can be obtained from Eq.~\eqref{electron_loop} as \cite{Fukushima:2015wck}
\begin{align}
\bar{\Pi}_{ab}^{\mu\nu}(p)&=\delta_{ab}\sum_{f}g_s^2\frac{|e_f B|}{8\pi^2}\exp\Big(-\frac{p^2_\perp}{2|e_f B|}\Big)\sum_{{\rm sgn}(k_z)=\pm1}\left.\frac{v_{\paral}^\mu v_{\paral}^l}{1+\xi_2}\Bigg[\eta_{\paral}^{\nu l}-\frac{\vpar^\nu p^l}{(\vpar\cdot\ppar+i \epsilon)}\Bigg]\right\vert_{l=3}~.\label{quark_loop}
\end{align}
In the static limit ($\omega=0, {\bm p}\rightarrow 0 $), the temporal component $\bar{\Pi}^{00}$ with $\xi_2=0$ becomes \cite{Karmakar:2018aig}
\begin{align}
\bar{m}_D^2&=\sum_{f}g_s^2\frac{|e_f B|}{4\pi^2}~,
\end{align}
which, together  with the magnetic field independent contribution $\tilde{m}_D^2$,  defines the Debye screening mass $\hat{m}_D=\sqrt{\tilde{m}_D^2+\bar{m}_D^2}$. Finally, the retarded gluon polarization function  is obtained from  the individual contributions given in Eq.~\eqref{glughost_part} and  Eq.~\eqref{quark_loop} as
\begin{align}
\Pi_{ab}^{\mu\nu}(p,eB,\xi,\Lambda_T)&=\tilde{\Pi}_{ab}^{\mu\nu}(p,\xi_1,\xi_2,\Lambda_T)+\bar{\Pi}_{ab}^{\mu\nu}(p, eB,\xi_2,\Lambda_T)~,
\label{pi}
\end{align}
where the dependence on the external parameters $p$, $eB$,  $\xi$ and $\Lambda_T$ has been shown explicitly in each case. The polarization function is symmetric in the Lorentz indices ($\Pi^{\mu\nu}=\Pi^{\nu\mu}$) and satisfies the transversality condition $p_\mu\Pi^{\mu\nu}=0$. Incorporating these constraint relations, the general structure of the polarization function can be constructed from the available basis tensors. A suitable choice in this regard is the basis set constructed for the ellipsoidal momentum anisotropy in Ref.~\cite{Ghosh:2020sng}. A list of the required basis tensors is provided in the Appendix \ref{list_tensor} for completeness. In that basis, the gluon polarization tensor can be expressed as 
\begin{align}
\Pi^{\mn}&=\alpha A^{\mn}+\beta B^{\mn}+\gamma C^{\mn}+\delta D^{\mn}+\sigma E^{\mn}+\lambda F^{\mn}~,
\end{align}
and the corresponding form factors can be extracted from 
Eq.~\eqref{pi} through suitable projections. Here we note 
that, all of the projection tensors are symmetric and transverse to the external momentum. Thus, the decomposition of the polarization function trivially  satisfies the symmetry constraint as well as the transversality condition. Now, the effective gluon propagator can be obtained from the Dyson--Schwinger equation 
\begin{align}
\mathcal{D}&=\mathcal{D}_0-\mathcal{D}_0\Pi \mathcal{D}~.
\end{align}
Here $\mathcal{D}_0$ is the bare propagator and its  inverse, with the gauge fixing parameter $\zeta$,  is given by 
\begin{align}
(\mathcal{D}_0^{-1})^{\mn}&=-p^2\eta^{\mn}-\frac{1-\zeta}{\zeta}p^\mu p^\nu~.
\end{align}
From the pole of the effective propagator, one can obtain the  gluon collective modes by solving 
\bea
p^2-\Omega_{0,\pm}(p)&=&0~.\label{disp}
\eea
Any deviation from the light-like dispersion is encoded in the mode functions  $\Omega_{0,\pm}$ which are given by  \cite{Ghosh:2020sng}
\bea
\Omega_0&=&\frac{1}{3}(\alpha + \beta + \delta)- \frac{1}{3}\frac{\vp}{\Big(\frac{\chi +\sqrt{4\vp^3+\chi^2}}{2}\Big)^{\frac{1}{3}}}+\frac{1}{3}\Big(\frac{\chi +\sqrt{4\vp^3+\chi^2}}{2}\Big)^{\frac{1}{3}},\label{om0}\\
\Omega_{\pm}&=&\frac{1}{3}(\alpha + \beta + \delta)+ \frac{1\pm i\sqrt{3}}{6}\frac{\vp}{\Big(\frac{\chi +\sqrt{4\vp^3+\chi^2}}{2}\Big)^{\frac{1}{3}}}-\frac{1\mp i\sqrt{3}}{6}\Big(\frac{\chi +\sqrt{4\vp^3+\chi^2}}{2}\Big)^{\frac{1}{3}},\label{ompm}
\eea
where, the  $\vp$ and $\chi$ in the expression are defined in terms of the form factors as
\bea
\vp&=&\alpha(\beta-\alpha)+\beta(\delta-\beta)+\delta(\alpha-\delta)-3(\gamma^2+\lambda^2+\sigma^2)~,\\
\chi&=&(2\alpha-\beta-\delta)(2\beta-\delta-\alpha)(2\delta-\alpha-\beta)+54\gamma\lambda\sigma\nn\\
&-&9\big[\alpha(2\lambda^2-\sigma^2-\gamma^2)+\beta(2\sigma^2-\gamma^2-\lambda^2)
+\delta(2\gamma^2-\lambda^2-\sigma^2)\big]~.
\eea
It should be noted here that among the six form factors, only the $\alpha$, $\beta$ and  $\gamma$  gets modified in presence of the external magnetic field. However, all of the form factors depend on the external anisotropy parameter $\xi$. Now, each of the mode functions being a nontrivial combination of all  six form factors, it is expected that, in addition to the anisotropy induced effects,  all  the  gluon collective modes will possess  magnetic modifications. We explore such anisotropic gluon collective modes in the following section.

\section{Results}
\begin{center}
\begin{figure}[tbh!]
 \begin{center}
 \includegraphics[width=5.5cm,scale=0.3]{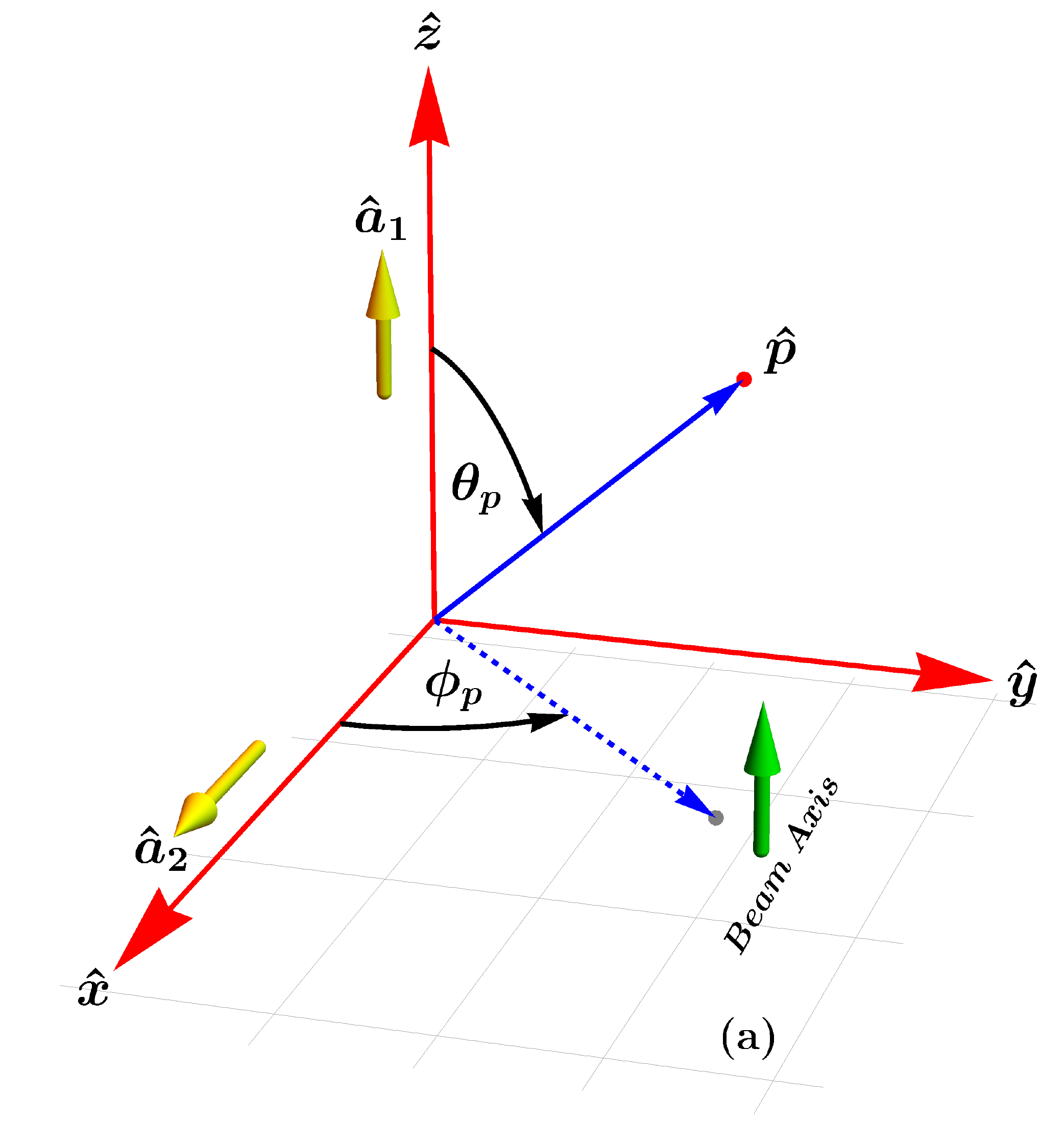} 
 \includegraphics[width=5.5cm,scale=0.3]{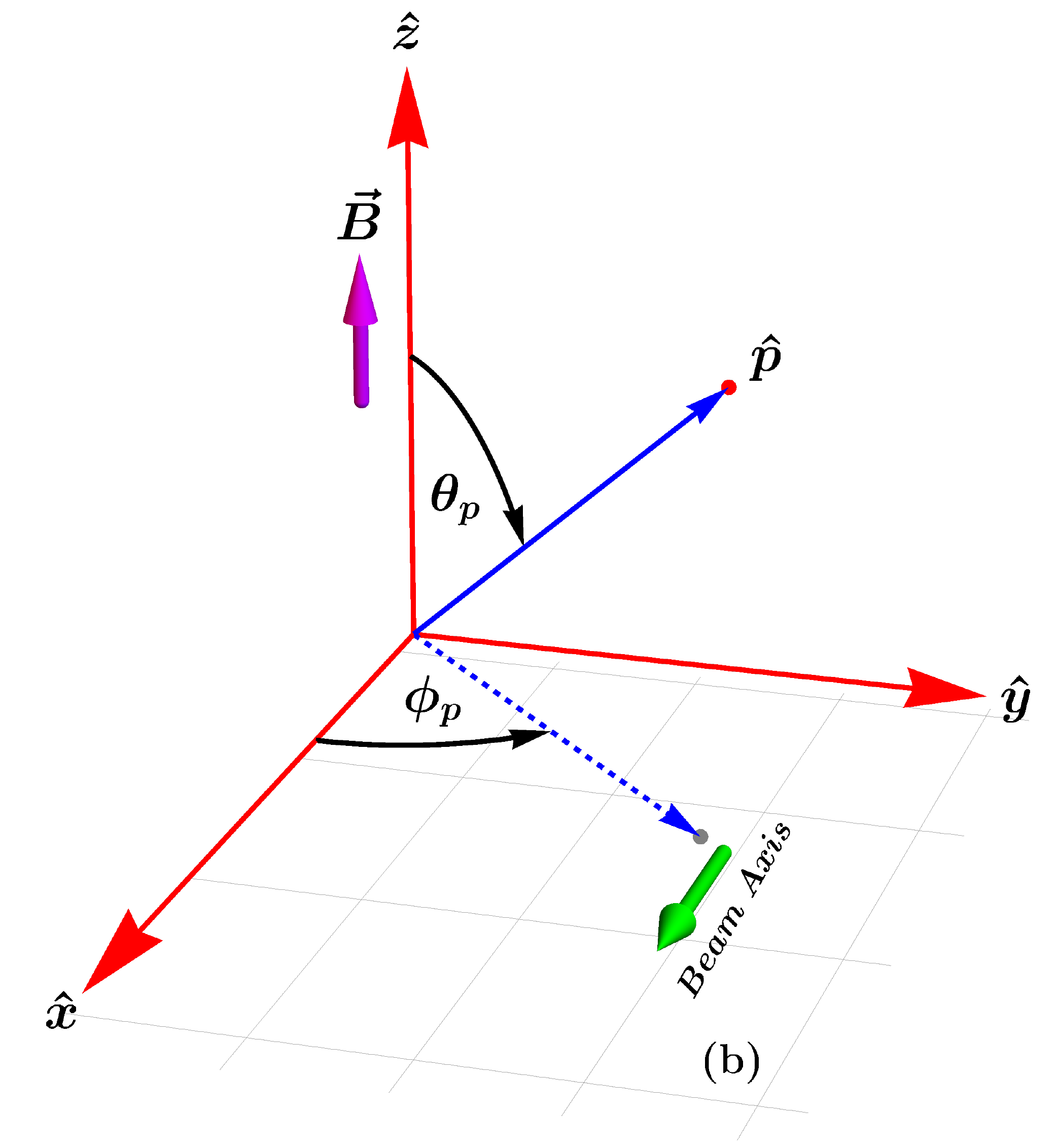} 
 \includegraphics[width=5.5cm,scale=0.3]{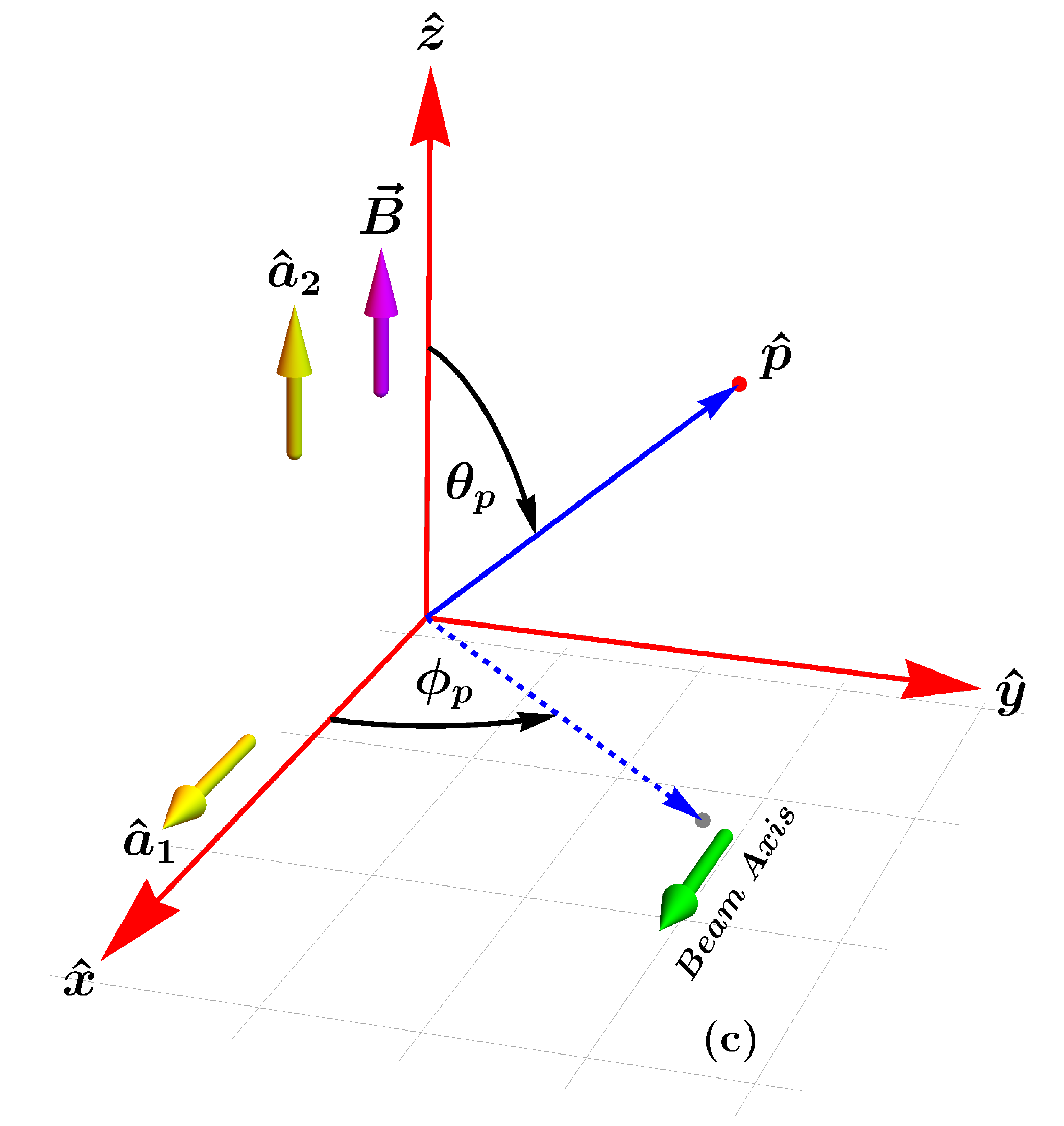} 
 \caption{A convenient choice of the reference frame is schematically shown for the following three cases: (a) the ellipsoidal momentum anisotropy without the external magnetic field, (b) the external magnetic field without the momentum space anisotropy and (c) the external magnetic field along with the momentum anisotropies as analyzed in the present study. }
  \label{coordinate_sys}
 \end{center}
\end{figure}
\end{center}
\begin{figure}[tbh!]
  \includegraphics[width=7 cm, scale=0.7]{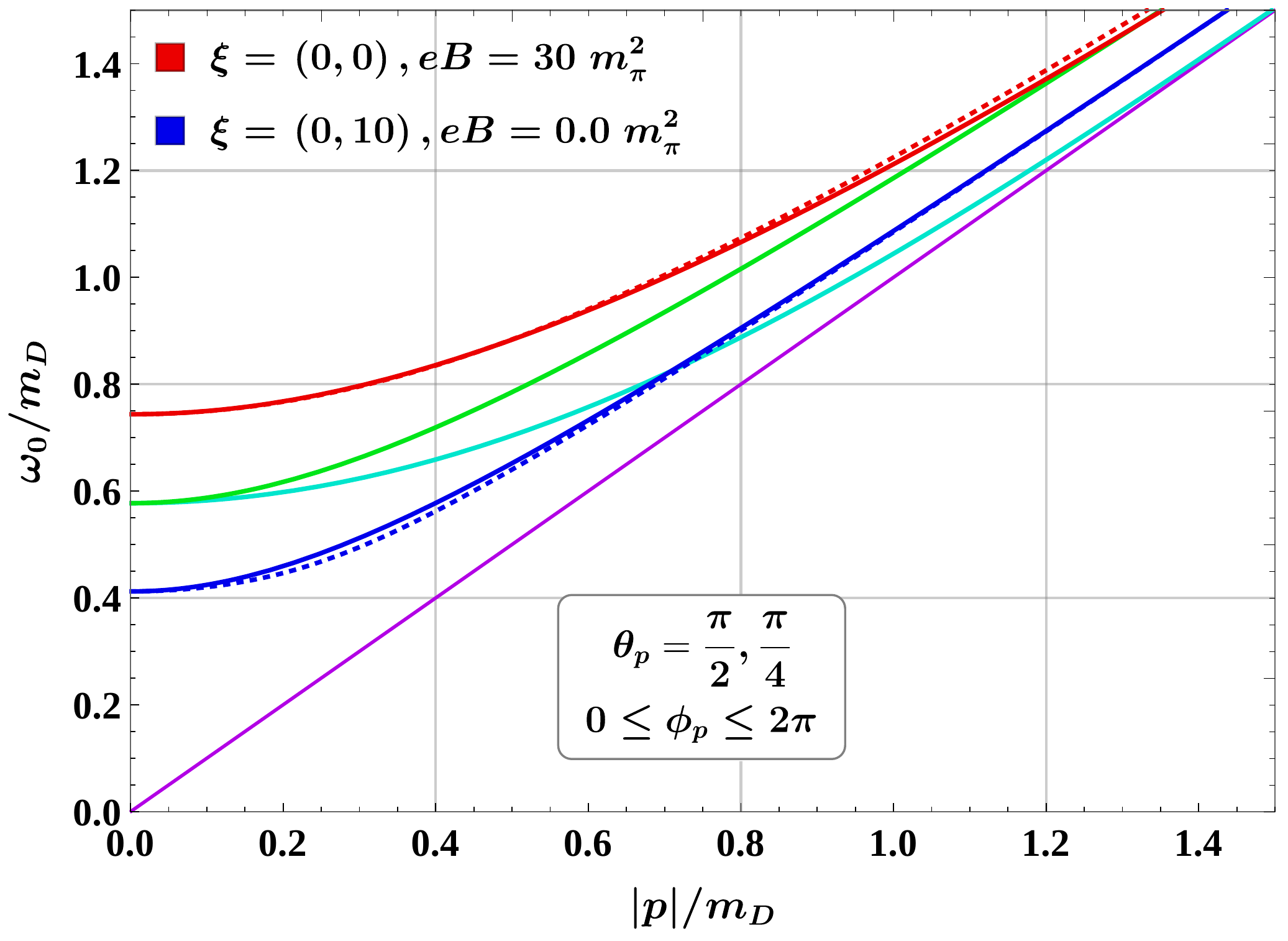} 
  \caption{The  collective mode of gluon  corresponding to the mode function $\Omega_0$ is shown at fixed momentum scale $\Lambda_T=0.2$ GeV and propagation angles $\theta_p=\pi/2$ (shown in solid style) and $\pi/4$ (shown in dotted style) for two different cases: (i) with  external magnetic field $eB=30 m_\pi^2$ (shown in red)  and (ii) with spheroidal anisotropy (shown in blue).  The light cone (magenta)  and the isotropic collective modes (green and cyan)  are also shown for comparison. }  
  \label{disp_plot_mo1}
\end{figure}
\begin{figure}[tbh!]
  \includegraphics[width=7 cm, scale=0.7]{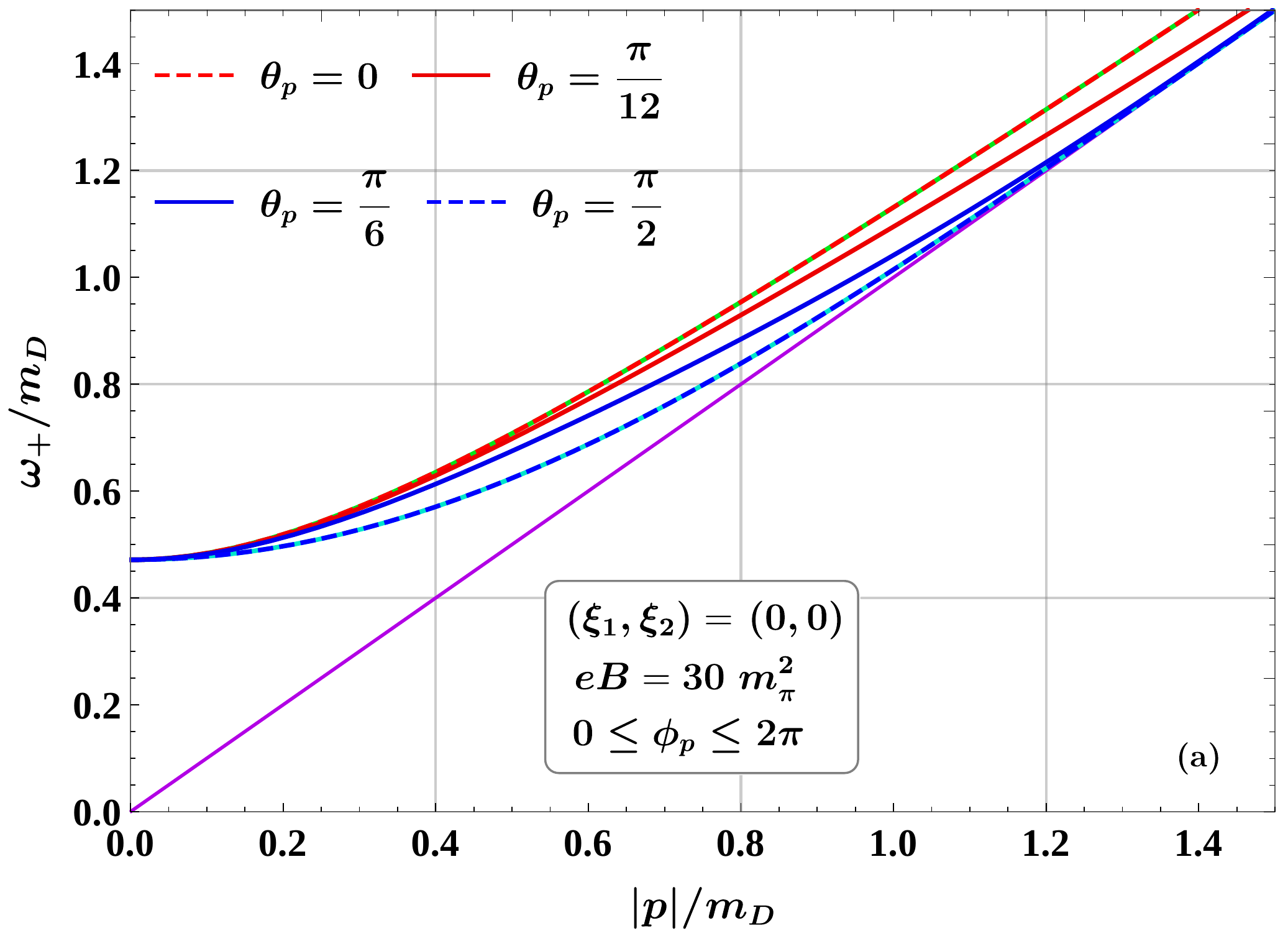} 
 \includegraphics[width=7cm, scale=0.7]{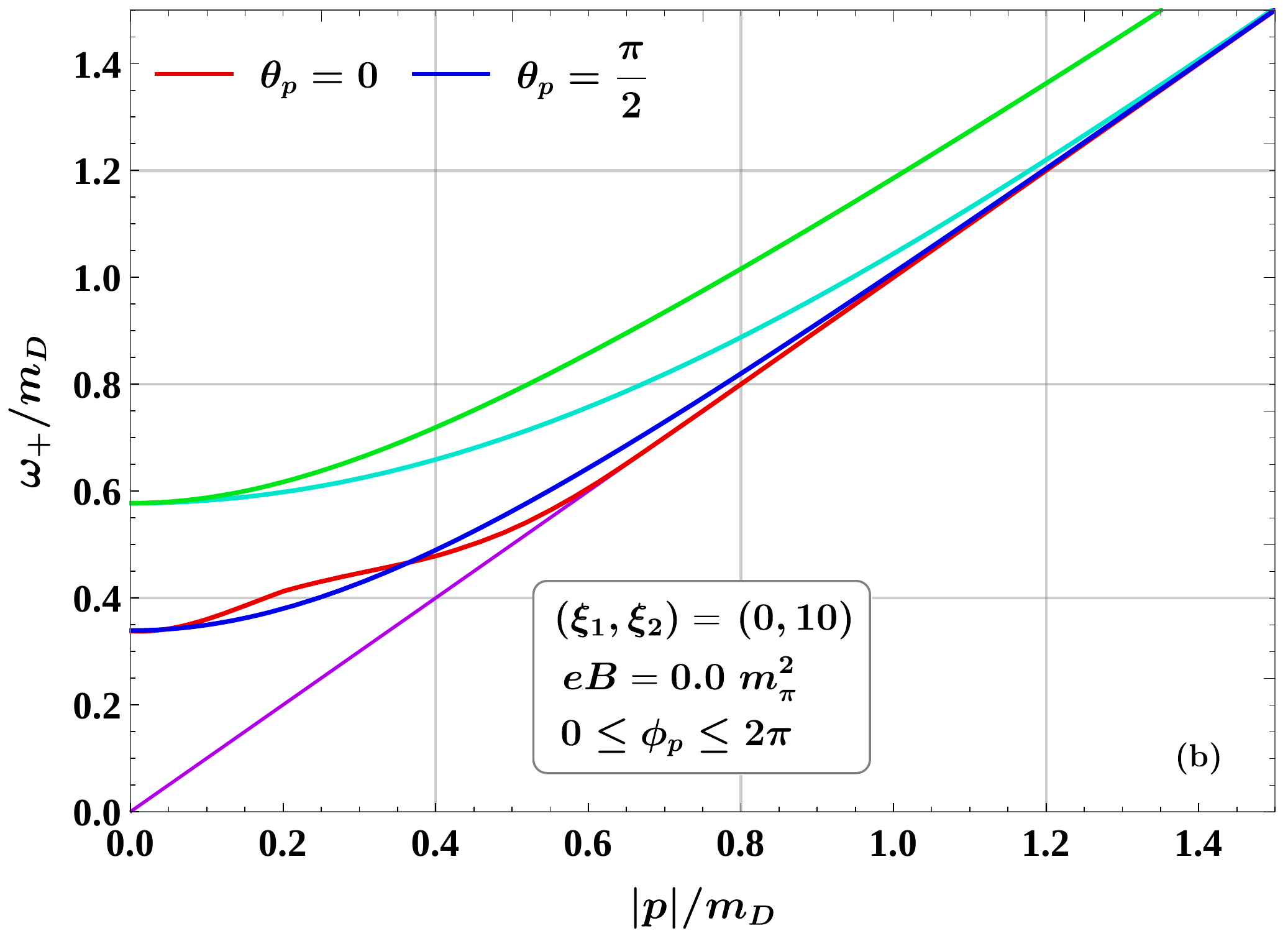} 
 \caption{Angular variation of the  collective mode of gluon  corresponding to the mode function $\Omega_+$ is shown at fixed momentum scale $\Lambda_T=0.2$ GeV  for two different cases: (a) with  external magnetic field $eB=30 m_\pi^2$   and (b) with spheroidal anisotropy.  The light cone is shown in continuous magenta style and the isotropic collective modes (with (a) $N_f=0$  and (b) $N_f=3$)  are  shown in continuous green and cyan  style for comparison. } 
  \label{disp_plot_mo2}
\end{figure}

As mentioned earlier, for a fixed set of external parameters, the gluon collective modes can be obtained by solving Eq.~\eqref{disp} which only requires the knowledge of  the momentum dependence of the form factors. Choosing a particular orientation of the reference frame, this momentum dependence can be obtained from  the components of the polarization tensor. Regarding the choice of the reference frame, it should be noted that usually in the studies concerned with a single momentum space anisotropy (say $\hat{a}_1$) without the background field, the anisotropy vector is chosen along the $\hat{z}$ direction which refers to the beam direction. In the ellipsoidal generalization (see for example Ref.~\cite{Ghosh:2020sng}),  a  convenient way to introduce another anisotropy direction (say $\hat{a}_2$) is to consider it to be  orthogonal to the previous anisotropy vector ($\hat{a}_1$). Without any loss of generality, one can reorient the reference frame to consider the $\hat{a}_2$ direction along the $\hat{x}$ axis as schematically shown in Fig.~\ref{coordinate_sys}(a). On the other hand, in the studies concerned with  the external magnetic field (see for example Refs.~\cite{Hattori:2017xoo,Karmakar:2018aig}), the $\hat{z}$ direction usually refers to the  direction of the  constant magnetic field  which is orthogonal to the reaction plane. A schematic representation of the corresponding  geometry is shown in Fig.~\ref{coordinate_sys}(b).  Now, to incorporate the two scenarios simultaneously, which is the main motivation of the present study, it is important to take into account the relative orientations of the anisotropy directions  with respect to the beam axis.  A convenient choice of the reference frame  for this purpose  is schematically shown in  Fig.~\ref{coordinate_sys}(c). In this geometry, the reaction plane is considered to be the X-Y plane, the polar angle $\theta_p$  is measured with respect to the external magnetic field direction ($\hat{z}$) whereas the azimuthal angle $\phi_p$ represents the angle between the anisotropy direction $\hat{a}_1$ chosen along the beam axis ($\hat{x}$) and the projection of the momentum vector $\hat{p}$ in the reaction plane. For the following numerical analysis, we fix the orientation of the reference frame according to Fig.~\ref{coordinate_sys}(c). Note that, the second anisotropy vector $\hat{a}_2$ is assumed to be parallel to the external field direction. As a consequence, when the two momentum anisotropies are switched off, the geometry reduces to the scenario shown in  Fig.~\ref{coordinate_sys}(b). Also it is evident from Fig.~\ref{coordinate_sys}(c) that even without the second anisotropy vector $\hat{a}_2$, one has to incorporate two anisotropy directions for the analysis of the collective modes. Consequently, the general structure of the polarization tensor in this case, as mentioned earlier, becomes similar to  the ellipsoidal momentum anisotropy scenario.
Here we note that,  without the background field, $\Lambda_T$ being  the only available energy scale for the anisotropic plasma, the  tensor components (and consequently the form factors and the mode functions) are proportional to the square of the thermal Debye screening mass given by 
\begin{align}
 m_D&=\sqrt{\frac{g_s^2 \Lambda_T^2}{3}\Bigg(N_c+\frac{N_f}{2}\Bigg)}~, 
\end{align}
and one can do  away with the $\Lambda_T$ dependence by simply expressing the dispersion in terms of the scaled variables $\omega/m_D$ and $|{\bm p}|/m_D$.  When the external magnetic field is turned on, the gluon and the quark loop contribution respectively becomes proportional to  $\tilde{m}_D^2$ and $\bar{m}_D^2$. However, to compare with the thermal case, here we consider the same $m_D$ for  scaling instead of  the thermo-magnetic Debye mass $\hat{m}_D$. With $N_c=3$, the ratio of the square of the Debye masses  arising from the  gluon   and the quark contribution can be expressed as
\begin{align}
\frac{\bar{m}_D^2}{\tilde{m}_D^2}&=\mathcal{R}^2\sum_f|q_f|~,
\end{align}
where the ratio of the two energy scales is set by $2\pi\mathcal{R}=\sqrt{|eB|}/\Lambda_T$. In the present study,  we consider  $eB=30 m_\pi^2 $ and $\Lambda_T=0.2$ GeV which gives $2\pi \mathcal{R}\sim 3$. 
 The value of the coupling $g_s$ at the fixed $\Lambda_T$ is determined considering the one loop running. For this purpose,  the $\overline{\rm{MS}}$ renormalization scale is set at 0.176 GeV by fixing the QCD fine structure constant $\alpha_s(1.5~ {\rm GeV},N_f=3)=0.326$ \cite{Bazavov:2012ka,Haque:2014rua}. With these fixed set of external parmeters, we now obtain the three stable gluon collective modes characterized by  the corresponding mode functions given in Eqs.~\eqref{om0} and \eqref{ompm}. At first we consider the case with one anisotropy direction. This can arise either due to the presence of  the external magnetic field  or by the  expansion of the medium resulting in an  anisotropic momentum distribution of the partons. Now, irrespective of the origin of the anisotropy, the gluon polarization tensor can be expressed in terms of four basis tensors and the pole of the effective propagator  gives rise to the  same  mode functions given by  \cite{Karmakar:2018aig,Ghosh:2020sng} 
 \bea
 \Omega_0&=&  \frac{1}{2}\bigg( \alpha + \beta+\sqrt{(\alpha - \beta)^2+4\gamma^2} \bigg), \label{RS1}\\
  \Omega_+&=& \frac{1}{2}\bigg( \alpha + \beta-\sqrt{(\alpha - \beta)^2+4\gamma^2} \bigg), \label{RS2}\\
   \Omega_-&=& \delta \label{RS3}.
 \eea
 However, it should be observed that the parameter dependences of the  form factors in the two cases are completely different and it is interesting to compare the two scenarios. In Fig.~\ref{disp_plot_mo1}, we consider the dispersion corresponding to the mode function $\Omega_0$ for two different values of  $\theta_p=\{\pi/2,\pi/4\}$ which represents the angle between the anisotropy vector and the external momentum. One can notice that the angular dependence is weak in both cases. In contrast to the magnetic field case, the mode corresponding to the spheroidal anisotropy shows more prominent  angular dependence  in the low momentum regime.   Also it can be noticed that the plasma frequency  in presence of the external magnetic field is significantly larger compared to  the spheroidal anisotropy scenario. The introduction of the  magnetic field  enhances the plasma frequency for this  mode compared to the isotropic case (also shown in the figure) whereas a spheroidal anisotropy decreases it.  
The scenario is quite different in case of the collective mode corresponding to $ \Omega_+$ as shown in Fig.~\ref{disp_plot_mo2}. In presence of a background magnetic field one can observe a prominent angular dependence in the dispersion shown in Fig.~\ref{disp_plot_mo2}(a). In the two limiting cases when  the propagation angle $\theta_p$ is  zero and $\pi/2$, the collective mode becomes identical respectively to the transverse ($\Pi_T$) and the longitudinal ($\Pi_L$) mode of the isotropic gluonic medium \cite{Karmakar:2018aig,Hattori:2017xoo}. It should be noted here that  as $\gamma$ vanishes in the isotropic case and  $\beta$ and $\delta=\Pi_T$ become degenerate, one obtains two distinct dispersive modes (also shown in the figure) corresponding to the mode functions \cite{Bellac:2011kqa} 
\bea
\Omega_0=\Pi_L=-\tilde{m}_D^2\frac{\omega^2-p^2}{ p^2}\bigg[ 1-\frac{\omega}{2p}\ln \frac{\omega+p}{\omega-p}\bigg]~,
\label{disp_iso_1}
\eea
and
\bea
\Omega_\pm=\Pi_T=\frac{\tilde{m}_D^2}{2}\frac{\omega^2}{p^2}\bigg[1-\frac{\omega^2-p^2}{2\omega p}\ln \frac{\omega+p}{\omega-p} \bigg]~.
\label{disp_iso_2}
\eea
For the intermediate angles (shown for $\theta_p=\pi/12,\pi/6$), the mode lies within the isotropic dispersion curves of the pure gluonic medium. On the other hand, in case of spheroidal momentum space anisotropy, the angular dependence of the collective mode is quite different from the magnetic field case as shown in Fig.~\ref{disp_plot_mo2}(b). Here we consider anisotropy tuple $\xi=(0,10)$. One can notice that the angular dependence is weaker. Moreover,  the isotropic dispersions can not be recovered by simply varying the propagation angle. It should be noted here that  in this case the isotropic mode functions are same as Eq.~\eqref{disp_iso_1} and Eq.~\eqref{disp_iso_2} however with the replacement of $\tilde{m}_D^2$  by $m_D^2$. Comparing the modes in Fig.~\ref{disp_plot_mo2}(a) and Fig.~\ref{disp_plot_mo2}(b), one can observe that in both cases the plasma frequency decreases compared to the  isotropic value  with $N_f=3$ and for the external magnetic field, it becomes equal to the plasma frequency of the isotropic pure gluonic medium ($ N_f=0$).  Interestingly, due to the similar decomposition of the basis tensor, in both cases the mode characterized by $\Omega_-$ becomes identical to the corresponding isotropic transverse mode (see Eq.~\eqref{RS3} where $\delta$ is respectively proportional to $\tilde{m}_D^2$ and $m_D^2$ for the magnetic and spheroidal anisotropy case) and consequently becomes independent of the propagation angle. 

\begin{figure}[tbh!]
  \includegraphics[width=7 cm, scale=0.7]{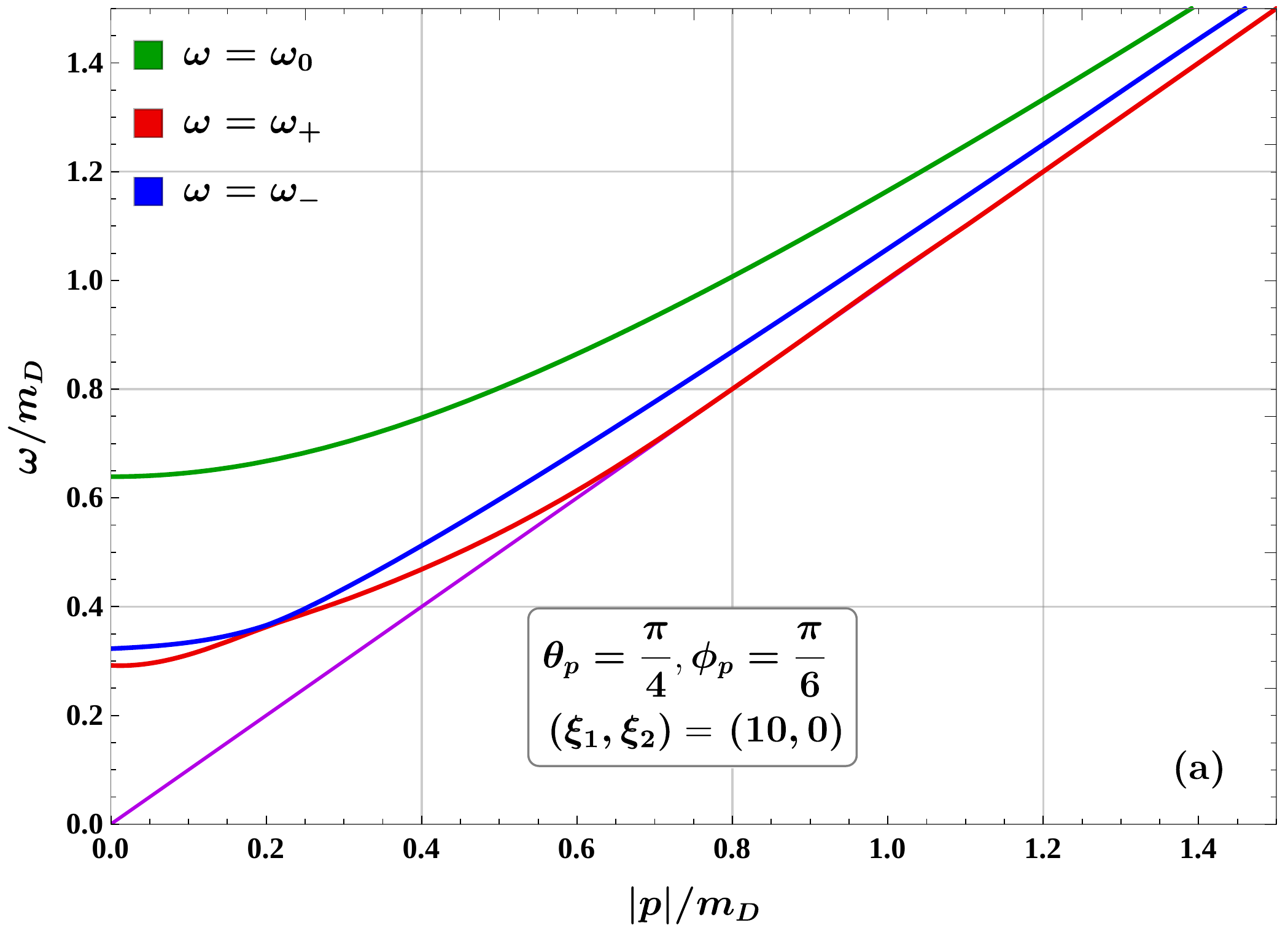} 
 \includegraphics[width=7cm, scale=0.7]{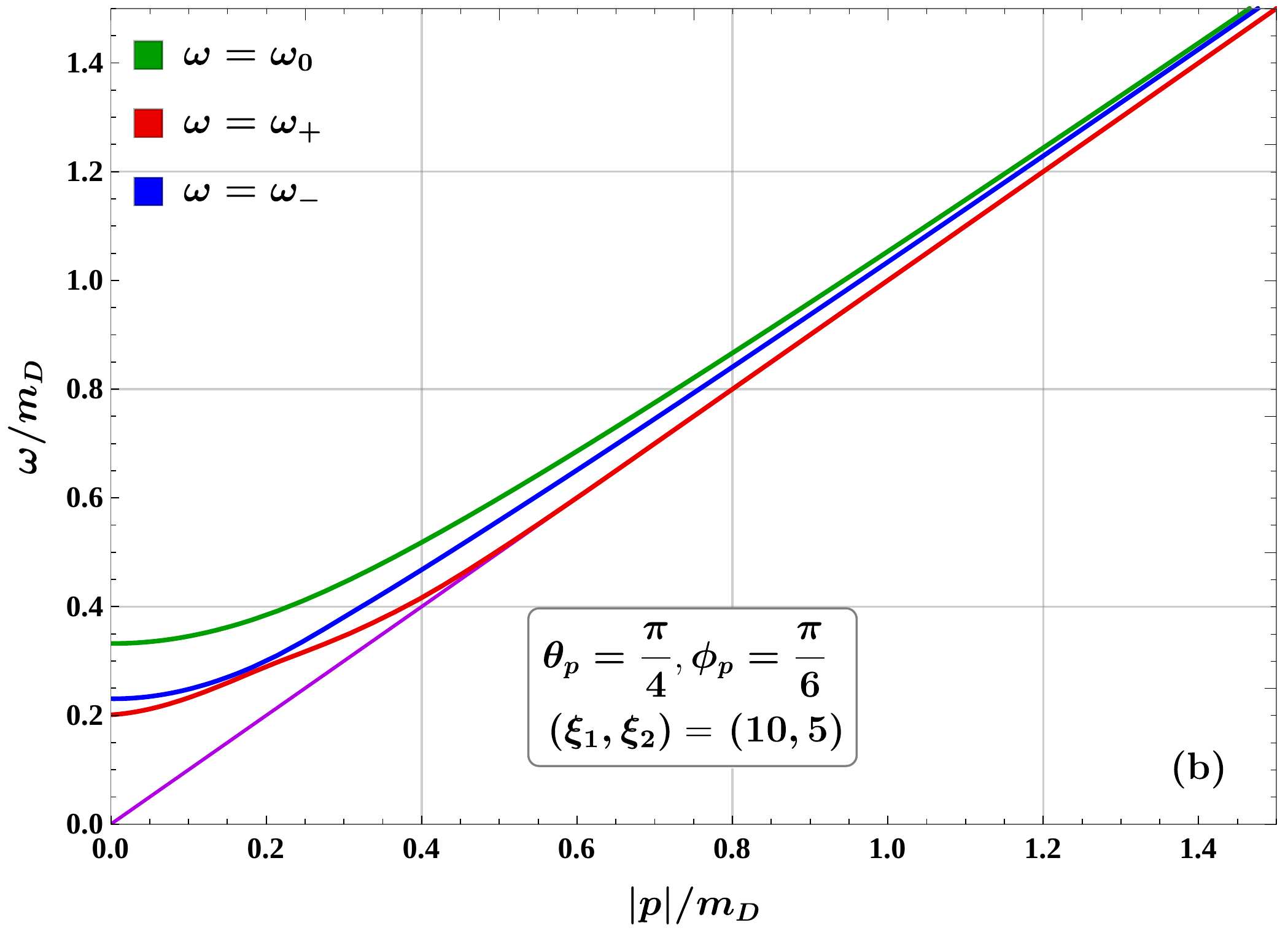} 
 \caption{The  collective modes of gluon with  (a) spheroidal and (b) ellipsoidal anisotropy  are shown for $\theta_p=\pi/4$ and $\phi_p=\pi/6$ at fixed momentum scale $\Lambda_T=0.2$ GeV and  magnetic field strength $30m_\pi^2$.  The light cone (magenta) is also shown for comparison.}
  \label{disp_plot}
\end{figure}

Finally, the dispersion relation for the three stable modes in presence of momentum space anisotropy as well as external magnetic field  is shown in Fig.~\ref{disp_plot}. In the left panel, we consider the momentum anisotropy along $\hat{x}$  which is  orthogonal to the magnetic field direction (along $\hat{z}$) and fix the  anisotropy tuple at $\xi=(10,0)$, whereas, in the right panel, the dispersion is shown for ellipsoidal momentum anisotropy with two anisotropy directions : one along the magnetic field ({\it{i.e.}} along $\hat{z}$) and the other  orthogonal to it ({\it{i.e.}} along $\hat{x}$). In this case the  anisotropy tuple is set at $\xi=(10,5)$. It should be noted that in the presence of either  magnetic field or spheroidal momentum anisotropy (say along $\hat{z}$), the rotational symmetry of the system is broken and  the dispersive modes  depend on the direction of propagation of the gluons which is characterized  by the polar angle $\theta_p$. However, when the two anisotropy directions are considered together, as long as they are not parallel to each other, the azimuthal symmetry of the system is also broken and consequently, the collective modes show azimuthal angular dependence. Here we consider a fixed propagation direction now characterized by $\theta_p=\pi/4$ and $\phi_p=\pi/6$. Unlike the magnetic field case discussed earlier (where the plasma frequencies of $\omega_+$ and $\omega_-$ were degenerate),  one can observe from Fig.~\ref{disp_plot}(a), that all the collective modes possess different plasma frequencies. Moreover, an overall decrease in the magnitude is  observed compared to the thermo-magnetic modes (shown in Figs.\ref{disp_plot_mo1} and ~\ref{disp_plot_mo2}(a)). Once the ellipsoidal anisotropy is considered, the plasma frequencies further decreases for all the modes as can be seen from Fig.~\ref{disp_plot}(b). This is in fact expected from  Eq.~\eqref{quark_loop} as the anisotropy parameter $\xi_2$ essentially suppresses the quark loop contribution thereby decreasing the overall magnitude.  
\begin{center}
\begin{figure}[tbh!]
 \begin{center}
 \includegraphics[scale=0.45]{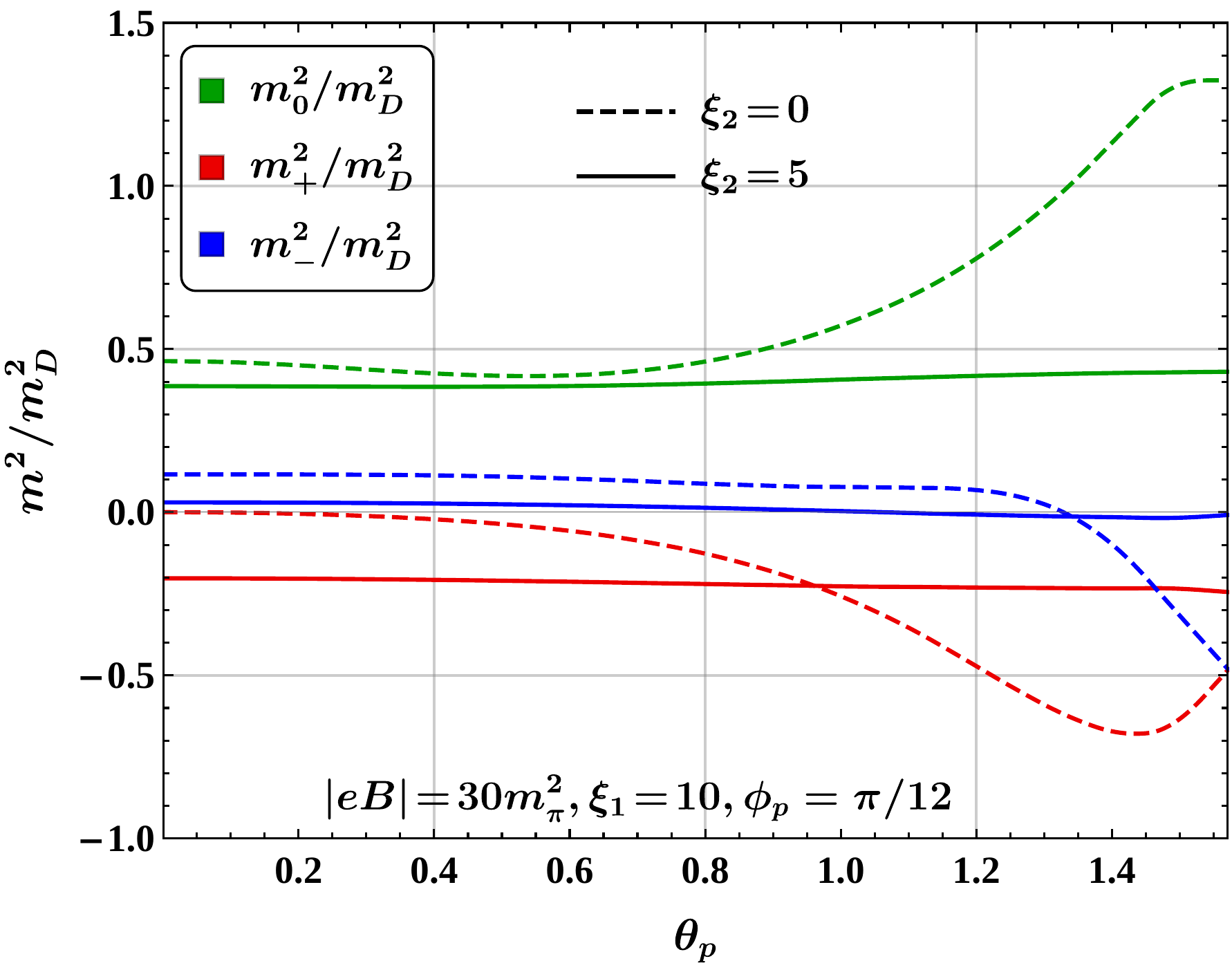} 
 \caption{Variation of the squared mass with polar angle $\theta_p$ is shown for each mode functions at fixed values of external parameters $\phi_p=\pi/12$, $\xi_1=10$, $\Lambda_T=0.2$ GeV and $eB=30m_\pi^2$. The continuous and the dashed curves represent $\xi_2=5$ and $\xi_2=0$ respectively.}
  \label{mass}
 \end{center}
\end{figure}
\end{center} 
Let us now consider the influence of the magnetic field on the unstable modes of the anisotropic medium. As in the case of spheroidal~\cite{Romatschke:2003ms} and ellipsoidal momentum anisotropy~\cite{Ghosh:2020sng}, in the  limit $\omega\rightarrow0$, one can define three mass scales ($m_0$ and $m_\pm$)  corresponding to the mode functions $\Omega_0$ and $\Omega_\pm$. A negative value of a given squared mass  indicates the existence of an unstable mode. It should be mentioned here that instead of considering $N_f=3$, if one considers a two flavour plasma, all the qualitative features remain the same and in the following, we study the mass scales and the instability growth rate considering $N_f=2$. 
\begin{center}
\begin{figure}[tbh!]
 \begin{center}
 \includegraphics[width=5.5cm,scale=0.35]{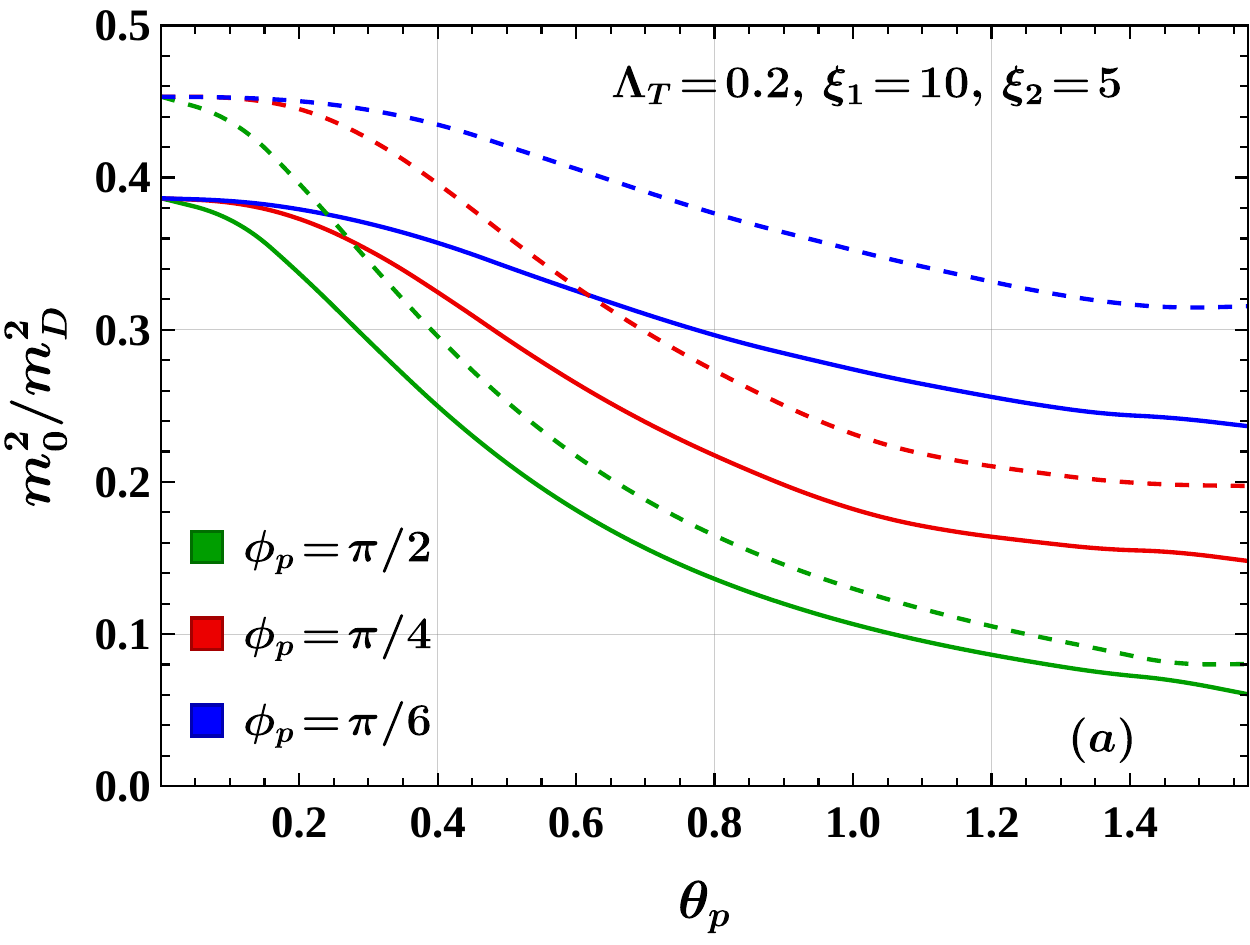} 
 \includegraphics[width=5.5cm,scale=0.35]{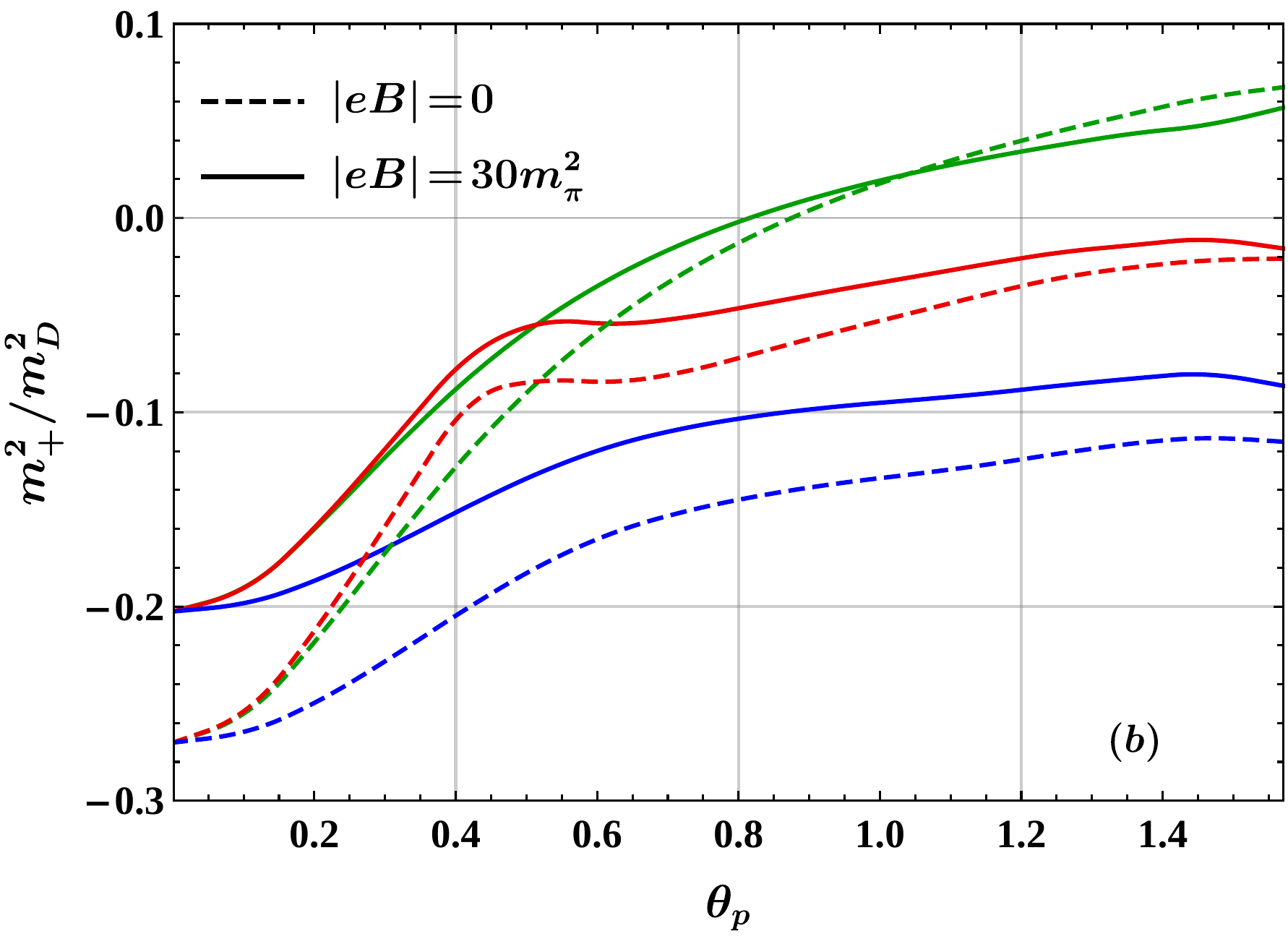} 
 \includegraphics[width=5.5cm,scale=0.35]{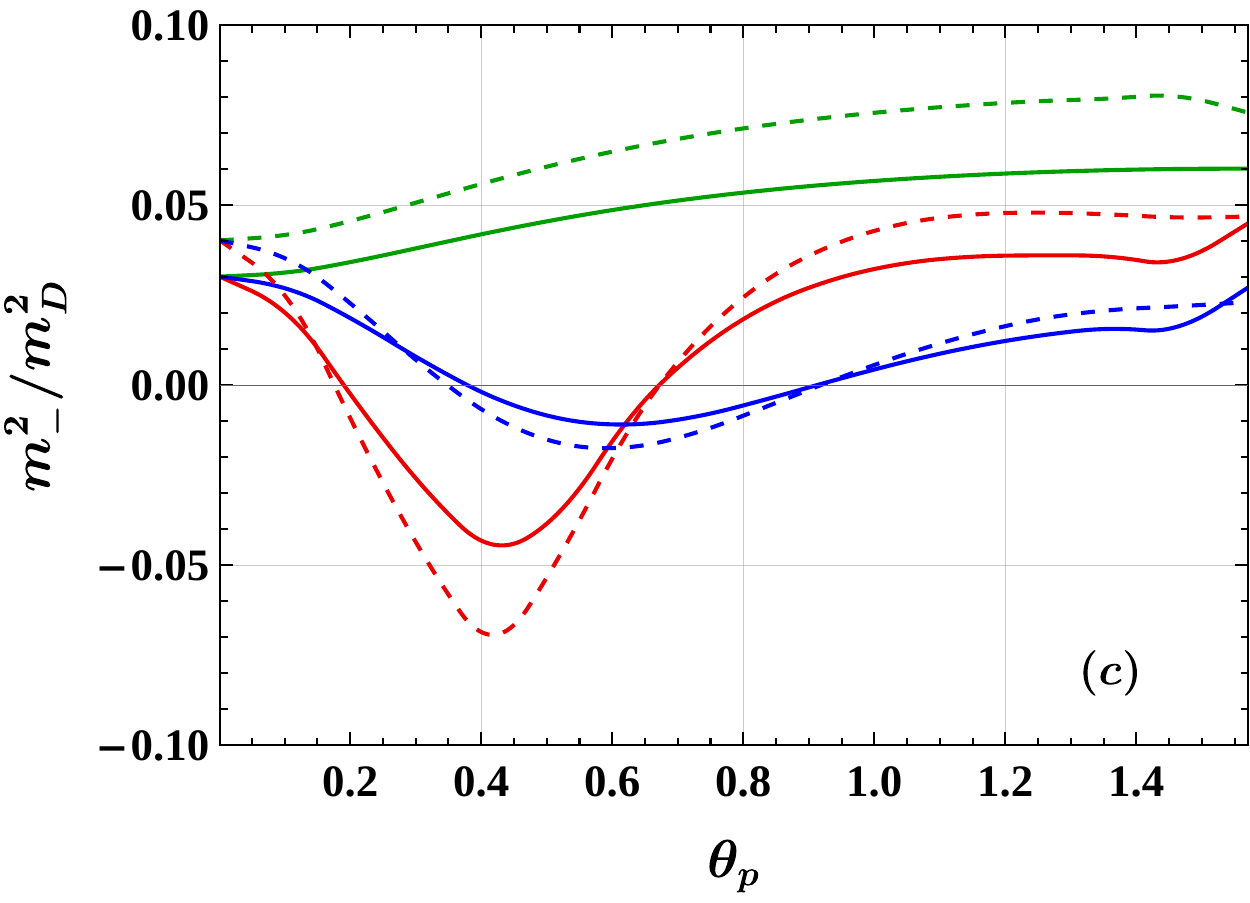} 
 \caption{Variation of the squared masses (a) $m_0^2$, (b) $m_+^2$ and (c) $m_-^2$ with the polar angle $\theta_p$ is shown at $\xi_1=10$ and $\xi_2=5$ with $\phi_p$ as a parameter. The continuous and the dashed  curves represent the magnetic field strength 30$m_\pi^2$  and $0$ respectively.}
  \label{mass_compare}
 \end{center}
\end{figure}
\end{center}
In Fig.~\ref{mass} we show the variation of  the squared mass with the propagation angle  of the  gluon with respect to the magnetic field direction.  For a  fixed $\phi_p=\pi/12$, we consider  two scenarios: one with $\xi=(10,0)$ and the other with $\xi=(10,5)$. In the former case, as we increase  $\theta_p$, $m^2_+$ and $m^2_-$ gradually become negative. However, a positive value value is observed for $m^2_0$ throughout the $\theta_p$ range. One should note    that, at small $\phi_p$ ( as considered here ), the higher values of $\theta_p$ indicates proximity to the anisotropy axis and the observed angular dependence of the mass scales  is similar to the spheroidal anisotropy case \cite{Romatschke:2003ms}. When the momentum anisotropy along the magnetic field direction is  turned on, all the mass scales become nearly independent of $\theta_p$. In this case, a prominent negative value for $m^2_+$ is observed for the entire range of the polar angle. It is interesting to compare the scenario with the ellipsoidal anisotropy results as obtained in Ref.~\cite{Ghosh:2020sng}.
 For this purpose, in Fig.~\ref{mass_compare}, we show the directional dependence of the  square mass scales  with and without  the external magnetic field. Here we consider  $\xi=(10,5)$. One can notice that the angular dependence of the mass scales are similar to the ellipsoidal anisotropy scenario showing a positive $m^2_0$ throughout the considered range of $\theta_p$ and $\phi_p$ along with instability windows for $m^2_\pm$.
 \begin{center}
\begin{figure}[tbh!]
 \begin{center}
 \includegraphics[width=7cm,scale=0.7]{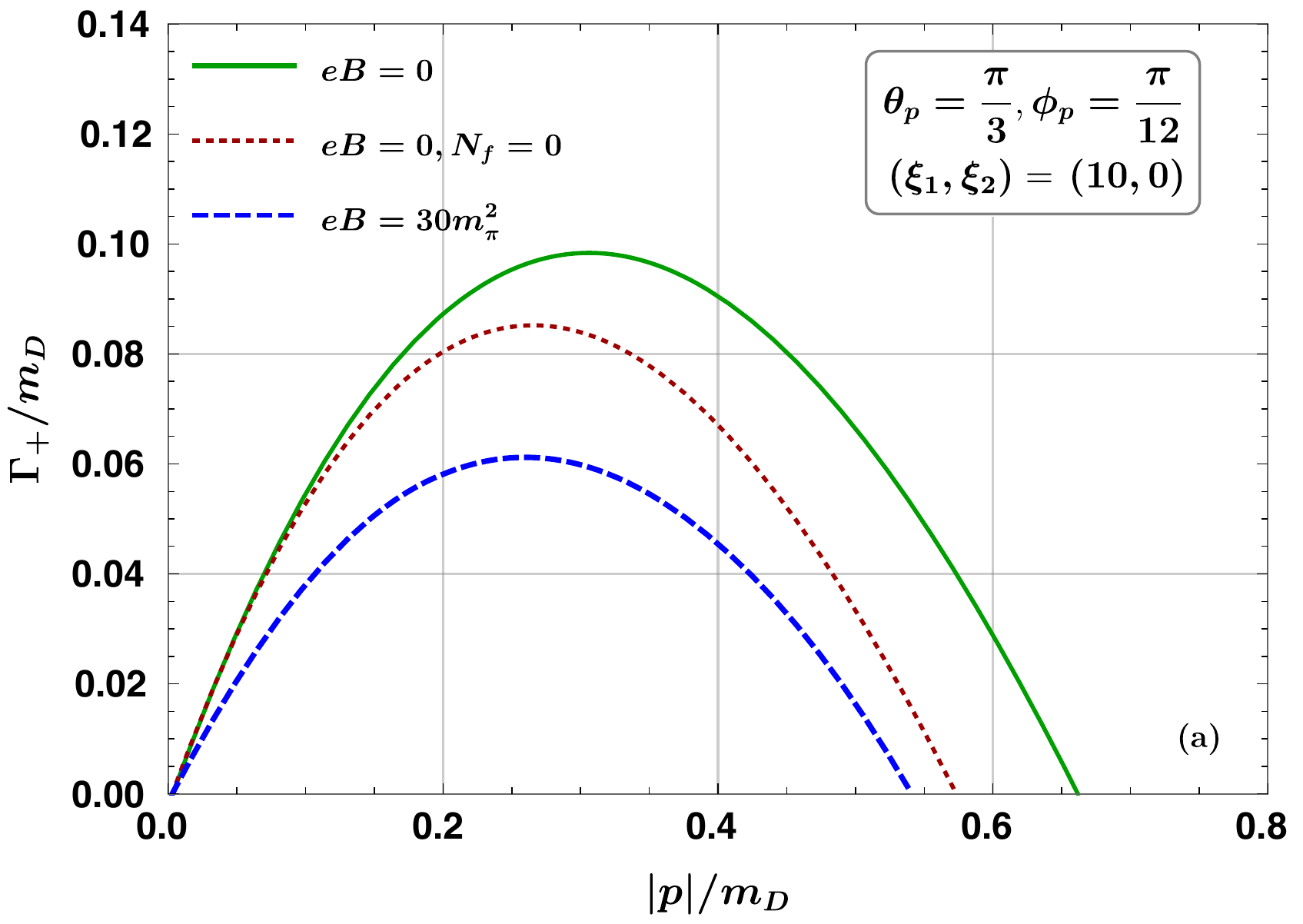}
 \includegraphics[width=7cm,scale=0.7]{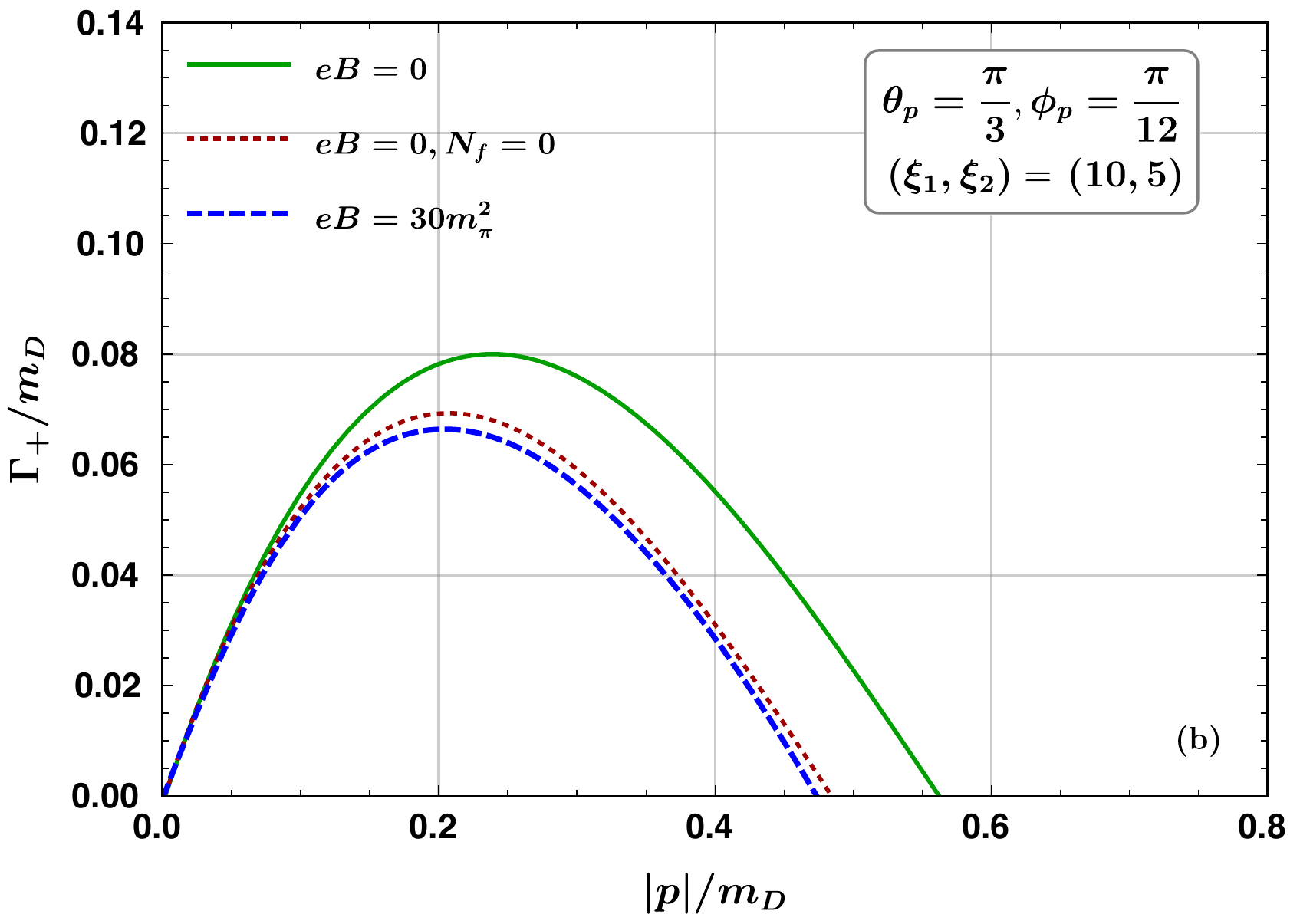}  
 \caption{The growth rate corresponding to $\Omega_+$ mode is plotted for  (a) $\xi=(10,0)$ and (b) $\xi=(10,5)$ at fixed angles $\theta_p=\pi/3$, $\phi_p=\pi/12$. The continuous and the dashed  curves correspond to the magnetic field strength  $0$ and 30$m_\pi^2$ respectively.  The pure gluon result is also shown for comparison.}
  \label{inst_mag}
 \end{center}
\end{figure}
\end{center}
As already mentioned, the  negative values in the square mass indicate the presence of unstable modes whose amplitude grows exponentially with time. The growth rate of such instabilities (that is the imaginary part of the mode frequency)  can  be obtained from the pole of the effective propagator. For this purpose,  the mode frequency ($p^0=\omega$) in Eq.~\eqref{disp}  is replaced by  $ i \Gamma_{0,\pm}$ and one  looks for the solution of $\Gamma$ corresponding to each mode functions ~\cite{Ghosh:2020sng,Kasmaei:2018yrr}. The numerical solution for $\Gamma_+$ is shown in Fig.~\ref{inst_mag} for a fixed propagation direction $(\theta_p,\phi_p)=(\pi/3,\pi/12)$. In the left panel, we consider the spheroidal momentum anisotropy  with $\xi=(10,0)$ whereas in the right panel, we take $\xi=(10,5)$ characterizing an ellipsoidal momentum space anisotropy. 
It can be observed that in both cases the amplitude of  the growth rate significantly  decreases in presence of the external magnetic field. It is also interesting to compare these modified growth rates with the pure gluon results. This is because, it  provides  information on the relative importance of the  field dependent part  compared to the field independent part of the  self energy. In the left panel of Fig.~\ref{inst_mag}, one can notice a prominent suppression compared to the  pure gluon result  where the anisotropy direction is considered orthogonal to the external field direction. On the other hand, with the same external parameter set, when the anisotropy along the magnetic field direction ($\xi_2$) is switched on (as shown in the right panel), the total $eB$ dependent quark loop contribution in the self energy  gets suppressed which results in a growth rate similar to the pure gluonic plasma.  It should be mentioned here that in the present analysis, the instability has been shown for some specific   values of the external parameters namely the anisotropy tuple $\xi=(\xi_1,\xi_2)$, the external angles $\theta_p$ and $\phi_p$, the external field strength $eB$ and the momentum scale $\Lambda_T$. While the magnitudes of the field strength and the momentum scale are adjusted to justify the LLL approximation, the values of the anisotropy tuple $\xi$ are chosen similar to  previous studies with the ellipsoidal anisotropic distribution \cite{Ghosh:2020sng,Kasmaei:2018yrr}. On the other hand, the propagation direction is set at  $(\theta_p,\phi_p)=(\pi/3,\pi/12)$ where, at the fixed magnetic field value, the instability for the $\Omega_+$ mode is expected for both $\xi_2=0$ and $\xi_2=5$ (see Fig.~\ref{mass} for the angular variation of the squared mass). The parameter set chosen for the present analysis is not unique and  different  parameter choices are possible. The growth rates being complicated functions of the angular variables, in general, a different propagation direction with unstable modes is expected to provide  a different amount of suppresion in the growth rate in the presence of the background magnetic field. However, based on the present analysis,  a large value of $\xi_2$ is expected to significantly suppress the dimensionally reduced quark loop contribution thereby leading to instabilities similar to the pure glue scenario. 

 It is interesting to note that for the spheroidal and ellipsoidal anisotropy without any magnetic background, there exists a critical value of the momentum beyond which the growth rate becomes negative and the instability ceases to exist. When the external magnetic field is turned on,  we observe a significant decrease in  the critical momentum providing a smaller momentum window for the positive growth rate. The situation may be compared to the instabilities  in  collisional plasma \cite{Schenke:2006xu,Jamal:2017dqs,Kumar:2017bja} where a critical collisional frequency exists beyond which the growth rate becomes negative for any value of external momentum. In a similar way, one may expect  a critical  magnetic field intensity   beyond which no instabilities occur.  Here we recall that in the present study we have considered the field intensity $\sqrt{eB}$ as high as three times the momentum scale $\Lambda_T$ to justify the lowest Landau level approximation. Now, for the anisotropic collisional plasma,  a small change in the collisional frequency significantly reduces the growth rate \cite{Schenke:2006xu}. However, in the present study we find that, even if one increases the magnetic field to several times the considered value,  the amplitude and the  critical momentum corresponding to the growth rate hardly decreases. Thus,  as long as the heavy ion collisions are concerned, a critical magnetic field intensity  is unlikely to be present in the realistic scenario.  

\section{Summary and Conclusion}
In this article, the collective modes of gluon in the presence of  momentum space anisotropy along with a constant background magnetic field have been studied using the hard-thermal loop perturbation theory. For this purpose, we have obtained the one loop  gluon self energy in the real time Schwinger-Keldysh formalism. The  contributions from the gluon and ghost loops remain unaffected by the external magnetic field  whereas the entire  modification arises from the quark loop contribution  which has been evaluated in the lowest Landau level approximation. To extract the Lorentz invariant form factors from the polarization tensor, we implement the basis decomposition obtained in Ref.~\cite{Ghosh:2020sng} which is originally  constructed for describing the ellipsoidal momentum anisotropy. From the pole of the effective gluon propagator, we obtain  three stable dispersive modes of gluon. At first we compare the collective modes of spheroidal anisotropy with that of isotropic thermal background along with external magnetic field. In both cases, the dispersion is governed by four non-vanishing form factors. Though the mode functions in terms of the form factors are identical in the two cases, the form factors themselves are different. Consequently,  significant differences are observed in the angular dependence of the collective modes  . When the external magnetic field is considered along with spheroidal or ellipsoidal momentum anisotropy,  the azimuthal symmetry of the system is lost. As a result, the collective modes  depend on the polar as well as on the azimuthal angles corresponding to the propagation direction. It is observed that due to the dimensional reduction in the LLL approximation, the parameter $\xi_2$ that characterizes the anisotropy along the magnetic field direction, appears in  the quark loop only in an overall suppressing factor. Thus, the momentum anisotropy along  the magnetic field direction essentially counterbalances the magnetic field effects. As the quark loop contribution is suppressed in this case, we  observe smaller plasma frequencies for all the collective modes.

To investigate the unstable modes, we have studied the angular dependence of the squared mass scales corresponding to each mode functions. Depending upon the propagation direction, we have observed negative values in the squared masses corresponding to $\Omega_\pm$ indicating instability in the collective modes. 
   Here we note that  no unstable gluon mode exists in an isotropic medium even in the presence of a background magnetic field. It is the momentum space anisotropy that gives rise to the instability. However,  the external magnetic field has a significant influence on  the growth rate of the unstable modes. In particular, the amplitude as well as the critical momentum corresponding to the growth rate of the unstable mode is significantly reduced in presence of strong magnetic background. This observation is similar to the instability growth rate in anisotropic collisional plasma \cite{Schenke:2006xu} where larger collisional frequency suppresses the growth rate and eventually, no unstable mode exists beyond a critical frequency. However, it has  been argued that the realistic collision frequencies usually lie within the critical value. Here also  we find that for  anisotropic thermal medium with realistic magnetic field intensity ( which is expected to be present  in heavy ion collisions ),  unstable collective modes do exist in certain propagation direction.  
  
 The present study has several interesting future directions. First of all, due to the lowest Landau level approximation,  only a 1+1 dimensional quark dynamics is considered here. Consequently, the momentum anisotropy  orthogonal to the magnetic field direction does not affect the quark  loop contribution at all. However, a non trivial influence  of such momentum space anisotropy is expected in the weak field limit where the energy eigen value of the quarks  have the usual three momentum dependence. Thus it is interesting to contrast such scenario with the strong field case as presented here. Also, the fermionic collective modes have recently been studied in presence of magnetic field \cite{Das:2017vfh} and also in case of ellipsoidal anisotropy \cite{Kasmaei:2016apv}.  Thus, the combined effect of the magnetic field and momentum space anisotropy  on the  fermionic collective modes deserves further investigation. Similar scenario also exists in the studies of heavy quark potential where the effect of the external magnetic field and the momentum anisotropy has been considered individually \cite{Dumitru:2007hy,Nopoush:2017zbu,Singh:2017nfa,Hasan:2020iwa,Ghosh:2022sxi} and their mutual influence remains to be explored. We intend to pursue such exploration in future.       
 \begin{acknowledgments}
A. M. would like to acknowledge fruitful discussions with Ashutosh Dash and Sunil Jaiswal. B. K.  acknowledges HORIZON 2020 European research council (ERC) 2016 Consolidation grant, ERC-2016-COG: 725741: QGP TOMOGRAPHY
(under contract with ERC).  R. G. is funded by University Grants Commission (UGC). A. M.  acknowledges Department of Science and Technology (DST), Government of India, for funding. 
\end{acknowledgments}

\appendix
\section{List of basis tensors}\label{list_tensor}
To obtain the general structure of the  gluon polarization tensor in presence of spheroidal or ellipsoidal momentum space anisotropy along with background magnetic field, we require a set of  six independent basis tensors. The basis tensors can be constructed using the  metric $\eta_{\mu\nu}$ and the available four vectors given by the fluid velocity $u^\mu$, gluon four momentum $p^\mu$ and the two anisotropy directions $a_1^\mu$ and $a_2^\mu$. In case of spheroidal anisotropy with external magnetic field, either $a_1^\mu$ or $a_2^\mu$ should be considered and the magnetic field vector $b^\mu$ plays the  role of the other anisotropy direction whereas for ellipsoidal scenario, $b^\mu$ becomes redundant. A suitable choice for the independent basis tensors is  given by  the following set:
\bea
A^{\mn}&=&\frac{\tilde u^\mu \tilde u^\nu}{\tilde u^2}~,\\
B^{\mn}&=&\frac{\tilde a_2^\mu \tilde a_2^\nu}{\tilde a_2^2}~,\\
C^{\mn}&=&\frac{\tilde u^\mu \tilde a_2^\nu +\tilde a_2^\mu \tilde u^\nu}{\sqrt{\tilde u^2}\sqrt{\tilde a_2^2}}~,\\
D^{\mn}&=& \frac{\tilde a_1^\mu \tilde a_1^\nu}{\tilde a_1^2}~,\\
E^{\mn}&=& \frac{\tilde u^\mu \tilde a_1^\nu +\tilde a_1^\mu \tilde u^\nu}{\sqrt{\tilde u^2}\sqrt{\tilde a_1^2}}~,\\
F^{\mn}&=& \frac{\tilde a_1^\mu \tilde a_2^\nu +\tilde a_2^\mu \tilde a_1^\nu}{\sqrt{\tilde a_1^2}\sqrt{\tilde a_2^2}}~,
\eea
where, we have defined 
\begin{align}
\tilde u^\mu&=\left(\eta^{\mn}-\frac{p^\mu p^\nu}{p^2}\right)u_\nu=V^{\mn}u_\nu~,\\
\tilde a_2^\mu&=\left(V^\mu_\nu-A^\mu_\nu\right)a_2^\nu=U^\mu_\nu a_2^\nu~,\\
\tilde a_1^\mu&=\left(U^\mu_\nu-B^\mu_\nu\right)a_1^\nu= R^\mu_\nu a_1^\nu~.
\end{align}
 Using the projection opeartor properties, one can obtain the form factors in terms of the self energy components. For example, it is easy to obtain the relations like
\begin{align}
 \alpha&=\frac{1}{\tilde{u}^2} u_\mu u_\nu\Pi^{\mu\nu}\,,\\
 \beta&=\frac{1}{\tilde{a}_2^2}\Big((a_2)_\mu(a_2)_\nu-2\frac{a_2\cdot\tilde{u}}{\tilde{u}^2}u_\mu (a_2)_\nu+\big(\frac{a_2\cdot\tilde{u}}{\tilde{u}^2}\big)^2 u_\mu u_\nu\Big)\Pi^{\mu\nu}\,,\\
 \gamma&=\frac{1}{\sqrt{\tilde{u}^2 \tilde{a}_2^2}}\Big(u_\mu (a_2)_\nu-\frac{a_2\cdot\tilde{u}}{\tilde{u}^2} u_\mu u_\nu\Big)\Pi^{\mu\nu}\,.
\end{align}
Together with similar relations for the other three form factors, one can express the mode functions in terms of the  self energy components. It is easy to show that due to the transversality condition and the symmetric nature of the polarization tensor,   explicit evaluation of only six spatial components is sufficient to obtain all the form factors and hence, the mode functions. In the present analysis, the modified quark loop contribution in presence of the external magnetic field has been obtained analytically whereas the anisotropic pure gluon contribution or the anisotropic gluon self energy  without $eB$ has been evaluated numerically. To obtain an analytic expression in the small-$\xi$ limit, it is convenient to rotate the lab reference axes to the parton frame \cite{Carrington:2021bnk,Kasmaei:2018yrr,Kasmaei:2016apv}. In that case  the rotated anisotropy and momentum directions are given by 
\begin{align}
 a_2^\mu&=\big(0,0,-\sin\theta_p,\cos\theta_p\big)\,,\\
 a_1^\mu&=\big(0,\sin\phi_p,\cos\theta_p\cos\phi_p,\sin\theta_p\cos\phi_p\big)\,,\\
 \hat{p}^\mu&=\big(\hat{p}_0,0,0,1\big)\,,
\end{align}
where $\hat{p}_0=p_0/|{\bm p}|$. Once the  scaled form factors are expanded in $\xi_1$ and $\xi_2$, the solid angle integral can be analytically performed. For example, keeping only up to the leading order terms in the small-$\xi$ expansion, one obtains  
 \begin{align}
 \alpha&=\frac{1}{24} \left(\hat{p}_0^2-1\right)\left(-3 \cos (2 \theta_p) (\xi_1-2 \xi_2) \left(6 \hat{p}_0^2+\left(3 \hat{p}_0^2-2\right) \hat{p}_0 (\log (\hat{p}_0-1)-\log (\hat{p}_0+1))-2\right)\right.\nn\\
 &\left.+6 \xi_1 \sin ^2\theta_p \left(6 \hat{p}_0^2+\left(3 \hat{p}_0^2-2\right) \hat{p}_0 (\log (\hat{p}_0-1)-\log (\hat{p}_0+1))-2\right) \cos (2 \phi_p)\right.\nn\\&\left.-3 \hat{p}_0 (\log (\hat{p}_0-1)-\log (\hat{p}_0+1)) \left(\xi_1 \left(\hat{p}_0^2-2\right)-2 \xi_2 \hat{p}_0^2+4\right)+\xi_1 \left(10-6 \hat{p}_0^2\right)+4 \left(\xi_2+3 \xi_2 \hat{p}_0^2-6\right)\right) \,\\
 \beta&=\frac{1}{192} \left(2 \left(\hat{p}_0^2-1\right) \left(\cos (2 \theta_p) \left(42 \hat{p}_0^2+3 \left(7 \hat{p}_0^2-5\right) \hat{p}_0 (\log (\hat{p}_0-1)-\log (\hat{p}_0+1))-16\right) (\xi_1 \cos (2 \phi_p)+\xi_1-2 \xi_2)\right.\right.\nn\\&\left.\left.+3 \hat{p}_0 (\log (\hat{p}_0-1)-\log (\hat{p}_0+1)) \left(\xi_1 \left(\hat{p}_0^2-3\right)-2 \xi_2 \left(\hat{p}_0^2+1\right)+8\right)-3 \xi_1 \hat{p}_0 \left(\left(3 \hat{p}_0^2-1\right) \log (\hat{p}_0-1)\right.\right.\right.\nn\\&\left.\left.\left.+\left(1-3 \hat{p}_0^2\right) \log (\hat{p}_0+1)+6 \hat{p}_0\right) \cos (2 \phi_p)\right)+4 \hat{p}_0^2 \left(-11 \xi_1-2 \xi_2+3 \xi_1 \hat{p}_0^2-6 \xi_2 \hat{p}_0^2+24\right)\right)\,,\\
 \gamma&=-\frac{1}{12}\sqrt{\hat{p}_0^2-1}\left(11 \hat{p}_0-12\hat{p}_0^3+3\left(1-5\hat{p}_0^2+4\hat{p}_0^4\right)\coth^{-1}(\hat{p}_0)\right)\cos\theta_p\left(\xi_1-2\xi_2+\xi_1 \cos(2\phi_p)\right)\sin\theta_p\,.
 \end{align}
  
In the limit $\xi_1=0$, one can recover the small-$\xi$ expanded known results given in Ref.~\cite{Romatschke:2003ms} once  the  differences in the  basis construction (see for example Ref.~\cite{Dumitru:2007hy} for the covariant formulation) is taken into account.

\bibliography{arxiv_submit_v3}

\begin{thebibliography}{90}%
\makeatletter
\providecommand \@ifxundefined [1]{%
 \@ifx{#1\undefined}
}%
\providecommand \@ifnum [1]{%
 \ifnum #1\expandafter \@firstoftwo
 \else \expandafter \@secondoftwo
 \fi
}%
\providecommand \@ifx [1]{%
 \ifx #1\expandafter \@firstoftwo
 \else \expandafter \@secondoftwo
 \fi
}%
\providecommand \natexlab [1]{#1}%
\providecommand \enquote  [1]{``#1''}%
\providecommand \bibnamefont  [1]{#1}%
\providecommand \bibfnamefont [1]{#1}%
\providecommand \citenamefont [1]{#1}%
\providecommand \href@noop [0]{\@secondoftwo}%
\providecommand \href [0]{\begingroup \@sanitize@url \@href}%
\providecommand \@href[1]{\@@startlink{#1}\@@href}%
\providecommand \@@href[1]{\endgroup#1\@@endlink}%
\providecommand \@sanitize@url [0]{\catcode `\\12\catcode `\$12\catcode
  `\&12\catcode `\#12\catcode `\^12\catcode `\_12\catcode `\%12\relax}%
\providecommand \@@startlink[1]{}%
\providecommand \@@endlink[0]{}%
\providecommand \url  [0]{\begingroup\@sanitize@url \@url }%
\providecommand \@url [1]{\endgroup\@href {#1}{\urlprefix }}%
\providecommand \urlprefix  [0]{URL }%
\providecommand \Eprint [0]{\href }%
\providecommand \doibase [0]{https://doi.org/}%
\providecommand \selectlanguage [0]{\@gobble}%
\providecommand \bibinfo  [0]{\@secondoftwo}%
\providecommand \bibfield  [0]{\@secondoftwo}%
\providecommand \translation [1]{[#1]}%
\providecommand \BibitemOpen [0]{}%
\providecommand \bibitemStop [0]{}%
\providecommand \bibitemNoStop [0]{.\EOS\space}%
\providecommand \EOS [0]{\spacefactor3000\relax}%
\providecommand \BibitemShut  [1]{\csname bibitem#1\endcsname}%
\let\auto@bib@innerbib\@empty
\bibitem [{\citenamefont {Busza}\ \emph {et~al.}(2018)\citenamefont {Busza},
  \citenamefont {Rajagopal},\ and\ \citenamefont {van~der
  Schee}}]{Busza:2018rrf}%
  \BibitemOpen
  \bibfield  {author} {\bibinfo {author} {\bibfnamefont {W.}~\bibnamefont
  {Busza}}, \bibinfo {author} {\bibfnamefont {K.}~\bibnamefont {Rajagopal}},\
  and\ \bibinfo {author} {\bibfnamefont {W.}~\bibnamefont {van~der Schee}},\
  }\bibfield  {title} {\bibinfo {title} {{Heavy Ion Collisions: The Big
  Picture, and the Big Questions}},\ }\href
  {https://doi.org/10.1146/annurev-nucl-101917-020852} {\bibfield  {journal}
  {\bibinfo  {journal} {Ann. Rev. Nucl. Part. Sci.}\ }\textbf {\bibinfo
  {volume} {68}},\ \bibinfo {pages} {339} (\bibinfo {year} {2018})},\ \Eprint
  {https://arxiv.org/abs/1802.04801} {arXiv:1802.04801 [hep-ph]} \BibitemShut
  {NoStop}%
\bibitem [{\citenamefont {Pasechnik}\ and\ \citenamefont
  {\v{S}umbera}(2017)}]{Pasechnik:2016wkt}%
  \BibitemOpen
  \bibfield  {author} {\bibinfo {author} {\bibfnamefont {R.}~\bibnamefont
  {Pasechnik}}\ and\ \bibinfo {author} {\bibfnamefont {M.}~\bibnamefont
  {\v{S}umbera}},\ }\bibfield  {title} {\bibinfo {title} {{Phenomenological
  Review on Quark\textendash{}Gluon Plasma: Concepts vs. Observations}},\
  }\href {https://doi.org/10.3390/universe3010007} {\bibfield  {journal}
  {\bibinfo  {journal} {Universe}\ }\textbf {\bibinfo {volume} {3}},\ \bibinfo
  {pages} {7} (\bibinfo {year} {2017})},\ \Eprint
  {https://arxiv.org/abs/1611.01533} {arXiv:1611.01533 [hep-ph]} \BibitemShut
  {NoStop}%
\bibitem [{\citenamefont {Heinz}(2004)}]{Heinz:2004pj}%
  \BibitemOpen
  \bibfield  {author} {\bibinfo {author} {\bibfnamefont {U.~W.}\ \bibnamefont
  {Heinz}},\ }\bibfield  {title} {\bibinfo {title} {{Thermalization at RHIC}},\
  }\href {https://doi.org/10.1063/1.1843595} {\bibfield  {journal} {\bibinfo
  {journal} {AIP Conf. Proc.}\ }\textbf {\bibinfo {volume} {739}},\ \bibinfo
  {pages} {163} (\bibinfo {year} {2004})},\ \Eprint
  {https://arxiv.org/abs/nucl-th/0407067} {arXiv:nucl-th/0407067} \BibitemShut
  {NoStop}%
\bibitem [{\citenamefont {Huovinen}\ \emph {et~al.}(2001)\citenamefont
  {Huovinen}, \citenamefont {Kolb}, \citenamefont {Heinz}, \citenamefont
  {Ruuskanen},\ and\ \citenamefont {Voloshin}}]{Huovinen:2001cy}%
  \BibitemOpen
  \bibfield  {author} {\bibinfo {author} {\bibfnamefont {P.}~\bibnamefont
  {Huovinen}}, \bibinfo {author} {\bibfnamefont {P.~F.}\ \bibnamefont {Kolb}},
  \bibinfo {author} {\bibfnamefont {U.~W.}\ \bibnamefont {Heinz}}, \bibinfo
  {author} {\bibfnamefont {P.~V.}\ \bibnamefont {Ruuskanen}},\ and\ \bibinfo
  {author} {\bibfnamefont {S.~A.}\ \bibnamefont {Voloshin}},\ }\bibfield
  {title} {\bibinfo {title} {{Radial and elliptic flow at RHIC: Further
  predictions}},\ }\href {https://doi.org/10.1016/S0370-2693(01)00219-2}
  {\bibfield  {journal} {\bibinfo  {journal} {Phys. Lett. B}\ }\textbf
  {\bibinfo {volume} {503}},\ \bibinfo {pages} {58} (\bibinfo {year} {2001})},\
  \Eprint {https://arxiv.org/abs/hep-ph/0101136} {arXiv:hep-ph/0101136}
  \BibitemShut {NoStop}%
\bibitem [{\citenamefont {Hirano}\ and\ \citenamefont
  {Tsuda}(2002)}]{Hirano:2002ds}%
  \BibitemOpen
  \bibfield  {author} {\bibinfo {author} {\bibfnamefont {T.}~\bibnamefont
  {Hirano}}\ and\ \bibinfo {author} {\bibfnamefont {K.}~\bibnamefont {Tsuda}},\
  }\bibfield  {title} {\bibinfo {title} {{Collective flow and two pion
  correlations from a relativistic hydrodynamic model with early chemical
  freezeout}},\ }\href {https://doi.org/10.1103/PhysRevC.66.054905} {\bibfield
  {journal} {\bibinfo  {journal} {Phys. Rev. C}\ }\textbf {\bibinfo {volume}
  {66}},\ \bibinfo {pages} {054905} (\bibinfo {year} {2002})},\ \Eprint
  {https://arxiv.org/abs/nucl-th/0205043} {arXiv:nucl-th/0205043} \BibitemShut
  {NoStop}%
\bibitem [{\citenamefont {Baier}\ \emph {et~al.}(2001)\citenamefont {Baier},
  \citenamefont {Mueller}, \citenamefont {Schiff},\ and\ \citenamefont
  {Son}}]{Baier:2000sb}%
  \BibitemOpen
  \bibfield  {author} {\bibinfo {author} {\bibfnamefont {R.}~\bibnamefont
  {Baier}}, \bibinfo {author} {\bibfnamefont {A.~H.}\ \bibnamefont {Mueller}},
  \bibinfo {author} {\bibfnamefont {D.}~\bibnamefont {Schiff}},\ and\ \bibinfo
  {author} {\bibfnamefont {D.~T.}\ \bibnamefont {Son}},\ }\bibfield  {title}
  {\bibinfo {title} {{`Bottom up' thermalization in heavy ion collisions}},\
  }\href {https://doi.org/10.1016/S0370-2693(01)00191-5} {\bibfield  {journal}
  {\bibinfo  {journal} {Phys. Lett. B}\ }\textbf {\bibinfo {volume} {502}},\
  \bibinfo {pages} {51} (\bibinfo {year} {2001})},\ \Eprint
  {https://arxiv.org/abs/hep-ph/0009237} {arXiv:hep-ph/0009237} \BibitemShut
  {NoStop}%
\bibitem [{\citenamefont {Berges}\ \emph {et~al.}(2021)\citenamefont {Berges},
  \citenamefont {Heller}, \citenamefont {Mazeliauskas},\ and\ \citenamefont
  {Venugopalan}}]{Berges:2020fwq}%
  \BibitemOpen
  \bibfield  {author} {\bibinfo {author} {\bibfnamefont {J.}~\bibnamefont
  {Berges}}, \bibinfo {author} {\bibfnamefont {M.~P.}\ \bibnamefont {Heller}},
  \bibinfo {author} {\bibfnamefont {A.}~\bibnamefont {Mazeliauskas}},\ and\
  \bibinfo {author} {\bibfnamefont {R.}~\bibnamefont {Venugopalan}},\
  }\bibfield  {title} {\bibinfo {title} {{QCD thermalization: Ab initio
  approaches and interdisciplinary connections}},\ }\href
  {https://doi.org/10.1103/RevModPhys.93.035003} {\bibfield  {journal}
  {\bibinfo  {journal} {Rev. Mod. Phys.}\ }\textbf {\bibinfo {volume} {93}},\
  \bibinfo {pages} {035003} (\bibinfo {year} {2021})},\ \Eprint
  {https://arxiv.org/abs/2005.12299} {arXiv:2005.12299 [hep-th]} \BibitemShut
  {NoStop}%
\bibitem [{\citenamefont {Epelbaum}\ and\ \citenamefont
  {Gelis}(2013)}]{Epelbaum:2013ekf}%
  \BibitemOpen
  \bibfield  {author} {\bibinfo {author} {\bibfnamefont {T.}~\bibnamefont
  {Epelbaum}}\ and\ \bibinfo {author} {\bibfnamefont {F.}~\bibnamefont
  {Gelis}},\ }\bibfield  {title} {\bibinfo {title} {{Pressure isotropization in
  high energy heavy ion collisions}},\ }\href
  {https://doi.org/10.1103/PhysRevLett.111.232301} {\bibfield  {journal}
  {\bibinfo  {journal} {Phys. Rev. Lett.}\ }\textbf {\bibinfo {volume} {111}},\
  \bibinfo {pages} {232301} (\bibinfo {year} {2013})},\ \Eprint
  {https://arxiv.org/abs/1307.2214} {arXiv:1307.2214 [hep-ph]} \BibitemShut
  {NoStop}%
\bibitem [{\citenamefont {Berges}\ \emph {et~al.}(2014)\citenamefont {Berges},
  \citenamefont {Boguslavski}, \citenamefont {Schlichting},\ and\ \citenamefont
  {Venugopalan}}]{Berges:2013fga}%
  \BibitemOpen
  \bibfield  {author} {\bibinfo {author} {\bibfnamefont {J.}~\bibnamefont
  {Berges}}, \bibinfo {author} {\bibfnamefont {K.}~\bibnamefont {Boguslavski}},
  \bibinfo {author} {\bibfnamefont {S.}~\bibnamefont {Schlichting}},\ and\
  \bibinfo {author} {\bibfnamefont {R.}~\bibnamefont {Venugopalan}},\
  }\bibfield  {title} {\bibinfo {title} {{Universal attractor in a highly
  occupied non-Abelian plasma}},\ }\href
  {https://doi.org/10.1103/PhysRevD.89.114007} {\bibfield  {journal} {\bibinfo
  {journal} {Phys. Rev. D}\ }\textbf {\bibinfo {volume} {89}},\ \bibinfo
  {pages} {114007} (\bibinfo {year} {2014})},\ \Eprint
  {https://arxiv.org/abs/1311.3005} {arXiv:1311.3005 [hep-ph]} \BibitemShut
  {NoStop}%
\bibitem [{\citenamefont {Strickland}(2015)}]{Strickland:2013uga}%
  \BibitemOpen
  \bibfield  {author} {\bibinfo {author} {\bibfnamefont {M.}~\bibnamefont
  {Strickland}},\ }\bibfield  {title} {\bibinfo {title} {{Thermalization and
  isotropization in heavy-ion collisions}},\ }\href
  {https://doi.org/10.1007/s12043-015-0972-1} {\bibfield  {journal} {\bibinfo
  {journal} {Pramana}\ }\textbf {\bibinfo {volume} {84}},\ \bibinfo {pages}
  {671} (\bibinfo {year} {2015})},\ \Eprint {https://arxiv.org/abs/1312.2285}
  {arXiv:1312.2285 [hep-ph]} \BibitemShut {NoStop}%
\bibitem [{\citenamefont {Chesler}\ and\ \citenamefont
  {Yaffe}(2009)}]{Chesler:2008hg}%
  \BibitemOpen
  \bibfield  {author} {\bibinfo {author} {\bibfnamefont {P.~M.}\ \bibnamefont
  {Chesler}}\ and\ \bibinfo {author} {\bibfnamefont {L.~G.}\ \bibnamefont
  {Yaffe}},\ }\bibfield  {title} {\bibinfo {title} {{Horizon formation and
  far-from-equilibrium isotropization in supersymmetric Yang-Mills plasma}},\
  }\href {https://doi.org/10.1103/PhysRevLett.102.211601} {\bibfield  {journal}
  {\bibinfo  {journal} {Phys. Rev. Lett.}\ }\textbf {\bibinfo {volume} {102}},\
  \bibinfo {pages} {211601} (\bibinfo {year} {2009})},\ \Eprint
  {https://arxiv.org/abs/0812.2053} {arXiv:0812.2053 [hep-th]} \BibitemShut
  {NoStop}%
\bibitem [{\citenamefont {Chesler}\ and\ \citenamefont
  {Yaffe}(2010)}]{Chesler:2009cy}%
  \BibitemOpen
  \bibfield  {author} {\bibinfo {author} {\bibfnamefont {P.~M.}\ \bibnamefont
  {Chesler}}\ and\ \bibinfo {author} {\bibfnamefont {L.~G.}\ \bibnamefont
  {Yaffe}},\ }\bibfield  {title} {\bibinfo {title} {{Boost invariant flow,
  black hole formation, and far-from-equilibrium dynamics in N = 4
  supersymmetric Yang-Mills theory}},\ }\href
  {https://doi.org/10.1103/PhysRevD.82.026006} {\bibfield  {journal} {\bibinfo
  {journal} {Phys. Rev. D}\ }\textbf {\bibinfo {volume} {82}},\ \bibinfo
  {pages} {026006} (\bibinfo {year} {2010})},\ \Eprint
  {https://arxiv.org/abs/0906.4426} {arXiv:0906.4426 [hep-th]} \BibitemShut
  {NoStop}%
\bibitem [{\citenamefont {Heller}\ \emph {et~al.}(2012)\citenamefont {Heller},
  \citenamefont {Janik},\ and\ \citenamefont {Witaszczyk}}]{Heller:2011ju}%
  \BibitemOpen
  \bibfield  {author} {\bibinfo {author} {\bibfnamefont {M.~P.}\ \bibnamefont
  {Heller}}, \bibinfo {author} {\bibfnamefont {R.~A.}\ \bibnamefont {Janik}},\
  and\ \bibinfo {author} {\bibfnamefont {P.}~\bibnamefont {Witaszczyk}},\
  }\bibfield  {title} {\bibinfo {title} {{The characteristics of thermalization
  of boost-invariant plasma from holography}},\ }\href
  {https://doi.org/10.1103/PhysRevLett.108.201602} {\bibfield  {journal}
  {\bibinfo  {journal} {Phys. Rev. Lett.}\ }\textbf {\bibinfo {volume} {108}},\
  \bibinfo {pages} {201602} (\bibinfo {year} {2012})},\ \Eprint
  {https://arxiv.org/abs/1103.3452} {arXiv:1103.3452 [hep-th]} \BibitemShut
  {NoStop}%
\bibitem [{\citenamefont {Casalderrey-Solana}\ \emph
  {et~al.}(2014)\citenamefont {Casalderrey-Solana}, \citenamefont {Liu},
  \citenamefont {Mateos}, \citenamefont {Rajagopal},\ and\ \citenamefont
  {Wiedemann}}]{Casalderrey-Solana:2011dxg}%
  \BibitemOpen
  \bibfield  {author} {\bibinfo {author} {\bibfnamefont {J.}~\bibnamefont
  {Casalderrey-Solana}}, \bibinfo {author} {\bibfnamefont {H.}~\bibnamefont
  {Liu}}, \bibinfo {author} {\bibfnamefont {D.}~\bibnamefont {Mateos}},
  \bibinfo {author} {\bibfnamefont {K.}~\bibnamefont {Rajagopal}},\ and\
  \bibinfo {author} {\bibfnamefont {U.~A.}\ \bibnamefont {Wiedemann}},\ }\href
  {https://doi.org/10.1017/CBO9781139136747} {\emph {\bibinfo {title}
  {{Gauge/String Duality, Hot QCD and Heavy Ion Collisions}}}}\ (\bibinfo
  {publisher} {Cambridge University Press},\ \bibinfo {year} {2014})\ \Eprint
  {https://arxiv.org/abs/1101.0618} {arXiv:1101.0618 [hep-th]} \BibitemShut
  {NoStop}%
\bibitem [{\citenamefont {Romatschke}(2017)}]{Romatschke:2016hle}%
  \BibitemOpen
  \bibfield  {author} {\bibinfo {author} {\bibfnamefont {P.}~\bibnamefont
  {Romatschke}},\ }\bibfield  {title} {\bibinfo {title} {{Do nuclear collisions
  create a locally equilibrated quark\textendash{}gluon plasma?}},\ }\href
  {https://doi.org/10.1140/epjc/s10052-016-4567-x} {\bibfield  {journal}
  {\bibinfo  {journal} {Eur. Phys. J. C}\ }\textbf {\bibinfo {volume} {77}},\
  \bibinfo {pages} {21} (\bibinfo {year} {2017})},\ \Eprint
  {https://arxiv.org/abs/1609.02820} {arXiv:1609.02820 [nucl-th]} \BibitemShut
  {NoStop}%
\bibitem [{\citenamefont {Romatschke}\ and\ \citenamefont
  {Romatschke}(2019)}]{Romatschke:2017ejr}%
  \BibitemOpen
  \bibfield  {author} {\bibinfo {author} {\bibfnamefont {P.}~\bibnamefont
  {Romatschke}}\ and\ \bibinfo {author} {\bibfnamefont {U.}~\bibnamefont
  {Romatschke}},\ }\href {https://doi.org/10.1017/9781108651998} {\emph
  {\bibinfo {title} {{Relativistic Fluid Dynamics In and Out of
  Equilibrium}}}},\ Cambridge Monographs on Mathematical Physics\ (\bibinfo
  {publisher} {Cambridge University Press},\ \bibinfo {year} {2019})\ \Eprint
  {https://arxiv.org/abs/1712.05815} {arXiv:1712.05815 [nucl-th]} \BibitemShut
  {NoStop}%
\bibitem [{\citenamefont {Florkowski}\ \emph {et~al.}(2018)\citenamefont
  {Florkowski}, \citenamefont {Heller},\ and\ \citenamefont
  {Spalinski}}]{Florkowski:2017olj}%
  \BibitemOpen
  \bibfield  {author} {\bibinfo {author} {\bibfnamefont {W.}~\bibnamefont
  {Florkowski}}, \bibinfo {author} {\bibfnamefont {M.~P.}\ \bibnamefont
  {Heller}},\ and\ \bibinfo {author} {\bibfnamefont {M.}~\bibnamefont
  {Spalinski}},\ }\bibfield  {title} {\bibinfo {title} {{New theories of
  relativistic hydrodynamics in the LHC era}},\ }\href
  {https://doi.org/10.1088/1361-6633/aaa091} {\bibfield  {journal} {\bibinfo
  {journal} {Rept. Prog. Phys.}\ }\textbf {\bibinfo {volume} {81}},\ \bibinfo
  {pages} {046001} (\bibinfo {year} {2018})},\ \Eprint
  {https://arxiv.org/abs/1707.02282} {arXiv:1707.02282 [hep-ph]} \BibitemShut
  {NoStop}%
\bibitem [{\citenamefont {Jaiswal}\ and\ \citenamefont
  {Roy}(2016)}]{Jaiswal:2016hex}%
  \BibitemOpen
  \bibfield  {author} {\bibinfo {author} {\bibfnamefont {A.}~\bibnamefont
  {Jaiswal}}\ and\ \bibinfo {author} {\bibfnamefont {V.}~\bibnamefont {Roy}},\
  }\bibfield  {title} {\bibinfo {title} {{Relativistic hydrodynamics in
  heavy-ion collisions: general aspects and recent developments}},\ }\href
  {https://doi.org/10.1155/2016/9623034} {\bibfield  {journal} {\bibinfo
  {journal} {Adv. High Energy Phys.}\ }\textbf {\bibinfo {volume} {2016}},\
  \bibinfo {pages} {9623034} (\bibinfo {year} {2016})},\ \Eprint
  {https://arxiv.org/abs/1605.08694} {arXiv:1605.08694 [nucl-th]} \BibitemShut
  {NoStop}%
\bibitem [{\citenamefont {Jeon}\ and\ \citenamefont
  {Heinz}(2015)}]{Jeon:2015dfa}%
  \BibitemOpen
  \bibfield  {author} {\bibinfo {author} {\bibfnamefont {S.}~\bibnamefont
  {Jeon}}\ and\ \bibinfo {author} {\bibfnamefont {U.}~\bibnamefont {Heinz}},\
  }\bibfield  {title} {\bibinfo {title} {{Introduction to Hydrodynamics}},\
  }\href {https://doi.org/10.1142/S0218301315300106} {\bibfield  {journal}
  {\bibinfo  {journal} {Int. J. Mod. Phys. E}\ }\textbf {\bibinfo {volume}
  {24}},\ \bibinfo {pages} {1530010} (\bibinfo {year} {2015})},\ \Eprint
  {https://arxiv.org/abs/1503.03931} {arXiv:1503.03931 [hep-ph]} \BibitemShut
  {NoStop}%
\bibitem [{\citenamefont {Kovtun}(2012)}]{Kovtun:2012rj}%
  \BibitemOpen
  \bibfield  {author} {\bibinfo {author} {\bibfnamefont {P.}~\bibnamefont
  {Kovtun}},\ }\bibfield  {title} {\bibinfo {title} {{Lectures on hydrodynamic
  fluctuations in relativistic theories}},\ }\href
  {https://doi.org/10.1088/1751-8113/45/47/473001} {\bibfield  {journal}
  {\bibinfo  {journal} {J. Phys. A}\ }\textbf {\bibinfo {volume} {45}},\
  \bibinfo {pages} {473001} (\bibinfo {year} {2012})},\ \Eprint
  {https://arxiv.org/abs/1205.5040} {arXiv:1205.5040 [hep-th]} \BibitemShut
  {NoStop}%
\bibitem [{\citenamefont {Mandal}\ and\ \citenamefont
  {Roy}(2013{\natexlab{a}})}]{Mandal:2013jkd}%
  \BibitemOpen
  \bibfield  {author} {\bibinfo {author} {\bibfnamefont {M.}~\bibnamefont
  {Mandal}}\ and\ \bibinfo {author} {\bibfnamefont {P.}~\bibnamefont {Roy}},\
  }\bibfield  {title} {\bibinfo {title} {{Some Aspects of Anisotropic
  Quark-Gluon Plasma}},\ }\href {https://doi.org/10.1155/2013/371908}
  {\bibfield  {journal} {\bibinfo  {journal} {Adv. High Energy Phys.}\ }\textbf
  {\bibinfo {volume} {2013}},\ \bibinfo {pages} {371908} (\bibinfo {year}
  {2013}{\natexlab{a}})}\BibitemShut {NoStop}%
\bibitem [{\citenamefont {Romatschke}\ and\ \citenamefont
  {Strickland}(2003)}]{Romatschke:2003ms}%
  \BibitemOpen
  \bibfield  {author} {\bibinfo {author} {\bibfnamefont {P.}~\bibnamefont
  {Romatschke}}\ and\ \bibinfo {author} {\bibfnamefont {M.}~\bibnamefont
  {Strickland}},\ }\bibfield  {title} {\bibinfo {title} {{Collective modes of
  an anisotropic quark gluon plasma}},\ }\href
  {https://doi.org/10.1103/PhysRevD.68.036004} {\bibfield  {journal} {\bibinfo
  {journal} {Phys. Rev. D}\ }\textbf {\bibinfo {volume} {68}},\ \bibinfo
  {pages} {036004} (\bibinfo {year} {2003})},\ \Eprint
  {https://arxiv.org/abs/hep-ph/0304092} {arXiv:hep-ph/0304092} \BibitemShut
  {NoStop}%
\bibitem [{\citenamefont {Arnold}\ \emph {et~al.}(2003)\citenamefont {Arnold},
  \citenamefont {Lenaghan},\ and\ \citenamefont {Moore}}]{Arnold:2003rq}%
  \BibitemOpen
  \bibfield  {author} {\bibinfo {author} {\bibfnamefont {P.~B.}\ \bibnamefont
  {Arnold}}, \bibinfo {author} {\bibfnamefont {J.}~\bibnamefont {Lenaghan}},\
  and\ \bibinfo {author} {\bibfnamefont {G.~D.}\ \bibnamefont {Moore}},\
  }\bibfield  {title} {\bibinfo {title} {{QCD plasma instabilities and bottom
  up thermalization}},\ }\href {https://doi.org/10.1088/1126-6708/2003/08/002}
  {\bibfield  {journal} {\bibinfo  {journal} {JHEP}\ }\textbf {\bibinfo
  {volume} {08}},\ \bibinfo {pages} {002}},\ \Eprint
  {https://arxiv.org/abs/hep-ph/0307325} {arXiv:hep-ph/0307325} \BibitemShut
  {NoStop}%
\bibitem [{\citenamefont {Mrowczynski}(1988)}]{Mrowczynski:1988dz}%
  \BibitemOpen
  \bibfield  {author} {\bibinfo {author} {\bibfnamefont {S.}~\bibnamefont
  {Mrowczynski}},\ }\bibfield  {title} {\bibinfo {title} {{Stream Instabilities
  of the Quark - Gluon Plasma}},\ }\href
  {https://doi.org/10.1016/0370-2693(88)90124-4} {\bibfield  {journal}
  {\bibinfo  {journal} {Phys. Lett. B}\ }\textbf {\bibinfo {volume} {214}},\
  \bibinfo {pages} {587} (\bibinfo {year} {1988})},\ \bibinfo {note} {[Erratum:
  Phys.Lett.B 656, 273 (2007)]}\BibitemShut {NoStop}%
\bibitem [{\citenamefont {Mrowczynski}(1993)}]{Mrowczynski:1993qm}%
  \BibitemOpen
  \bibfield  {author} {\bibinfo {author} {\bibfnamefont {S.}~\bibnamefont
  {Mrowczynski}},\ }\bibfield  {title} {\bibinfo {title} {{Plasma instability
  at the initial stage of ultrarelativistic heavy ion collisions}},\ }\href
  {https://doi.org/10.1016/0370-2693(93)91330-P} {\bibfield  {journal}
  {\bibinfo  {journal} {Phys. Lett. B}\ }\textbf {\bibinfo {volume} {314}},\
  \bibinfo {pages} {118} (\bibinfo {year} {1993})}\BibitemShut {NoStop}%
\bibitem [{\citenamefont {Mrowczynski}\ and\ \citenamefont
  {Thoma}(2000)}]{Mrowczynski:2000ed}%
  \BibitemOpen
  \bibfield  {author} {\bibinfo {author} {\bibfnamefont {S.}~\bibnamefont
  {Mrowczynski}}\ and\ \bibinfo {author} {\bibfnamefont {M.~H.}\ \bibnamefont
  {Thoma}},\ }\bibfield  {title} {\bibinfo {title} {{Hard loop approach to
  anisotropic systems}},\ }\href {https://doi.org/10.1103/PhysRevD.62.036011}
  {\bibfield  {journal} {\bibinfo  {journal} {Phys. Rev. D}\ }\textbf {\bibinfo
  {volume} {62}},\ \bibinfo {pages} {036011} (\bibinfo {year} {2000})},\
  \Eprint {https://arxiv.org/abs/hep-ph/0001164} {arXiv:hep-ph/0001164}
  \BibitemShut {NoStop}%
\bibitem [{\citenamefont {Randrup}\ and\ \citenamefont
  {Mrowczynski}(2003)}]{Randrup:2003cw}%
  \BibitemOpen
  \bibfield  {author} {\bibinfo {author} {\bibfnamefont {J.}~\bibnamefont
  {Randrup}}\ and\ \bibinfo {author} {\bibfnamefont {S.}~\bibnamefont
  {Mrowczynski}},\ }\bibfield  {title} {\bibinfo {title} {{Chromodynamic Weibel
  instabilities in relativistic nuclear collisions}},\ }\href
  {https://doi.org/10.1103/PhysRevC.68.034909} {\bibfield  {journal} {\bibinfo
  {journal} {Phys. Rev. C}\ }\textbf {\bibinfo {volume} {68}},\ \bibinfo
  {pages} {034909} (\bibinfo {year} {2003})},\ \Eprint
  {https://arxiv.org/abs/nucl-th/0303021} {arXiv:nucl-th/0303021} \BibitemShut
  {NoStop}%
\bibitem [{\citenamefont {Strickland}(2014)}]{Strickland:2014pga}%
  \BibitemOpen
  \bibfield  {author} {\bibinfo {author} {\bibfnamefont {M.}~\bibnamefont
  {Strickland}},\ }\bibfield  {title} {\bibinfo {title} {{Anisotropic
  Hydrodynamics: Three lectures}},\ }\href
  {https://doi.org/10.5506/APhysPolB.45.2355} {\bibfield  {journal} {\bibinfo
  {journal} {Acta Phys. Polon. B}\ }\textbf {\bibinfo {volume} {45}},\ \bibinfo
  {pages} {2355} (\bibinfo {year} {2014})},\ \Eprint
  {https://arxiv.org/abs/1410.5786} {arXiv:1410.5786 [nucl-th]} \BibitemShut
  {NoStop}%
\bibitem [{\citenamefont {Alqahtani}\ \emph {et~al.}(2018)\citenamefont
  {Alqahtani}, \citenamefont {Nopoush},\ and\ \citenamefont
  {Strickland}}]{Alqahtani:2017mhy}%
  \BibitemOpen
  \bibfield  {author} {\bibinfo {author} {\bibfnamefont {M.}~\bibnamefont
  {Alqahtani}}, \bibinfo {author} {\bibfnamefont {M.}~\bibnamefont {Nopoush}},\
  and\ \bibinfo {author} {\bibfnamefont {M.}~\bibnamefont {Strickland}},\
  }\bibfield  {title} {\bibinfo {title} {{Relativistic anisotropic
  hydrodynamics}},\ }\href {https://doi.org/10.1016/j.ppnp.2018.05.004}
  {\bibfield  {journal} {\bibinfo  {journal} {Prog. Part. Nucl. Phys.}\
  }\textbf {\bibinfo {volume} {101}},\ \bibinfo {pages} {204} (\bibinfo {year}
  {2018})},\ \Eprint {https://arxiv.org/abs/1712.03282} {arXiv:1712.03282
  [nucl-th]} \BibitemShut {NoStop}%
\bibitem [{\citenamefont {Ghiglieri}\ \emph {et~al.}(2020)\citenamefont
  {Ghiglieri}, \citenamefont {Kurkela}, \citenamefont {Strickland},\ and\
  \citenamefont {Vuorinen}}]{Ghiglieri:2020dpq}%
  \BibitemOpen
  \bibfield  {author} {\bibinfo {author} {\bibfnamefont {J.}~\bibnamefont
  {Ghiglieri}}, \bibinfo {author} {\bibfnamefont {A.}~\bibnamefont {Kurkela}},
  \bibinfo {author} {\bibfnamefont {M.}~\bibnamefont {Strickland}},\ and\
  \bibinfo {author} {\bibfnamefont {A.}~\bibnamefont {Vuorinen}},\ }\bibfield
  {title} {\bibinfo {title} {{Perturbative Thermal QCD: Formalism and
  Applications}},\ }\href {https://doi.org/10.1016/j.physrep.2020.07.004}
  {\bibfield  {journal} {\bibinfo  {journal} {Phys. Rept.}\ }\textbf {\bibinfo
  {volume} {880}},\ \bibinfo {pages} {1} (\bibinfo {year} {2020})},\ \Eprint
  {https://arxiv.org/abs/2002.10188} {arXiv:2002.10188 [hep-ph]} \BibitemShut
  {NoStop}%
\bibitem [{\citenamefont {Su}(2012)}]{Su:2012iy}%
  \BibitemOpen
  \bibfield  {author} {\bibinfo {author} {\bibfnamefont {N.}~\bibnamefont
  {Su}},\ }\bibfield  {title} {\bibinfo {title} {{A brief overview of
  hard-thermal-loop perturbation theory}},\ }\href
  {https://doi.org/10.1088/0253-6102/57/3/12} {\bibfield  {journal} {\bibinfo
  {journal} {Commun. Theor. Phys.}\ }\textbf {\bibinfo {volume} {57}},\
  \bibinfo {pages} {409} (\bibinfo {year} {2012})},\ \Eprint
  {https://arxiv.org/abs/1204.0260} {arXiv:1204.0260 [hep-ph]} \BibitemShut
  {NoStop}%
\bibitem [{\citenamefont {Nopoush}\ \emph {et~al.}(2017)\citenamefont
  {Nopoush}, \citenamefont {Guo},\ and\ \citenamefont
  {Strickland}}]{Nopoush:2017zbu}%
  \BibitemOpen
  \bibfield  {author} {\bibinfo {author} {\bibfnamefont {M.}~\bibnamefont
  {Nopoush}}, \bibinfo {author} {\bibfnamefont {Y.}~\bibnamefont {Guo}},\ and\
  \bibinfo {author} {\bibfnamefont {M.}~\bibnamefont {Strickland}},\ }\bibfield
   {title} {\bibinfo {title} {{The static hard-loop gluon propagator to all
  orders in anisotropy}},\ }\href {https://doi.org/10.1007/JHEP09(2017)063}
  {\bibfield  {journal} {\bibinfo  {journal} {JHEP}\ }\textbf {\bibinfo
  {volume} {09}},\ \bibinfo {pages} {063}},\ \Eprint
  {https://arxiv.org/abs/1706.08091} {arXiv:1706.08091 [hep-ph]} \BibitemShut
  {NoStop}%
\bibitem [{\citenamefont {Romatschke}\ and\ \citenamefont
  {Strickland}(2004)}]{Romatschke:2004jh}%
  \BibitemOpen
  \bibfield  {author} {\bibinfo {author} {\bibfnamefont {P.}~\bibnamefont
  {Romatschke}}\ and\ \bibinfo {author} {\bibfnamefont {M.}~\bibnamefont
  {Strickland}},\ }\bibfield  {title} {\bibinfo {title} {{Collective modes of
  an anisotropic quark-gluon plasma II}},\ }\href
  {https://doi.org/10.1103/PhysRevD.70.116006} {\bibfield  {journal} {\bibinfo
  {journal} {Phys. Rev. D}\ }\textbf {\bibinfo {volume} {70}},\ \bibinfo
  {pages} {116006} (\bibinfo {year} {2004})},\ \Eprint
  {https://arxiv.org/abs/hep-ph/0406188} {arXiv:hep-ph/0406188} \BibitemShut
  {NoStop}%
\bibitem [{\citenamefont {Dumitru}\ \emph {et~al.}(2008)\citenamefont
  {Dumitru}, \citenamefont {Guo},\ and\ \citenamefont
  {Strickland}}]{Dumitru:2007hy}%
  \BibitemOpen
  \bibfield  {author} {\bibinfo {author} {\bibfnamefont {A.}~\bibnamefont
  {Dumitru}}, \bibinfo {author} {\bibfnamefont {Y.}~\bibnamefont {Guo}},\ and\
  \bibinfo {author} {\bibfnamefont {M.}~\bibnamefont {Strickland}},\ }\bibfield
   {title} {\bibinfo {title} {{The Heavy-quark potential in an anisotropic
  (viscous) plasma}},\ }\href {https://doi.org/10.1016/j.physletb.2008.02.048}
  {\bibfield  {journal} {\bibinfo  {journal} {Phys. Lett. B}\ }\textbf
  {\bibinfo {volume} {662}},\ \bibinfo {pages} {37} (\bibinfo {year} {2008})},\
  \Eprint {https://arxiv.org/abs/0711.4722} {arXiv:0711.4722 [hep-ph]}
  \BibitemShut {NoStop}%
\bibitem [{\citenamefont {Burnier}\ \emph {et~al.}(2009)\citenamefont
  {Burnier}, \citenamefont {Laine},\ and\ \citenamefont
  {Vepsalainen}}]{Burnier:2009yu}%
  \BibitemOpen
  \bibfield  {author} {\bibinfo {author} {\bibfnamefont {Y.}~\bibnamefont
  {Burnier}}, \bibinfo {author} {\bibfnamefont {M.}~\bibnamefont {Laine}},\
  and\ \bibinfo {author} {\bibfnamefont {M.}~\bibnamefont {Vepsalainen}},\
  }\bibfield  {title} {\bibinfo {title} {{Quarkonium dissociation in the
  presence of a small momentum space anisotropy}},\ }\href
  {https://doi.org/10.1016/j.physletb.2009.05.067} {\bibfield  {journal}
  {\bibinfo  {journal} {Phys. Lett. B}\ }\textbf {\bibinfo {volume} {678}},\
  \bibinfo {pages} {86} (\bibinfo {year} {2009})},\ \Eprint
  {https://arxiv.org/abs/0903.3467} {arXiv:0903.3467 [hep-ph]} \BibitemShut
  {NoStop}%
\bibitem [{\citenamefont {Strickland}(2011)}]{Strickland:2011mw}%
  \BibitemOpen
  \bibfield  {author} {\bibinfo {author} {\bibfnamefont {M.}~\bibnamefont
  {Strickland}},\ }\bibfield  {title} {\bibinfo {title} {{Thermal
  $\upsilon_{1s}$ and chi\_b1 suppression in $\sqrt{s_{NN}}=2.76$ TeV Pb-Pb
  collisions at the LHC}},\ }\href
  {https://doi.org/10.1103/PhysRevLett.107.132301} {\bibfield  {journal}
  {\bibinfo  {journal} {Phys. Rev. Lett.}\ }\textbf {\bibinfo {volume} {107}},\
  \bibinfo {pages} {132301} (\bibinfo {year} {2011})},\ \Eprint
  {https://arxiv.org/abs/1106.2571} {arXiv:1106.2571 [hep-ph]} \BibitemShut
  {NoStop}%
\bibitem [{\citenamefont {Strickland}\ and\ \citenamefont
  {Bazow}(2012)}]{Strickland:2011aa}%
  \BibitemOpen
  \bibfield  {author} {\bibinfo {author} {\bibfnamefont {M.}~\bibnamefont
  {Strickland}}\ and\ \bibinfo {author} {\bibfnamefont {D.}~\bibnamefont
  {Bazow}},\ }\bibfield  {title} {\bibinfo {title} {{Thermal Bottomonium
  Suppression at RHIC and LHC}},\ }\href
  {https://doi.org/10.1016/j.nuclphysa.2012.02.003} {\bibfield  {journal}
  {\bibinfo  {journal} {Nucl. Phys. A}\ }\textbf {\bibinfo {volume} {879}},\
  \bibinfo {pages} {25} (\bibinfo {year} {2012})},\ \Eprint
  {https://arxiv.org/abs/1112.2761} {arXiv:1112.2761 [nucl-th]} \BibitemShut
  {NoStop}%
\bibitem [{\citenamefont {Krouppa}\ \emph {et~al.}(2015)\citenamefont
  {Krouppa}, \citenamefont {Ryblewski},\ and\ \citenamefont
  {Strickland}}]{Krouppa:2015yoa}%
  \BibitemOpen
  \bibfield  {author} {\bibinfo {author} {\bibfnamefont {B.}~\bibnamefont
  {Krouppa}}, \bibinfo {author} {\bibfnamefont {R.}~\bibnamefont {Ryblewski}},\
  and\ \bibinfo {author} {\bibfnamefont {M.}~\bibnamefont {Strickland}},\
  }\bibfield  {title} {\bibinfo {title} {{Bottomonia suppression in 2.76 TeV
  Pb-Pb collisions}},\ }\href {https://doi.org/10.1103/PhysRevC.92.061901}
  {\bibfield  {journal} {\bibinfo  {journal} {Phys. Rev. C}\ }\textbf {\bibinfo
  {volume} {92}},\ \bibinfo {pages} {061901} (\bibinfo {year} {2015})},\
  \Eprint {https://arxiv.org/abs/1507.03951} {arXiv:1507.03951 [hep-ph]}
  \BibitemShut {NoStop}%
\bibitem [{\citenamefont {Schenke}\ and\ \citenamefont
  {Strickland}(2007)}]{Schenke:2006yp}%
  \BibitemOpen
  \bibfield  {author} {\bibinfo {author} {\bibfnamefont {B.}~\bibnamefont
  {Schenke}}\ and\ \bibinfo {author} {\bibfnamefont {M.}~\bibnamefont
  {Strickland}},\ }\bibfield  {title} {\bibinfo {title} {{Photon production
  from an anisotropic quark-gluon plasma}},\ }\href
  {https://doi.org/10.1103/PhysRevD.76.025023} {\bibfield  {journal} {\bibinfo
  {journal} {Phys. Rev. D}\ }\textbf {\bibinfo {volume} {76}},\ \bibinfo
  {pages} {025023} (\bibinfo {year} {2007})},\ \Eprint
  {https://arxiv.org/abs/hep-ph/0611332} {arXiv:hep-ph/0611332} \BibitemShut
  {NoStop}%
\bibitem [{\citenamefont {Bhattacharya}\ \emph {et~al.}(2016)\citenamefont
  {Bhattacharya}, \citenamefont {Ryblewski},\ and\ \citenamefont
  {Strickland}}]{Bhattacharya:2015ada}%
  \BibitemOpen
  \bibfield  {author} {\bibinfo {author} {\bibfnamefont {L.}~\bibnamefont
  {Bhattacharya}}, \bibinfo {author} {\bibfnamefont {R.}~\bibnamefont
  {Ryblewski}},\ and\ \bibinfo {author} {\bibfnamefont {M.}~\bibnamefont
  {Strickland}},\ }\bibfield  {title} {\bibinfo {title} {{Photon production
  from a nonequilibrium quark-gluon plasma}},\ }\href
  {https://doi.org/10.1103/PhysRevD.93.065005} {\bibfield  {journal} {\bibinfo
  {journal} {Phys. Rev. D}\ }\textbf {\bibinfo {volume} {93}},\ \bibinfo
  {pages} {065005} (\bibinfo {year} {2016})},\ \Eprint
  {https://arxiv.org/abs/1507.06605} {arXiv:1507.06605 [hep-ph]} \BibitemShut
  {NoStop}%
\bibitem [{\citenamefont {Mandal}\ and\ \citenamefont
  {Roy}(2013{\natexlab{b}})}]{Mandal:2013jla}%
  \BibitemOpen
  \bibfield  {author} {\bibinfo {author} {\bibfnamefont {M.}~\bibnamefont
  {Mandal}}\ and\ \bibinfo {author} {\bibfnamefont {P.}~\bibnamefont {Roy}},\
  }\bibfield  {title} {\bibinfo {title} {{Wake potential in collisional
  anisotropic quark-gluon plasma}},\ }\href
  {https://doi.org/10.1103/PhysRevD.88.074013} {\bibfield  {journal} {\bibinfo
  {journal} {Phys. Rev. D}\ }\textbf {\bibinfo {volume} {88}},\ \bibinfo
  {pages} {074013} (\bibinfo {year} {2013}{\natexlab{b}})},\ \Eprint
  {https://arxiv.org/abs/1310.4660} {arXiv:1310.4660 [hep-ph]} \BibitemShut
  {NoStop}%
\bibitem [{\citenamefont {Tinti}\ and\ \citenamefont
  {Florkowski}(2014)}]{Tinti:2013vba}%
  \BibitemOpen
  \bibfield  {author} {\bibinfo {author} {\bibfnamefont {L.}~\bibnamefont
  {Tinti}}\ and\ \bibinfo {author} {\bibfnamefont {W.}~\bibnamefont
  {Florkowski}},\ }\bibfield  {title} {\bibinfo {title} {{Projection method and
  new formulation of leading-order anisotropic hydrodynamics}},\ }\href
  {https://doi.org/10.1103/PhysRevC.89.034907} {\bibfield  {journal} {\bibinfo
  {journal} {Phys. Rev. C}\ }\textbf {\bibinfo {volume} {89}},\ \bibinfo
  {pages} {034907} (\bibinfo {year} {2014})},\ \Eprint
  {https://arxiv.org/abs/1312.6614} {arXiv:1312.6614 [nucl-th]} \BibitemShut
  {NoStop}%
\bibitem [{\citenamefont {Kasmaei}\ and\ \citenamefont
  {Strickland}(2018)}]{Kasmaei:2018yrr}%
  \BibitemOpen
  \bibfield  {author} {\bibinfo {author} {\bibfnamefont {B.~S.}\ \bibnamefont
  {Kasmaei}}\ and\ \bibinfo {author} {\bibfnamefont {M.}~\bibnamefont
  {Strickland}},\ }\bibfield  {title} {\bibinfo {title} {{Parton self-energies
  for general momentum-space anisotropy}},\ }\href
  {https://doi.org/10.1103/PhysRevD.97.054022} {\bibfield  {journal} {\bibinfo
  {journal} {Phys. Rev. D}\ }\textbf {\bibinfo {volume} {97}},\ \bibinfo
  {pages} {054022} (\bibinfo {year} {2018})},\ \Eprint
  {https://arxiv.org/abs/1801.00863} {arXiv:1801.00863 [hep-ph]} \BibitemShut
  {NoStop}%
\bibitem [{\citenamefont {Ghosh}\ \emph {et~al.}(2020)\citenamefont {Ghosh},
  \citenamefont {Karmakar},\ and\ \citenamefont {Mukherjee}}]{Ghosh:2020sng}%
  \BibitemOpen
  \bibfield  {author} {\bibinfo {author} {\bibfnamefont {R.}~\bibnamefont
  {Ghosh}}, \bibinfo {author} {\bibfnamefont {B.}~\bibnamefont {Karmakar}},\
  and\ \bibinfo {author} {\bibfnamefont {A.}~\bibnamefont {Mukherjee}},\
  }\bibfield  {title} {\bibinfo {title} {{Covariant formulation of gluon
  self-energy in presence of ellipsoidal anisotropy}},\ }\href
  {https://doi.org/10.1103/PhysRevD.102.114002} {\bibfield  {journal} {\bibinfo
   {journal} {Phys. Rev. D}\ }\textbf {\bibinfo {volume} {102}},\ \bibinfo
  {pages} {114002} (\bibinfo {year} {2020})},\ \Eprint
  {https://arxiv.org/abs/2011.03374} {arXiv:2011.03374 [hep-ph]} \BibitemShut
  {NoStop}%
\bibitem [{\citenamefont {Carrington}\ \emph {et~al.}(2021)\citenamefont
  {Carrington}, \citenamefont {Forster},\ and\ \citenamefont
  {Makar}}]{Carrington:2021bnk}%
  \BibitemOpen
  \bibfield  {author} {\bibinfo {author} {\bibfnamefont {M.~E.}\ \bibnamefont
  {Carrington}}, \bibinfo {author} {\bibfnamefont {B.~M.}\ \bibnamefont
  {Forster}},\ and\ \bibinfo {author} {\bibfnamefont {S.}~\bibnamefont
  {Makar}},\ }\bibfield  {title} {\bibinfo {title} {{Collective modes in
  anisotropic systems}},\ }\href {https://doi.org/10.1103/PhysRevC.104.064908}
  {\bibfield  {journal} {\bibinfo  {journal} {Phys. Rev. C}\ }\textbf {\bibinfo
  {volume} {104}},\ \bibinfo {pages} {064908} (\bibinfo {year} {2021})},\
  \Eprint {https://arxiv.org/abs/2107.08229} {arXiv:2107.08229 [hep-ph]}
  \BibitemShut {NoStop}%
\bibitem [{\citenamefont {Kharzeev}\ \emph {et~al.}(2008)\citenamefont
  {Kharzeev}, \citenamefont {McLerran},\ and\ \citenamefont
  {Warringa}}]{Kharzeev:2007jp}%
  \BibitemOpen
  \bibfield  {author} {\bibinfo {author} {\bibfnamefont {D.~E.}\ \bibnamefont
  {Kharzeev}}, \bibinfo {author} {\bibfnamefont {L.~D.}\ \bibnamefont
  {McLerran}},\ and\ \bibinfo {author} {\bibfnamefont {H.~J.}\ \bibnamefont
  {Warringa}},\ }\bibfield  {title} {\bibinfo {title} {{The Effects of
  topological charge change in heavy ion collisions: 'Event by event P and CP
  violation'}},\ }\href {https://doi.org/10.1016/j.nuclphysa.2008.02.298}
  {\bibfield  {journal} {\bibinfo  {journal} {Nucl. Phys. A}\ }\textbf
  {\bibinfo {volume} {803}},\ \bibinfo {pages} {227} (\bibinfo {year}
  {2008})},\ \Eprint {https://arxiv.org/abs/0711.0950} {arXiv:0711.0950
  [hep-ph]} \BibitemShut {NoStop}%
\bibitem [{\citenamefont {Skokov}\ \emph {et~al.}(2009)\citenamefont {Skokov},
  \citenamefont {Illarionov},\ and\ \citenamefont {Toneev}}]{Skokov:2009qp}%
  \BibitemOpen
  \bibfield  {author} {\bibinfo {author} {\bibfnamefont {V.}~\bibnamefont
  {Skokov}}, \bibinfo {author} {\bibfnamefont {A.~Y.}\ \bibnamefont
  {Illarionov}},\ and\ \bibinfo {author} {\bibfnamefont {V.}~\bibnamefont
  {Toneev}},\ }\bibfield  {title} {\bibinfo {title} {{Estimate of the magnetic
  field strength in heavy-ion collisions}},\ }\href
  {https://doi.org/10.1142/S0217751X09047570} {\bibfield  {journal} {\bibinfo
  {journal} {Int. J. Mod. Phys. A}\ }\textbf {\bibinfo {volume} {24}},\
  \bibinfo {pages} {5925} (\bibinfo {year} {2009})},\ \Eprint
  {https://arxiv.org/abs/0907.1396} {arXiv:0907.1396 [nucl-th]} \BibitemShut
  {NoStop}%
\bibitem [{\citenamefont {Huang}(2016)}]{Huang:2015oca}%
  \BibitemOpen
  \bibfield  {author} {\bibinfo {author} {\bibfnamefont {X.-G.}\ \bibnamefont
  {Huang}},\ }\bibfield  {title} {\bibinfo {title} {{Electromagnetic fields and
  anomalous transports in heavy-ion collisions --- A pedagogical review}},\
  }\href {https://doi.org/10.1088/0034-4885/79/7/076302} {\bibfield  {journal}
  {\bibinfo  {journal} {Rept. Prog. Phys.}\ }\textbf {\bibinfo {volume} {79}},\
  \bibinfo {pages} {076302} (\bibinfo {year} {2016})},\ \Eprint
  {https://arxiv.org/abs/1509.04073} {arXiv:1509.04073 [nucl-th]} \BibitemShut
  {NoStop}%
\bibitem [{\citenamefont {Miransky}\ and\ \citenamefont
  {Shovkovy}(2015)}]{Miransky:2015ava}%
  \BibitemOpen
  \bibfield  {author} {\bibinfo {author} {\bibfnamefont {V.~A.}\ \bibnamefont
  {Miransky}}\ and\ \bibinfo {author} {\bibfnamefont {I.~A.}\ \bibnamefont
  {Shovkovy}},\ }\bibfield  {title} {\bibinfo {title} {{Quantum field theory in
  a magnetic field: From quantum chromodynamics to graphene and Dirac
  semimetals}},\ }\href {https://doi.org/10.1016/j.physrep.2015.02.003}
  {\bibfield  {journal} {\bibinfo  {journal} {Phys. Rept.}\ }\textbf {\bibinfo
  {volume} {576}},\ \bibinfo {pages} {1} (\bibinfo {year} {2015})},\ \Eprint
  {https://arxiv.org/abs/1503.00732} {arXiv:1503.00732 [hep-ph]} \BibitemShut
  {NoStop}%
\bibitem [{\citenamefont {McLerran}\ and\ \citenamefont
  {Skokov}(2014)}]{McLerran:2013hla}%
  \BibitemOpen
  \bibfield  {author} {\bibinfo {author} {\bibfnamefont {L.}~\bibnamefont
  {McLerran}}\ and\ \bibinfo {author} {\bibfnamefont {V.}~\bibnamefont
  {Skokov}},\ }\bibfield  {title} {\bibinfo {title} {{Comments About the
  Electromagnetic Field in Heavy-Ion Collisions}},\ }\href
  {https://doi.org/10.1016/j.nuclphysa.2014.05.008} {\bibfield  {journal}
  {\bibinfo  {journal} {Nucl. Phys. A}\ }\textbf {\bibinfo {volume} {929}},\
  \bibinfo {pages} {184} (\bibinfo {year} {2014})},\ \Eprint
  {https://arxiv.org/abs/1305.0774} {arXiv:1305.0774 [hep-ph]} \BibitemShut
  {NoStop}%
\bibitem [{\citenamefont {Roy}\ \emph {et~al.}(2017)\citenamefont {Roy},
  \citenamefont {Pu}, \citenamefont {Rezzolla},\ and\ \citenamefont
  {Rischke}}]{Roy:2017yvg}%
  \BibitemOpen
  \bibfield  {author} {\bibinfo {author} {\bibfnamefont {V.}~\bibnamefont
  {Roy}}, \bibinfo {author} {\bibfnamefont {S.}~\bibnamefont {Pu}}, \bibinfo
  {author} {\bibfnamefont {L.}~\bibnamefont {Rezzolla}},\ and\ \bibinfo
  {author} {\bibfnamefont {D.~H.}\ \bibnamefont {Rischke}},\ }\bibfield
  {title} {\bibinfo {title} {{Effect of intense magnetic fields on reduced-MHD
  evolution in $\sqrt{s_{\rm NN}}$ = 200 GeV Au+Au collisions}},\ }\href
  {https://doi.org/10.1103/PhysRevC.96.054909} {\bibfield  {journal} {\bibinfo
  {journal} {Phys. Rev. C}\ }\textbf {\bibinfo {volume} {96}},\ \bibinfo
  {pages} {054909} (\bibinfo {year} {2017})},\ \Eprint
  {https://arxiv.org/abs/1706.05326} {arXiv:1706.05326 [nucl-th]} \BibitemShut
  {NoStop}%
\bibitem [{\citenamefont {Tuchin}(2013{\natexlab{a}})}]{Tuchin:2013apa}%
  \BibitemOpen
  \bibfield  {author} {\bibinfo {author} {\bibfnamefont {K.}~\bibnamefont
  {Tuchin}},\ }\bibfield  {title} {\bibinfo {title} {{Time and space dependence
  of the electromagnetic field in relativistic heavy-ion collisions}},\ }\href
  {https://doi.org/10.1103/PhysRevC.88.024911} {\bibfield  {journal} {\bibinfo
  {journal} {Phys. Rev. C}\ }\textbf {\bibinfo {volume} {88}},\ \bibinfo
  {pages} {024911} (\bibinfo {year} {2013}{\natexlab{a}})},\ \Eprint
  {https://arxiv.org/abs/1305.5806} {arXiv:1305.5806 [hep-ph]} \BibitemShut
  {NoStop}%
\bibitem [{\citenamefont {Tuchin}(2013{\natexlab{b}})}]{Tuchin:2013ie}%
  \BibitemOpen
  \bibfield  {author} {\bibinfo {author} {\bibfnamefont {K.}~\bibnamefont
  {Tuchin}},\ }\bibfield  {title} {\bibinfo {title} {{Particle production in
  strong electromagnetic fields in relativistic heavy-ion collisions}},\ }\href
  {https://doi.org/10.1155/2013/490495} {\bibfield  {journal} {\bibinfo
  {journal} {Adv. High Energy Phys.}\ }\textbf {\bibinfo {volume} {2013}},\
  \bibinfo {pages} {490495} (\bibinfo {year} {2013}{\natexlab{b}})},\ \Eprint
  {https://arxiv.org/abs/1301.0099} {arXiv:1301.0099 [hep-ph]} \BibitemShut
  {NoStop}%
\bibitem [{\citenamefont {Fukushima}\ \emph {et~al.}(2008)\citenamefont
  {Fukushima}, \citenamefont {Kharzeev},\ and\ \citenamefont
  {Warringa}}]{Fukushima:2008xe}%
  \BibitemOpen
  \bibfield  {author} {\bibinfo {author} {\bibfnamefont {K.}~\bibnamefont
  {Fukushima}}, \bibinfo {author} {\bibfnamefont {D.~E.}\ \bibnamefont
  {Kharzeev}},\ and\ \bibinfo {author} {\bibfnamefont {H.~J.}\ \bibnamefont
  {Warringa}},\ }\bibfield  {title} {\bibinfo {title} {{The Chiral Magnetic
  Effect}},\ }\href {https://doi.org/10.1103/PhysRevD.78.074033} {\bibfield
  {journal} {\bibinfo  {journal} {Phys. Rev. D}\ }\textbf {\bibinfo {volume}
  {78}},\ \bibinfo {pages} {074033} (\bibinfo {year} {2008})},\ \Eprint
  {https://arxiv.org/abs/0808.3382} {arXiv:0808.3382 [hep-ph]} \BibitemShut
  {NoStop}%
\bibitem [{\citenamefont {Lee}\ \emph {et~al.}(1997)\citenamefont {Lee},
  \citenamefont {Leung},\ and\ \citenamefont {Ng}}]{Lee:1997zj}%
  \BibitemOpen
  \bibfield  {author} {\bibinfo {author} {\bibfnamefont {D.~S.}\ \bibnamefont
  {Lee}}, \bibinfo {author} {\bibfnamefont {C.~N.}\ \bibnamefont {Leung}},\
  and\ \bibinfo {author} {\bibfnamefont {Y.~J.}\ \bibnamefont {Ng}},\
  }\bibfield  {title} {\bibinfo {title} {{Chiral symmetry breaking in a uniform
  external magnetic field}},\ }\href {https://doi.org/10.1103/PhysRevD.55.6504}
  {\bibfield  {journal} {\bibinfo  {journal} {Phys. Rev. D}\ }\textbf {\bibinfo
  {volume} {55}},\ \bibinfo {pages} {6504} (\bibinfo {year} {1997})},\ \Eprint
  {https://arxiv.org/abs/hep-th/9701172} {arXiv:hep-th/9701172} \BibitemShut
  {NoStop}%
\bibitem [{\citenamefont {Bali}\ \emph {et~al.}(2012)\citenamefont {Bali},
  \citenamefont {Bruckmann}, \citenamefont {Endrodi}, \citenamefont {Fodor},
  \citenamefont {Katz}, \citenamefont {Krieg}, \citenamefont {Schafer},\ and\
  \citenamefont {Szabo}}]{Bali:2011qj}%
  \BibitemOpen
  \bibfield  {author} {\bibinfo {author} {\bibfnamefont {G.~S.}\ \bibnamefont
  {Bali}}, \bibinfo {author} {\bibfnamefont {F.}~\bibnamefont {Bruckmann}},
  \bibinfo {author} {\bibfnamefont {G.}~\bibnamefont {Endrodi}}, \bibinfo
  {author} {\bibfnamefont {Z.}~\bibnamefont {Fodor}}, \bibinfo {author}
  {\bibfnamefont {S.~D.}\ \bibnamefont {Katz}}, \bibinfo {author}
  {\bibfnamefont {S.}~\bibnamefont {Krieg}}, \bibinfo {author} {\bibfnamefont
  {A.}~\bibnamefont {Schafer}},\ and\ \bibinfo {author} {\bibfnamefont {K.~K.}\
  \bibnamefont {Szabo}},\ }\bibfield  {title} {\bibinfo {title} {{The QCD phase
  diagram for external magnetic fields}},\ }\href
  {https://doi.org/10.1007/JHEP02(2012)044} {\bibfield  {journal} {\bibinfo
  {journal} {JHEP}\ }\textbf {\bibinfo {volume} {02}},\ \bibinfo {pages}
  {044}},\ \Eprint {https://arxiv.org/abs/1111.4956} {arXiv:1111.4956
  [hep-lat]} \BibitemShut {NoStop}%
\bibitem [{\citenamefont {Ayala}\ \emph {et~al.}(2014)\citenamefont {Ayala},
  \citenamefont {Loewe}, \citenamefont {Mizher},\ and\ \citenamefont
  {Zamora}}]{Ayala:2014iba}%
  \BibitemOpen
  \bibfield  {author} {\bibinfo {author} {\bibfnamefont {A.}~\bibnamefont
  {Ayala}}, \bibinfo {author} {\bibfnamefont {M.}~\bibnamefont {Loewe}},
  \bibinfo {author} {\bibfnamefont {A.~J.}\ \bibnamefont {Mizher}},\ and\
  \bibinfo {author} {\bibfnamefont {R.}~\bibnamefont {Zamora}},\ }\bibfield
  {title} {\bibinfo {title} {{Inverse magnetic catalysis for the chiral
  transition induced by thermo-magnetic effects on the coupling constant}},\
  }\href {https://doi.org/10.1103/PhysRevD.90.036001} {\bibfield  {journal}
  {\bibinfo  {journal} {Phys. Rev. D}\ }\textbf {\bibinfo {volume} {90}},\
  \bibinfo {pages} {036001} (\bibinfo {year} {2014})},\ \Eprint
  {https://arxiv.org/abs/1406.3885} {arXiv:1406.3885 [hep-ph]} \BibitemShut
  {NoStop}%
\bibitem [{\citenamefont {Andersen}(2012)}]{Andersen:2012dz}%
  \BibitemOpen
  \bibfield  {author} {\bibinfo {author} {\bibfnamefont {J.~O.}\ \bibnamefont
  {Andersen}},\ }\bibfield  {title} {\bibinfo {title} {{Thermal pions in a
  magnetic background}},\ }\href {https://doi.org/10.1103/PhysRevD.86.025020}
  {\bibfield  {journal} {\bibinfo  {journal} {Phys. Rev. D}\ }\textbf {\bibinfo
  {volume} {86}},\ \bibinfo {pages} {025020} (\bibinfo {year} {2012})},\
  \Eprint {https://arxiv.org/abs/1202.2051} {arXiv:1202.2051 [hep-ph]}
  \BibitemShut {NoStop}%
\bibitem [{\citenamefont {Avancini}\ \emph {et~al.}(2017)\citenamefont
  {Avancini}, \citenamefont {Farias}, \citenamefont {Benghi~Pinto},
  \citenamefont {Tavares},\ and\ \citenamefont {Tim\'oteo}}]{Avancini:2016fgq}%
  \BibitemOpen
  \bibfield  {author} {\bibinfo {author} {\bibfnamefont {S.~S.}\ \bibnamefont
  {Avancini}}, \bibinfo {author} {\bibfnamefont {R.~L.~S.}\ \bibnamefont
  {Farias}}, \bibinfo {author} {\bibfnamefont {M.}~\bibnamefont
  {Benghi~Pinto}}, \bibinfo {author} {\bibfnamefont {W.~R.}\ \bibnamefont
  {Tavares}},\ and\ \bibinfo {author} {\bibfnamefont {V.~S.}\ \bibnamefont
  {Tim\'oteo}},\ }\bibfield  {title} {\bibinfo {title} {{$\pi_0$ pole mass
  calculation in a strong magnetic field and lattice constraints}},\ }\href
  {https://doi.org/10.1016/j.physletb.2017.02.002} {\bibfield  {journal}
  {\bibinfo  {journal} {Phys. Lett. B}\ }\textbf {\bibinfo {volume} {767}},\
  \bibinfo {pages} {247} (\bibinfo {year} {2017})},\ \Eprint
  {https://arxiv.org/abs/1606.05754} {arXiv:1606.05754 [hep-ph]} \BibitemShut
  {NoStop}%
\bibitem [{\citenamefont {Wang}\ \emph {et~al.}(2020)\citenamefont {Wang},
  \citenamefont {Shovkovy}, \citenamefont {Yu},\ and\ \citenamefont
  {Huang}}]{Wang:2020dsr}%
  \BibitemOpen
  \bibfield  {author} {\bibinfo {author} {\bibfnamefont {X.}~\bibnamefont
  {Wang}}, \bibinfo {author} {\bibfnamefont {I.~A.}\ \bibnamefont {Shovkovy}},
  \bibinfo {author} {\bibfnamefont {L.}~\bibnamefont {Yu}},\ and\ \bibinfo
  {author} {\bibfnamefont {M.}~\bibnamefont {Huang}},\ }\bibfield  {title}
  {\bibinfo {title} {{Ellipticity of photon emission from strongly magnetized
  hot QCD plasma}},\ }\href {https://doi.org/10.1103/PhysRevD.102.076010}
  {\bibfield  {journal} {\bibinfo  {journal} {Phys. Rev. D}\ }\textbf {\bibinfo
  {volume} {102}},\ \bibinfo {pages} {076010} (\bibinfo {year} {2020})},\
  \Eprint {https://arxiv.org/abs/2006.16254} {arXiv:2006.16254 [hep-ph]}
  \BibitemShut {NoStop}%
\bibitem [{\citenamefont {Tuchin}(2013{\natexlab{c}})}]{Tuchin:2013bda}%
  \BibitemOpen
  \bibfield  {author} {\bibinfo {author} {\bibfnamefont {K.}~\bibnamefont
  {Tuchin}},\ }\bibfield  {title} {\bibinfo {title} {{Magnetic contribution to
  dilepton production in heavy-ion collisions}},\ }\href
  {https://doi.org/10.1103/PhysRevC.88.024910} {\bibfield  {journal} {\bibinfo
  {journal} {Phys. Rev. C}\ }\textbf {\bibinfo {volume} {88}},\ \bibinfo
  {pages} {024910} (\bibinfo {year} {2013}{\natexlab{c}})},\ \Eprint
  {https://arxiv.org/abs/1305.0545} {arXiv:1305.0545 [nucl-th]} \BibitemShut
  {NoStop}%
\bibitem [{\citenamefont {Bandyopadhyay}\ \emph {et~al.}(2016)\citenamefont
  {Bandyopadhyay}, \citenamefont {Islam},\ and\ \citenamefont
  {Mustafa}}]{Bandyopadhyay:2016fyd}%
  \BibitemOpen
  \bibfield  {author} {\bibinfo {author} {\bibfnamefont {A.}~\bibnamefont
  {Bandyopadhyay}}, \bibinfo {author} {\bibfnamefont {C.~A.}\ \bibnamefont
  {Islam}},\ and\ \bibinfo {author} {\bibfnamefont {M.~G.}\ \bibnamefont
  {Mustafa}},\ }\bibfield  {title} {\bibinfo {title} {{Electromagnetic spectral
  properties and Debye screening of a strongly magnetized hot medium}},\ }\href
  {https://doi.org/10.1103/PhysRevD.94.114034} {\bibfield  {journal} {\bibinfo
  {journal} {Phys. Rev. D}\ }\textbf {\bibinfo {volume} {94}},\ \bibinfo
  {pages} {114034} (\bibinfo {year} {2016})},\ \Eprint
  {https://arxiv.org/abs/1602.06769} {arXiv:1602.06769 [hep-ph]} \BibitemShut
  {NoStop}%
\bibitem [{\citenamefont {Das}\ \emph {et~al.}(2021)\citenamefont {Das},
  \citenamefont {Bandyopadhyay},\ and\ \citenamefont {Islam}}]{Das:2021fma}%
  \BibitemOpen
  \bibfield  {author} {\bibinfo {author} {\bibfnamefont {A.}~\bibnamefont
  {Das}}, \bibinfo {author} {\bibfnamefont {A.}~\bibnamefont {Bandyopadhyay}},\
  and\ \bibinfo {author} {\bibfnamefont {C.~A.}\ \bibnamefont {Islam}},\
  }\bibfield  {title} {\bibinfo {title} {{Lepton pair production from a hot and
  dense QCD medium in presence of an arbitrary magnetic field}},\ }\href@noop
  {} {\  (\bibinfo {year} {2021})},\ \Eprint {https://arxiv.org/abs/2109.00019}
  {arXiv:2109.00019 [hep-ph]} \BibitemShut {NoStop}%
\bibitem [{\citenamefont {Ghosh}\ and\ \citenamefont
  {Chandra}(2018)}]{Ghosh:2018xhh}%
  \BibitemOpen
  \bibfield  {author} {\bibinfo {author} {\bibfnamefont {S.}~\bibnamefont
  {Ghosh}}\ and\ \bibinfo {author} {\bibfnamefont {V.}~\bibnamefont
  {Chandra}},\ }\bibfield  {title} {\bibinfo {title} {{Electromagnetic spectral
  function and dilepton rate in a hot magnetized QCD medium}},\ }\href
  {https://doi.org/10.1103/PhysRevD.98.076006} {\bibfield  {journal} {\bibinfo
  {journal} {Phys. Rev. D}\ }\textbf {\bibinfo {volume} {98}},\ \bibinfo
  {pages} {076006} (\bibinfo {year} {2018})},\ \Eprint
  {https://arxiv.org/abs/1808.05176} {arXiv:1808.05176 [hep-ph]} \BibitemShut
  {NoStop}%
\bibitem [{\citenamefont {Hattori}\ \emph {et~al.}(2021)\citenamefont
  {Hattori}, \citenamefont {Taya},\ and\ \citenamefont
  {Yoshida}}]{Hattori:2020htm}%
  \BibitemOpen
  \bibfield  {author} {\bibinfo {author} {\bibfnamefont {K.}~\bibnamefont
  {Hattori}}, \bibinfo {author} {\bibfnamefont {H.}~\bibnamefont {Taya}},\ and\
  \bibinfo {author} {\bibfnamefont {S.}~\bibnamefont {Yoshida}},\ }\bibfield
  {title} {\bibinfo {title} {{Di-lepton production from a single photon in
  strong magnetic fields: vacuum dichroism}},\ }\href
  {https://doi.org/10.1007/JHEP01(2021)093} {\bibfield  {journal} {\bibinfo
  {journal} {JHEP}\ }\textbf {\bibinfo {volume} {01}},\ \bibinfo {pages}
  {093}},\ \Eprint {https://arxiv.org/abs/2010.13492} {arXiv:2010.13492
  [hep-ph]} \BibitemShut {NoStop}%
\bibitem [{\citenamefont {Rath}\ and\ \citenamefont
  {Patra}(2017)}]{Rath:2017fdv}%
  \BibitemOpen
  \bibfield  {author} {\bibinfo {author} {\bibfnamefont {S.}~\bibnamefont
  {Rath}}\ and\ \bibinfo {author} {\bibfnamefont {B.~K.}\ \bibnamefont
  {Patra}},\ }\bibfield  {title} {\bibinfo {title} {{One-loop QCD
  thermodynamics in a strong homogeneous and static magnetic field}},\ }\href
  {https://doi.org/10.1007/JHEP12(2017)098} {\bibfield  {journal} {\bibinfo
  {journal} {JHEP}\ }\textbf {\bibinfo {volume} {12}},\ \bibinfo {pages}
  {098}},\ \Eprint {https://arxiv.org/abs/1707.02890} {arXiv:1707.02890
  [hep-th]} \BibitemShut {NoStop}%
\bibitem [{\citenamefont {Karmakar}\ \emph
  {et~al.}(2019{\natexlab{a}})\citenamefont {Karmakar}, \citenamefont {Ghosh},
  \citenamefont {Bandyopadhyay}, \citenamefont {Haque},\ and\ \citenamefont
  {Mustafa}}]{Karmakar:2019tdp}%
  \BibitemOpen
  \bibfield  {author} {\bibinfo {author} {\bibfnamefont {B.}~\bibnamefont
  {Karmakar}}, \bibinfo {author} {\bibfnamefont {R.}~\bibnamefont {Ghosh}},
  \bibinfo {author} {\bibfnamefont {A.}~\bibnamefont {Bandyopadhyay}}, \bibinfo
  {author} {\bibfnamefont {N.}~\bibnamefont {Haque}},\ and\ \bibinfo {author}
  {\bibfnamefont {M.~G.}\ \bibnamefont {Mustafa}},\ }\bibfield  {title}
  {\bibinfo {title} {{Anisotropic pressure of deconfined QCD matter in presence
  of strong magnetic field within one-loop approximation}},\ }\href
  {https://doi.org/10.1103/PhysRevD.99.094002} {\bibfield  {journal} {\bibinfo
  {journal} {Phys. Rev. D}\ }\textbf {\bibinfo {volume} {99}},\ \bibinfo
  {pages} {094002} (\bibinfo {year} {2019}{\natexlab{a}})},\ \Eprint
  {https://arxiv.org/abs/1902.02607} {arXiv:1902.02607 [hep-ph]} \BibitemShut
  {NoStop}%
\bibitem [{\citenamefont {Bandyopadhyay}\ \emph {et~al.}(2019)\citenamefont
  {Bandyopadhyay}, \citenamefont {Karmakar}, \citenamefont {Haque},\ and\
  \citenamefont {Mustafa}}]{Bandyopadhyay:2017cle}%
  \BibitemOpen
  \bibfield  {author} {\bibinfo {author} {\bibfnamefont {A.}~\bibnamefont
  {Bandyopadhyay}}, \bibinfo {author} {\bibfnamefont {B.}~\bibnamefont
  {Karmakar}}, \bibinfo {author} {\bibfnamefont {N.}~\bibnamefont {Haque}},\
  and\ \bibinfo {author} {\bibfnamefont {M.~G.}\ \bibnamefont {Mustafa}},\
  }\bibfield  {title} {\bibinfo {title} {{Pressure of a weakly magnetized hot
  and dense deconfined QCD matter in one-loop hard-thermal-loop perturbation
  theory}},\ }\href {https://doi.org/10.1103/PhysRevD.100.034031} {\bibfield
  {journal} {\bibinfo  {journal} {Phys. Rev. D}\ }\textbf {\bibinfo {volume}
  {100}},\ \bibinfo {pages} {034031} (\bibinfo {year} {2019})},\ \Eprint
  {https://arxiv.org/abs/1702.02875} {arXiv:1702.02875 [hep-ph]} \BibitemShut
  {NoStop}%
\bibitem [{\citenamefont {Singh}\ \emph {et~al.}(2018)\citenamefont {Singh},
  \citenamefont {Thakur},\ and\ \citenamefont {Mishra}}]{Singh:2017nfa}%
  \BibitemOpen
  \bibfield  {author} {\bibinfo {author} {\bibfnamefont {B.}~\bibnamefont
  {Singh}}, \bibinfo {author} {\bibfnamefont {L.}~\bibnamefont {Thakur}},\ and\
  \bibinfo {author} {\bibfnamefont {H.}~\bibnamefont {Mishra}},\ }\bibfield
  {title} {\bibinfo {title} {{Heavy quark complex potential in a strongly
  magnetized hot QGP medium}},\ }\href
  {https://doi.org/10.1103/PhysRevD.97.096011} {\bibfield  {journal} {\bibinfo
  {journal} {Phys. Rev. D}\ }\textbf {\bibinfo {volume} {97}},\ \bibinfo
  {pages} {096011} (\bibinfo {year} {2018})},\ \Eprint
  {https://arxiv.org/abs/1711.03071} {arXiv:1711.03071 [hep-ph]} \BibitemShut
  {NoStop}%
\bibitem [{\citenamefont {Kurian}\ \emph {et~al.}(2019)\citenamefont {Kurian},
  \citenamefont {Mitra}, \citenamefont {Ghosh},\ and\ \citenamefont
  {Chandra}}]{Kurian:2018qwb}%
  \BibitemOpen
  \bibfield  {author} {\bibinfo {author} {\bibfnamefont {M.}~\bibnamefont
  {Kurian}}, \bibinfo {author} {\bibfnamefont {S.}~\bibnamefont {Mitra}},
  \bibinfo {author} {\bibfnamefont {S.}~\bibnamefont {Ghosh}},\ and\ \bibinfo
  {author} {\bibfnamefont {V.}~\bibnamefont {Chandra}},\ }\bibfield  {title}
  {\bibinfo {title} {{Transport coefficients of hot magnetized QCD matter
  beyond the lowest Landau level approximation}},\ }\href
  {https://doi.org/10.1140/epjc/s10052-019-6649-z} {\bibfield  {journal}
  {\bibinfo  {journal} {Eur. Phys. J. C}\ }\textbf {\bibinfo {volume} {79}},\
  \bibinfo {pages} {134} (\bibinfo {year} {2019})},\ \Eprint
  {https://arxiv.org/abs/1805.07313} {arXiv:1805.07313 [nucl-th]} \BibitemShut
  {NoStop}%
\bibitem [{\citenamefont {Kurian}\ and\ \citenamefont
  {Chandra}(2018)}]{Kurian:2018dbn}%
  \BibitemOpen
  \bibfield  {author} {\bibinfo {author} {\bibfnamefont {M.}~\bibnamefont
  {Kurian}}\ and\ \bibinfo {author} {\bibfnamefont {V.}~\bibnamefont
  {Chandra}},\ }\bibfield  {title} {\bibinfo {title} {{Bulk viscosity of a hot
  QCD medium in a strong magnetic field within the relaxation-time
  approximation}},\ }\href {https://doi.org/10.1103/PhysRevD.97.116008}
  {\bibfield  {journal} {\bibinfo  {journal} {Phys. Rev. D}\ }\textbf {\bibinfo
  {volume} {97}},\ \bibinfo {pages} {116008} (\bibinfo {year} {2018})},\
  \Eprint {https://arxiv.org/abs/1802.07904} {arXiv:1802.07904 [nucl-th]}
  \BibitemShut {NoStop}%
\bibitem [{\citenamefont {Hattori}\ and\ \citenamefont
  {Satow}(2018)}]{Hattori:2017xoo}%
  \BibitemOpen
  \bibfield  {author} {\bibinfo {author} {\bibfnamefont {K.}~\bibnamefont
  {Hattori}}\ and\ \bibinfo {author} {\bibfnamefont {D.}~\bibnamefont
  {Satow}},\ }\bibfield  {title} {\bibinfo {title} {{Gluon spectrum in a
  quark-gluon plasma under strong magnetic fields}},\ }\href
  {https://doi.org/10.1103/PhysRevD.97.014023} {\bibfield  {journal} {\bibinfo
  {journal} {Phys. Rev. D}\ }\textbf {\bibinfo {volume} {97}},\ \bibinfo
  {pages} {014023} (\bibinfo {year} {2018})},\ \Eprint
  {https://arxiv.org/abs/1704.03191} {arXiv:1704.03191 [hep-ph]} \BibitemShut
  {NoStop}%
\bibitem [{\citenamefont {Karmakar}\ \emph
  {et~al.}(2019{\natexlab{b}})\citenamefont {Karmakar}, \citenamefont
  {Bandyopadhyay}, \citenamefont {Haque},\ and\ \citenamefont
  {Mustafa}}]{Karmakar:2018aig}%
  \BibitemOpen
  \bibfield  {author} {\bibinfo {author} {\bibfnamefont {B.}~\bibnamefont
  {Karmakar}}, \bibinfo {author} {\bibfnamefont {A.}~\bibnamefont
  {Bandyopadhyay}}, \bibinfo {author} {\bibfnamefont {N.}~\bibnamefont
  {Haque}},\ and\ \bibinfo {author} {\bibfnamefont {M.~G.}\ \bibnamefont
  {Mustafa}},\ }\bibfield  {title} {\bibinfo {title} {{General structure of
  gauge boson propagator and its spectra in a hot magnetized medium}},\ }\href
  {https://doi.org/10.1140/epjc/s10052-019-7154-0} {\bibfield  {journal}
  {\bibinfo  {journal} {Eur. Phys. J. C}\ }\textbf {\bibinfo {volume} {79}},\
  \bibinfo {pages} {658} (\bibinfo {year} {2019}{\natexlab{b}})},\ \Eprint
  {https://arxiv.org/abs/1804.11336} {arXiv:1804.11336 [hep-ph]} \BibitemShut
  {NoStop}%
\bibitem [{\citenamefont {Dumitru}\ \emph {et~al.}(2009)\citenamefont
  {Dumitru}, \citenamefont {Guo},\ and\ \citenamefont
  {Strickland}}]{Dumitru:2009fy}%
  \BibitemOpen
  \bibfield  {author} {\bibinfo {author} {\bibfnamefont {A.}~\bibnamefont
  {Dumitru}}, \bibinfo {author} {\bibfnamefont {Y.}~\bibnamefont {Guo}},\ and\
  \bibinfo {author} {\bibfnamefont {M.}~\bibnamefont {Strickland}},\ }\bibfield
   {title} {\bibinfo {title} {{The Imaginary part of the static gluon
  propagator in an anisotropic (viscous) QCD plasma}},\ }\href
  {https://doi.org/10.1103/PhysRevD.79.114003} {\bibfield  {journal} {\bibinfo
  {journal} {Phys. Rev. D}\ }\textbf {\bibinfo {volume} {79}},\ \bibinfo
  {pages} {114003} (\bibinfo {year} {2009})},\ \Eprint
  {https://arxiv.org/abs/0903.4703} {arXiv:0903.4703 [hep-ph]} \BibitemShut
  {NoStop}%
\bibitem [{\citenamefont {Carrington}\ \emph {et~al.}(1999)\citenamefont
  {Carrington}, \citenamefont {Hou},\ and\ \citenamefont
  {Thoma}}]{Carrington:1997sq}%
  \BibitemOpen
  \bibfield  {author} {\bibinfo {author} {\bibfnamefont {M.~E.}\ \bibnamefont
  {Carrington}}, \bibinfo {author} {\bibfnamefont {D.-f.}\ \bibnamefont
  {Hou}},\ and\ \bibinfo {author} {\bibfnamefont {M.~H.}\ \bibnamefont
  {Thoma}},\ }\bibfield  {title} {\bibinfo {title} {{Equilibrium and
  nonequilibrium hard thermal loop resummation in the real time formalism}},\
  }\href {https://doi.org/10.1007/s100520050412} {\bibfield  {journal}
  {\bibinfo  {journal} {Eur. Phys. J. C}\ }\textbf {\bibinfo {volume} {7}},\
  \bibinfo {pages} {347} (\bibinfo {year} {1999})},\ \Eprint
  {https://arxiv.org/abs/hep-ph/9708363} {arXiv:hep-ph/9708363} \BibitemShut
  {NoStop}%
\bibitem [{\citenamefont {Carrington}\ \emph {et~al.}(1998)\citenamefont
  {Carrington}, \citenamefont {Hou},\ and\ \citenamefont
  {Thoma}}]{Carrington:1998jj}%
  \BibitemOpen
  \bibfield  {author} {\bibinfo {author} {\bibfnamefont {M.~E.}\ \bibnamefont
  {Carrington}}, \bibinfo {author} {\bibfnamefont {D.-f.}\ \bibnamefont
  {Hou}},\ and\ \bibinfo {author} {\bibfnamefont {M.~H.}\ \bibnamefont
  {Thoma}},\ }\bibfield  {title} {\bibinfo {title} {{Ward identities in
  nonequilibrium QED}},\ }\href {https://doi.org/10.1103/PhysRevD.58.085025}
  {\bibfield  {journal} {\bibinfo  {journal} {Phys. Rev. D}\ }\textbf {\bibinfo
  {volume} {58}},\ \bibinfo {pages} {085025} (\bibinfo {year} {1998})},\
  \Eprint {https://arxiv.org/abs/hep-th/9801103} {arXiv:hep-th/9801103}
  \BibitemShut {NoStop}%
\bibitem [{\citenamefont {Mrowczynski}\ \emph {et~al.}(2017)\citenamefont
  {Mrowczynski}, \citenamefont {Schenke},\ and\ \citenamefont
  {Strickland}}]{Mrowczynski:2016etf}%
  \BibitemOpen
  \bibfield  {author} {\bibinfo {author} {\bibfnamefont {S.}~\bibnamefont
  {Mrowczynski}}, \bibinfo {author} {\bibfnamefont {B.}~\bibnamefont
  {Schenke}},\ and\ \bibinfo {author} {\bibfnamefont {M.}~\bibnamefont
  {Strickland}},\ }\bibfield  {title} {\bibinfo {title} {{Color instabilities
  in the quark\textendash{}gluon plasma}},\ }\href
  {https://doi.org/10.1016/j.physrep.2017.03.003} {\bibfield  {journal}
  {\bibinfo  {journal} {Phys. Rept.}\ }\textbf {\bibinfo {volume} {682}},\
  \bibinfo {pages} {1} (\bibinfo {year} {2017})},\ \Eprint
  {https://arxiv.org/abs/1603.08946} {arXiv:1603.08946 [hep-ph]} \BibitemShut
  {NoStop}%
\bibitem [{\citenamefont {Kasmaei}\ \emph {et~al.}(2016)\citenamefont
  {Kasmaei}, \citenamefont {Nopoush},\ and\ \citenamefont
  {Strickland}}]{Kasmaei:2016apv}%
  \BibitemOpen
  \bibfield  {author} {\bibinfo {author} {\bibfnamefont {B.~S.}\ \bibnamefont
  {Kasmaei}}, \bibinfo {author} {\bibfnamefont {M.}~\bibnamefont {Nopoush}},\
  and\ \bibinfo {author} {\bibfnamefont {M.}~\bibnamefont {Strickland}},\
  }\bibfield  {title} {\bibinfo {title} {{Quark self-energy in an ellipsoidally
  anisotropic quark-gluon plasma}},\ }\href
  {https://doi.org/10.1103/PhysRevD.94.125001} {\bibfield  {journal} {\bibinfo
  {journal} {Phys. Rev. D}\ }\textbf {\bibinfo {volume} {94}},\ \bibinfo
  {pages} {125001} (\bibinfo {year} {2016})},\ \Eprint
  {https://arxiv.org/abs/1608.06018} {arXiv:1608.06018 [hep-ph]} \BibitemShut
  {NoStop}%
\bibitem [{\citenamefont {Schwinger}(1951)}]{Schwinger:1951nm}%
  \BibitemOpen
  \bibfield  {author} {\bibinfo {author} {\bibfnamefont {J.~S.}\ \bibnamefont
  {Schwinger}},\ }\bibfield  {title} {\bibinfo {title} {{On gauge invariance
  and vacuum polarization}},\ }\href {https://doi.org/10.1103/PhysRev.82.664}
  {\bibfield  {journal} {\bibinfo  {journal} {Phys. Rev.}\ }\textbf {\bibinfo
  {volume} {82}},\ \bibinfo {pages} {664} (\bibinfo {year} {1951})}\BibitemShut
  {NoStop}%
\bibitem [{\citenamefont {Shovkovy}(2013)}]{Shovkovy:2012zn}%
  \BibitemOpen
  \bibfield  {author} {\bibinfo {author} {\bibfnamefont {I.~A.}\ \bibnamefont
  {Shovkovy}},\ }\bibfield  {title} {\bibinfo {title} {{Magnetic Catalysis: A
  Review}},\ }\href {https://doi.org/10.1007/978-3-642-37305-3_2} {\bibfield
  {journal} {\bibinfo  {journal} {Lect. Notes Phys.}\ }\textbf {\bibinfo
  {volume} {871}},\ \bibinfo {pages} {13} (\bibinfo {year} {2013})},\ \Eprint
  {https://arxiv.org/abs/1207.5081} {arXiv:1207.5081 [hep-ph]} \BibitemShut
  {NoStop}%
\bibitem [{\citenamefont {Fukushima}\ \emph {et~al.}(2016)\citenamefont
  {Fukushima}, \citenamefont {Hattori}, \citenamefont {Yee},\ and\
  \citenamefont {Yin}}]{Fukushima:2015wck}%
  \BibitemOpen
  \bibfield  {author} {\bibinfo {author} {\bibfnamefont {K.}~\bibnamefont
  {Fukushima}}, \bibinfo {author} {\bibfnamefont {K.}~\bibnamefont {Hattori}},
  \bibinfo {author} {\bibfnamefont {H.-U.}\ \bibnamefont {Yee}},\ and\ \bibinfo
  {author} {\bibfnamefont {Y.}~\bibnamefont {Yin}},\ }\bibfield  {title}
  {\bibinfo {title} {{Heavy Quark Diffusion in Strong Magnetic Fields at Weak
  Coupling and Implications for Elliptic Flow}},\ }\href
  {https://doi.org/10.1103/PhysRevD.93.074028} {\bibfield  {journal} {\bibinfo
  {journal} {Phys. Rev. D}\ }\textbf {\bibinfo {volume} {93}},\ \bibinfo
  {pages} {074028} (\bibinfo {year} {2016})},\ \Eprint
  {https://arxiv.org/abs/1512.03689} {arXiv:1512.03689 [hep-ph]} \BibitemShut
  {NoStop}%
\bibitem [{\citenamefont {Bazavov}\ \emph {et~al.}(2012)\citenamefont
  {Bazavov}, \citenamefont {Brambilla}, \citenamefont {Garcia~i Tormo},
  \citenamefont {Petreczky}, \citenamefont {Soto},\ and\ \citenamefont
  {Vairo}}]{Bazavov:2012ka}%
  \BibitemOpen
  \bibfield  {author} {\bibinfo {author} {\bibfnamefont {A.}~\bibnamefont
  {Bazavov}}, \bibinfo {author} {\bibfnamefont {N.}~\bibnamefont {Brambilla}},
  \bibinfo {author} {\bibfnamefont {X.}~\bibnamefont {Garcia~i Tormo}},
  \bibinfo {author} {\bibfnamefont {P.}~\bibnamefont {Petreczky}}, \bibinfo
  {author} {\bibfnamefont {J.}~\bibnamefont {Soto}},\ and\ \bibinfo {author}
  {\bibfnamefont {A.}~\bibnamefont {Vairo}},\ }\bibfield  {title} {\bibinfo
  {title} {{Determination of $\alpha_s$ from the QCD static energy}},\ }\href
  {https://doi.org/10.1103/PhysRevD.86.114031} {\bibfield  {journal} {\bibinfo
  {journal} {Phys. Rev. D}\ }\textbf {\bibinfo {volume} {86}},\ \bibinfo
  {pages} {114031} (\bibinfo {year} {2012})},\ \Eprint
  {https://arxiv.org/abs/1205.6155} {arXiv:1205.6155 [hep-ph]} \BibitemShut
  {NoStop}%
\bibitem [{\citenamefont {Haque}\ \emph {et~al.}(2014)\citenamefont {Haque},
  \citenamefont {Bandyopadhyay}, \citenamefont {Andersen}, \citenamefont
  {Mustafa}, \citenamefont {Strickland},\ and\ \citenamefont
  {Su}}]{Haque:2014rua}%
  \BibitemOpen
  \bibfield  {author} {\bibinfo {author} {\bibfnamefont {N.}~\bibnamefont
  {Haque}}, \bibinfo {author} {\bibfnamefont {A.}~\bibnamefont
  {Bandyopadhyay}}, \bibinfo {author} {\bibfnamefont {J.~O.}\ \bibnamefont
  {Andersen}}, \bibinfo {author} {\bibfnamefont {M.~G.}\ \bibnamefont
  {Mustafa}}, \bibinfo {author} {\bibfnamefont {M.}~\bibnamefont
  {Strickland}},\ and\ \bibinfo {author} {\bibfnamefont {N.}~\bibnamefont
  {Su}},\ }\bibfield  {title} {\bibinfo {title} {{Three-loop HTLpt
  thermodynamics at finite temperature and chemical potential}},\ }\href
  {https://doi.org/10.1007/JHEP05(2014)027} {\bibfield  {journal} {\bibinfo
  {journal} {JHEP}\ }\textbf {\bibinfo {volume} {05}},\ \bibinfo {pages}
  {027}},\ \Eprint {https://arxiv.org/abs/1402.6907} {arXiv:1402.6907 [hep-ph]}
  \BibitemShut {NoStop}%
\bibitem [{\citenamefont {Bellac}(2011)}]{Bellac:2011kqa}%
  \BibitemOpen
  \bibfield  {author} {\bibinfo {author} {\bibfnamefont {M.~L.}\ \bibnamefont
  {Bellac}},\ }\href {https://doi.org/10.1017/CBO9780511721700} {\emph
  {\bibinfo {title} {{Thermal Field Theory}}}},\ Cambridge Monographs on
  Mathematical Physics\ (\bibinfo  {publisher} {Cambridge University Press},\
  \bibinfo {year} {2011})\BibitemShut {NoStop}%
\bibitem [{\citenamefont {Schenke}\ \emph {et~al.}(2006)\citenamefont
  {Schenke}, \citenamefont {Strickland}, \citenamefont {Greiner},\ and\
  \citenamefont {Thoma}}]{Schenke:2006xu}%
  \BibitemOpen
  \bibfield  {author} {\bibinfo {author} {\bibfnamefont {B.}~\bibnamefont
  {Schenke}}, \bibinfo {author} {\bibfnamefont {M.}~\bibnamefont {Strickland}},
  \bibinfo {author} {\bibfnamefont {C.}~\bibnamefont {Greiner}},\ and\ \bibinfo
  {author} {\bibfnamefont {M.~H.}\ \bibnamefont {Thoma}},\ }\bibfield  {title}
  {\bibinfo {title} {{A Model of the effect of collisions on QCD plasma
  instabilities}},\ }\href {https://doi.org/10.1103/PhysRevD.73.125004}
  {\bibfield  {journal} {\bibinfo  {journal} {Phys. Rev. D}\ }\textbf {\bibinfo
  {volume} {73}},\ \bibinfo {pages} {125004} (\bibinfo {year} {2006})},\
  \Eprint {https://arxiv.org/abs/hep-ph/0603029} {arXiv:hep-ph/0603029}
  \BibitemShut {NoStop}%
\bibitem [{\citenamefont {Jamal}\ \emph {et~al.}(2017)\citenamefont {Jamal},
  \citenamefont {Mitra},\ and\ \citenamefont {Chandra}}]{Jamal:2017dqs}%
  \BibitemOpen
  \bibfield  {author} {\bibinfo {author} {\bibfnamefont {M.~Y.}\ \bibnamefont
  {Jamal}}, \bibinfo {author} {\bibfnamefont {S.}~\bibnamefont {Mitra}},\ and\
  \bibinfo {author} {\bibfnamefont {V.}~\bibnamefont {Chandra}},\ }\bibfield
  {title} {\bibinfo {title} {{Collective excitations of hot QCD medium in a
  quasiparticle description}},\ }\href
  {https://doi.org/10.1103/PhysRevD.95.094022} {\bibfield  {journal} {\bibinfo
  {journal} {Phys. Rev. D}\ }\textbf {\bibinfo {volume} {95}},\ \bibinfo
  {pages} {094022} (\bibinfo {year} {2017})},\ \Eprint
  {https://arxiv.org/abs/1701.06162} {arXiv:1701.06162 [nucl-th]} \BibitemShut
  {NoStop}%
\bibitem [{\citenamefont {Kumar}\ \emph {et~al.}(2018)\citenamefont {Kumar},
  \citenamefont {Jamal}, \citenamefont {Chandra},\ and\ \citenamefont
  {Bhatt}}]{Kumar:2017bja}%
  \BibitemOpen
  \bibfield  {author} {\bibinfo {author} {\bibfnamefont {A.}~\bibnamefont
  {Kumar}}, \bibinfo {author} {\bibfnamefont {M.~Y.}\ \bibnamefont {Jamal}},
  \bibinfo {author} {\bibfnamefont {V.}~\bibnamefont {Chandra}},\ and\ \bibinfo
  {author} {\bibfnamefont {J.~R.}\ \bibnamefont {Bhatt}},\ }\bibfield  {title}
  {\bibinfo {title} {{Collective excitations of a hot anisotropic QCD medium
  with the Bhatnagar-Gross-Krook collisional kernel within an effective
  description}},\ }\href {https://doi.org/10.1103/PhysRevD.97.034007}
  {\bibfield  {journal} {\bibinfo  {journal} {Phys. Rev. D}\ }\textbf {\bibinfo
  {volume} {97}},\ \bibinfo {pages} {034007} (\bibinfo {year} {2018})},\
  \Eprint {https://arxiv.org/abs/1709.01032} {arXiv:1709.01032 [nucl-th]}
  \BibitemShut {NoStop}%
\bibitem [{\citenamefont {Das}\ \emph {et~al.}(2018)\citenamefont {Das},
  \citenamefont {Bandyopadhyay}, \citenamefont {Roy},\ and\ \citenamefont
  {Mustafa}}]{Das:2017vfh}%
  \BibitemOpen
  \bibfield  {author} {\bibinfo {author} {\bibfnamefont {A.}~\bibnamefont
  {Das}}, \bibinfo {author} {\bibfnamefont {A.}~\bibnamefont {Bandyopadhyay}},
  \bibinfo {author} {\bibfnamefont {P.~K.}\ \bibnamefont {Roy}},\ and\ \bibinfo
  {author} {\bibfnamefont {M.~G.}\ \bibnamefont {Mustafa}},\ }\bibfield
  {title} {\bibinfo {title} {{General structure of fermion two-point function
  and its spectral representation in a hot magnetized medium}},\ }\href
  {https://doi.org/10.1103/PhysRevD.97.034024} {\bibfield  {journal} {\bibinfo
  {journal} {Phys. Rev. D}\ }\textbf {\bibinfo {volume} {97}},\ \bibinfo
  {pages} {034024} (\bibinfo {year} {2018})},\ \Eprint
  {https://arxiv.org/abs/1709.08365} {arXiv:1709.08365 [hep-ph]} \BibitemShut
  {NoStop}%
\bibitem [{\citenamefont {Hasan}\ and\ \citenamefont
  {Patra}(2020)}]{Hasan:2020iwa}%
  \BibitemOpen
  \bibfield  {author} {\bibinfo {author} {\bibfnamefont {M.}~\bibnamefont
  {Hasan}}\ and\ \bibinfo {author} {\bibfnamefont {B.~K.}\ \bibnamefont
  {Patra}},\ }\bibfield  {title} {\bibinfo {title} {{Dissociation of heavy
  quarkonia in a weak magnetic field}},\ }\href
  {https://doi.org/10.1103/PhysRevD.102.036020} {\bibfield  {journal} {\bibinfo
   {journal} {Phys. Rev. D}\ }\textbf {\bibinfo {volume} {102}},\ \bibinfo
  {pages} {036020} (\bibinfo {year} {2020})},\ \Eprint
  {https://arxiv.org/abs/2004.12857} {arXiv:2004.12857 [hep-ph]} \BibitemShut
  {NoStop}%
\bibitem [{\citenamefont {Ghosh}\ \emph {et~al.}(2022)\citenamefont {Ghosh},
  \citenamefont {Bandyopadhyay}, \citenamefont {Nilima},\ and\ \citenamefont
  {Ghosh}}]{Ghosh:2022sxi}%
  \BibitemOpen
  \bibfield  {author} {\bibinfo {author} {\bibfnamefont {R.}~\bibnamefont
  {Ghosh}}, \bibinfo {author} {\bibfnamefont {A.}~\bibnamefont
  {Bandyopadhyay}}, \bibinfo {author} {\bibfnamefont {I.}~\bibnamefont
  {Nilima}},\ and\ \bibinfo {author} {\bibfnamefont {S.}~\bibnamefont
  {Ghosh}},\ }\bibfield  {title} {\bibinfo {title} {{Anisotropic tomography of
  heavy quark dissociation by using general propagator structure at finite
  magnetic field}},\ }\href@noop {} {\  (\bibinfo {year} {2022})},\ \Eprint
  {https://arxiv.org/abs/2204.02312} {arXiv:2204.02312 [hep-ph]} \BibitemShut
  {NoStop}%
\end{thebibliography}%

\end{document}